\documentclass[12pt]{iopart}
\expandafter\let\csname equation*\endcsname\relax
\expandafter\let\csname endequation*\endcsname\relax
\usepackage{epsfig}
\usepackage{graphicx}
\usepackage{amssymb}
\usepackage{amsmath}
\usepackage{mathtools}
\usepackage{multicol}
\usepackage [english]{babel}
\usepackage{enumitem}
\usepackage{gensymb}
\usepackage{geometry}
\usepackage{cancel}
\usepackage[titletoc,title]{appendix}
\usepackage{cite}
\usepackage{epstopdf}

\renewcommand\vec\mathbf

\begin{document}
\title{Solution to a collisionless shallow-angle magnetic presheath with kinetic ions}
\author{A Geraldini$^{1,2}$, F I Parra$^{1,2}$ and F Militello$^{2}$}
\address{$^{1}$ Rudolf Peierls Centre for Theoretical Physics, University of Oxford, Oxford, OX1 3PU, UK}
\address{$^2$ CCFE, Culham Science Centre, Abingdon, OX14 3DB, UK}
\ead{alessandro.geraldini@merton.ox.ac.uk}   
 
\begin{abstract}
Using a kinetic model for the ions and adiabatic electrons, we solve a steady state, electron-repelling magnetic presheath in which a uniform magnetic field makes a small angle $\alpha \ll 1$ (in radians) with the wall. 
The presheath characteristic thickness is the typical ion gyroradius $\rho_{\text{i}}$. 
The Debye length $\lambda_{\text{D}}$ and the collisional mean free path of an ion $\lambda_{\text{mfp}}$ satisfy the ordering $\lambda_{\text{D}}  \ll \rho_{\text{i}} \ll \alpha \lambda_{\text{mfp}}$, so a quasineutral and collisionless model is used. We assume that the electrostatic potential is a function only of distance from the wall, and it varies over the scale $\rho_{\text{i}}$. 
Using the expansion in $\alpha \ll 1$, we derive an analytical expression for the ion density that only depends on the ion distribution function at the entrance of the magnetic presheath and the electrostatic potential profile.
Importantly, we have added the crucial contribution of the orbits in the region near the wall.
By imposing the quasineutrality equation, we derive a condition that the ion distribution function must satisfy at the magnetic presheath entrance --- the kinetic equivalent of the Chodura condition. 
Using an ion distribution function at the entrance of the magnetic presheath that satisfies the kinetic Chodura condition, we find numerical solutions for the self-consistent electrostatic potential, ion density and flow across the magnetic presheath for several values of $\alpha$.
Our numerical results also include the distribution of ion velocities at the Debye sheath entrance.
We find that at small values of $\alpha$ there are substantially fewer ions travelling with a large normal component of the velocity into the wall. 
\end{abstract}

\section{Introduction}

In a typical fusion plasma device, the interaction between the confined plasma and the wall of the device happens at specified locations called divertor or limiter targets \cite{Stangeby-book}. 
The magnetic field usually makes a small angle $\alpha \ll 1$ (in radians) with the surface tangent to the target in order to minimize the heat flux onto the wall materials \cite{Loarte-2007}. 
Hence, an appropriate model of plasma-wall interaction in a fusion device must accurately describe the effect of such small angles. Such a model could be applicable to other areas where plasma-wall interaction is important, such as thrusters \cite{Martinez-1998}, probes \cite{Hutchinson-book} and magnetic filters \cite{Anders-1994-filters, Anders-1995-filters}.  

When a steady-state plasma is in contact with a wall, a potential difference between the bulk plasma and the wall develops which depends on the density and temperature of the plasma and on the current flowing from the plasma to the wall. 
This potential drop forms due to the difference in mobility between ions and electrons, with the electrons usually reaching the wall faster and hence charging it negatively. %We focus on plasma-wall interaction in which the wall is negatively charged and the typical time it takes for electrons to reach the wall is so much less than than the time it takes for ions to reach it that most electrons are repelled by the wall. 
A thin layer of plasma called Debye sheath, with a thickness of several Debye lengths $\lambda_{\text{D}} = \sqrt{e^2 n_{\text{e}}/\epsilon_0 T_{\text{e}}}$, charges positively because of the net loss of electrons to the wall. Here $e$ is the proton charge, $n_{\text{e}}$ is the number density of electrons in the plasma, $\epsilon_0$ is the permittivity of free space and $T_{\text{e}}$ is the electron temperature (measured in energy units throughout this paper). The Debye sheath shields most of the wall potential from the bulk plasma. 
The rest of the potential difference between wall and plasma occurs in a quasineutral presheath, of size $\lambda_{\text{ps}} \gg \lambda_{\text{D}}$. Usually $\lambda_{\text{ps}} \ll L_{\text{s}}$, where $L_{\text{s}}$ is the scale of the device (for example, the minor radius of a tokamak), which implies that the presheath can be treated as a thin boundary layer with respect to the bulk plasma in the device. 

\begin{figure}
\centering
\label{fig-boundary-layers}
\includegraphics[width = 0.7\textwidth]{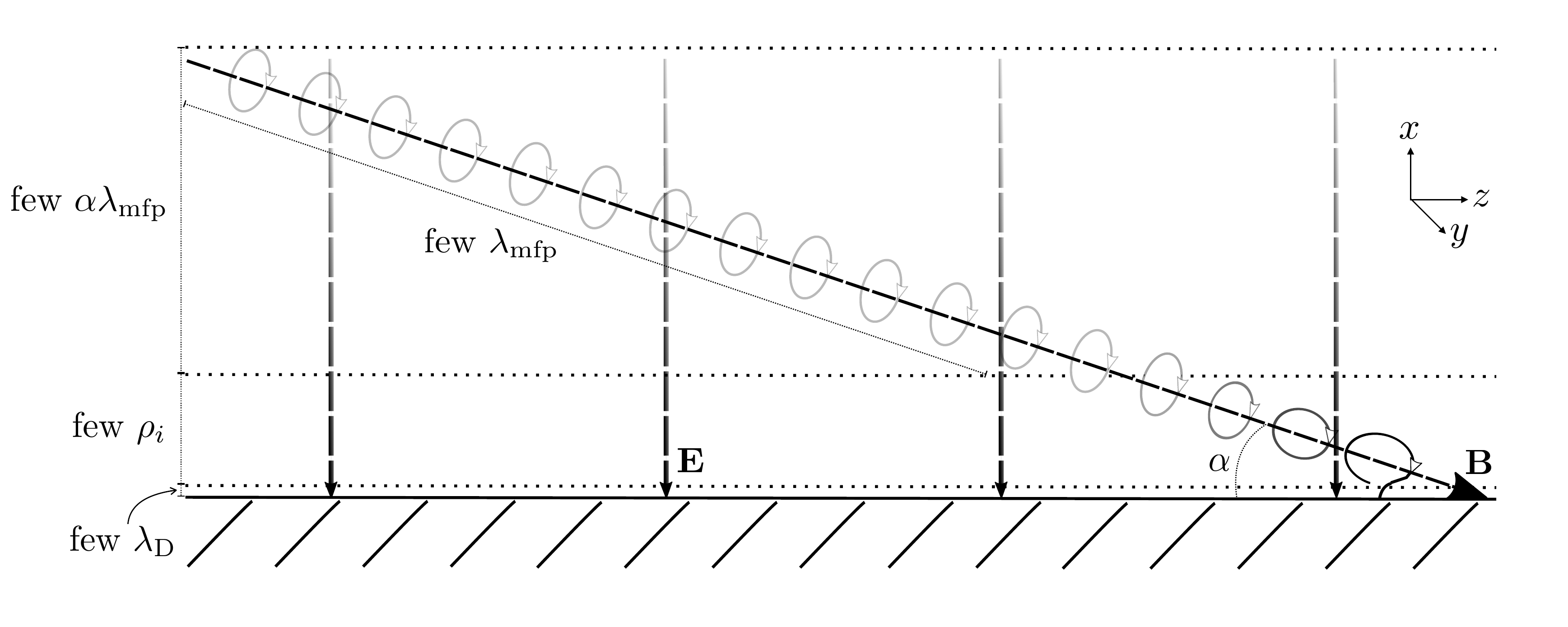}
\caption{Cartoon of ion orbits in the neighbourhood of the divertor target of a tokamak plasma, with $\lambda_{\text{D}} \ll \rho_{\text{i}} \ll \alpha \lambda_{\text{mfp}}$. The orbits have a size $\rho_{\text{i}}$ and are tied to a dashed line representing the magnetic field $\vec{B}$, which is inclined at a small angle $\alpha$ with the wall. The electric field $\vec{E}$ is shown as a dashed vertical line, and is shaded darker nearer to the wall, where it is stronger. Highly distorted orbits in the magnetic presheath are black, while circular orbits in the collisional presheath are light grey.  }
\end{figure}

We consider a presheath in which the ion collisional mean free path $\lambda_{\text{mfp}}$ projected in the direction normal to the wall, $\lambda_{\text{mfp}} \sin \alpha \simeq \alpha \lambda_{\text{mfp}}$, is much larger than the ion gyroradius $\rho_{\text{i}}$. Hence, we assume 
\begin{align} \label{scale-sep}
\lambda_{\text{D}} \ll \rho_{\text{i}} \ll  \alpha \lambda_{\text{mfp}} \text{.}
\end{align}
 This is consistent with the value of these quantities near a divertor target in attached divertor regimes: $\lambda_{\text{D}} \sim 0.02 \text{mm}$, $\rho_{\text{i}} \sim 0.7 \text{mm}$, $\alpha \lambda_{\text{mfp}} \sim 100 \text{mm}$ \cite{Geraldini-2017}. 
In detached regimes, the magnetic presheath can become collisional because the mean free path $\lambda_{\text{mfp}}$ (for both Coulomb and charge-exchange collisions) can be substantially smaller than the quoted value \cite{Tskhakaya-2017}.
 With the scale separation (\ref{scale-sep}), we can split the boundary layer into three separate layers: a collisional presheath of size $\alpha \lambda_{\text{mfp}}$, a collisionless magnetic presheath of size $\rho_{\text{i}}$ and a non-neutral Debye sheath \cite{Loizu-2012}. 
The ion motion is of a very different nature in the three layers: in the collisional layer ions are magnetized in circular gyro-orbits and stream parallel to the magnetic field, in the magnetic presheath ion gyro-orbits are distorted % initially and then demagnetized 
by increasingly strong electric fields, and finally in the Debye sheath ions are accelerated towards the wall by an electric force much larger than the magnetic force.
A cartoon of the ion motion across all boundary layers is shown in Figure \ref{fig-boundary-layers}.

In this paper we focus on the magnetic presheath, which was first studied by Chodura \cite{Chodura-1982}. 
By using fluid equations for the electrons and ions, which are valid provided ions are much colder than electrons, Chodura found a solution for the electrostatic potential and ion flow across the magnetic presheath. 
He also found that, for cold ions, the ion flow parallel to the magnetic field at the presheath entrance must at least be equal to the Bohm speed
\begin{align} \label{Bohm-speed}
v_{\text{B}} = \sqrt{\frac{ZT_{\text{e}}}{m_{\text{i}}}} \text{,}
\end{align} 
which is known as the Chodura (or Bohm-Chodura) condition \cite{Chodura-1982, Riemann-1994}. 
In equation (\ref{Bohm-speed}), $Z$ is the ionic charge state of the ion species, $T_{\text{e}}$ is the electron temperature and $m_{\text{i}}$ is the ion mass.
%Bohm \cite{Bohm-1949} first obtained a similar condition at the entrance of the Debye sheath, where he concluded that the ions enter with a normal velocity component at least equal to $v_{\text{B}}$.
%Hence, the ions are expected to enter the magnetic presheath with a parallel velocity component equal to (at least) the sound speed, which is then
%Hence, the ion fluid velocity in the magnetic presheath is turned by the strong electric field from being parallel to the magnetic field to being normal to the wall. % (and enter the Debye sheath).
%Chodura also predicted that the electrostatic potential drop across the magnetic presheath is approximately $\left( T_{\text{e}} / e \right) \ln\left( 1 / \sin \alpha \right)$.
%His predictions were supported by PIC simulations with a kinetic model.
%By assuming that the ions were much colder than electrons, Chodura could use fluid equations for the electrons and ions, and found a solution for the electrostatic potential and ion flow across the magnetic presheath. 
%Chodura also found that the ion fluid velocity parallel to the magnetic field at the presheath entrance must at least be equal to the Bohm speed
%\begin{align} \label{Bohm-speed}
%v_{\text{B}} = \sqrt{\frac{ZT_{\text{e}}}{m_{\text{i}}}} \text{,}
%\end{align} 
%which is known as Chodura's condition \cite{Chodura-1982, Riemann-1994}. In equation (\ref{Bohm-speed}), $Z$ is the ionic charge state of the ion species and $m_{\text{i}}$ is the ion mass.

Chodura's results prepared the ground for several other studies of the magnetic presheath, many of which also used fluid equations to model the ion species \cite{Ahedo-1997, Ahedo-2009, Ahedo-Carralero-2009, Riemann-1994}. However, the assumption that a fluid model is adequate for ions in the magnetic presheath is not well motivated, because their Larmor orbits are highly distorted with a characteristic radius equal to the characteristic thickness of the layer \cite{Siddiqui-Hershkowitz-2016}. The fluid model can only correctly describe cold ions with $T_{\text{i}} \ll T_{\text{e}}$, where $T_{\text{i}}$ is the ion temperature, because such ions can be treated as mono-energetic. 
Treatments of the magnetic presheath which take into account the kinetic nature of the ions are less common and are mostly numerical \cite{Tskhakaya-2002, Tskhakaya-2003, Tskhakaya-2004, Kovacic-2009, Khaziev-Curreli-2015, Devaux-2006, Coulette-2014, Coulette-Manfredi-2016}, although some analytical contributions have been made \cite{Cohen-Ryutov-1998, Cohen-Ryutov-2004-sheath-boundary-conditions, Daube-Riemann-1999, Holland-Fried-Morales-1993, Daybelge-Bein-1981}. 
In this paper, we extend the analytical work carried out in \cite{Geraldini-2017} and we numerically solve a grazing-angle collisionless magnetic presheath assuming Boltzmann electrons and using a fully kinetic model for the ions. 
The wall is assumed to be perfectly absorbing and non-emitting.

As in references \cite{Geraldini-2017, Cohen-Ryutov-1998, Cohen-Ryutov-2004-sheath-boundary-conditions}, we perform an asymptotic expansion in $\alpha$ of the ion trajectories in the magnetic presheath. 
This approach is equivalent to a ``gyrokinetic'' separation of timescales. 
Most of the time, an ion trajectory is well approximated to lowest order in $\alpha$ by a non-circular periodic orbit with a fast gyration timescale $\sim 1/\Omega$. 
Here, $\Omega = ZeB/m_{\text{i}}$ is the typical ion gyrofrequency and $B$ is the magnitude of the magnetic field.
To higher order, the trajectory is a sequence of approximately ``closed'' orbits: it can be described by varying some of the parameters of the periodic motion over the long characteristic time $1/\alpha \Omega$. 
In reference \cite{Geraldini-2017} we obtained an expression for the density of ions in approximately periodic orbits in the magnetic presheath.
A short time $\sim 1/\Omega$ before the ion reaches the wall, its trajectory cannot be considered approximately periodic and is therefore an ``open'' orbit.
% ions gyrate around the magnetic field with a characteristic period of $1/\Omega$ while moving towards the wall in a characteristic time $1/\alpha \Omega$. Here $\Omega = ZeB/m_{\text{i}}$ is the frequency of gyration of an ion moving in a system with a constant magnetic field $B$. 
%In a previous paper \cite{Geraldini-2017}, we studied in detail the motion of ions in approximately closed orbits in a more general magnetic presheath system (with weak turbulent electric fields parallel to the wall) and obtained an expression for the ion density. 
%Here, we extend that work by studying the open orbit piece of the ion trajectory in the simple case where no turbulence is present. 
In this work, we show that the contribution to the density of ions in open orbits is crucial and we derive an analytical expression for it. 

Using the equations presented in this paper, we numerically find a self-consistent solution for the electrostatic potential in the magnetic presheath. We rely on a boundary condition at the magnetic presheath entrance that satisfies a condition, derived herein, which is the kinetic generalization of Chodura's condition \cite{Chodura-1982}. The numerical solution we obtain for the electrostatic potential is used to evaluate the ion density and flow across the magnetic presheath. %, as well as the ion distribution function reaching the Debye sheath. 
Moreover, we obtain the distribution of ion velocities at the entrance of the Debye sheath, and find that the kinetic Bohm condition \cite{Riemann-review} is satisfied, as we also predict analytically. The results of our model indicate that the number of ions entering the Debye sheath travelling with a large normal component of the velocity towards the wall is substantially reduced at smaller values of the angle $\alpha$. %, which could reduce damage to the tiles of divertor targets. 

This paper is structured as follows. In Section 2, we explain the orderings that we use in our model. In Section 3, we expand the ion trajectories in the small parameter $\alpha \ll 1$. %, is solved to lowest order over the fast timescale $1/\Omega$. 
In Section 4 we obtain an expression for the density of ions across the magnetic presheath in terms of their distribution function at the magnetic presheath entrance, including the contribution of open orbits. 
%We also obtain an expression for the distribution function of ions reaching the Debye sheath entrance. 
In Section 5 we analytically expand the quasineutrality equation near the magnetic presheath entrance and near the Debye sheath entrance.
%This is the generalization of the well-known Chodura (or Bohm-Chodura) condition \cite{Chodura-1982} to include the effect of kinetic ions. 
One of the analytical results of these expansions is a solvability condition that the ion distribution function must satisfy at the magnetic presheath entrance. In Section 6 we state the ion distribution function used as an entrance boundary condition, explain the numerical procedure used to solve the quasineutrality equation and present the numerical solutions. In Section 7, we summarize our main results and make some concluding remarks.

\section{Orderings and assumptions}

In this work, we consider a steady state plasma at $x \geqslant 0$, which is magnetized by a uniform and constant magnetic field $\vec{B} = B \cos \alpha \hat{\vec{z}} - B \sin \alpha \hat{\vec{x}} $, where $B = |\vec{B}|$, $ \hat{\vec{x}}, \hat{\vec{y}}, \hat{\vec{z}}  $ are the unit vectors along the $ x,y, z$ axes and $\alpha$ is a small angle (see \cite{Geraldini-2017} for a discussion of when $\vec{B}$ can be assumed to be constant in time and space). The coordinate system we use is shown in Figure \ref{fig-boundary-layers}. We assume no gradients in the two directions parallel to the wall, $y$ and $z$ (note that in \cite{Geraldini-2017} we allowed for gradients in $y$). Distances from the wall are ordered 
\begin{align} \label{x-order}
x \sim \rho_{\text{i}} = \frac{v_{\text{t,i}}}{\Omega}
\end{align}
and the magnitude of the ion velocity $\vec{v} = \left( v_x, v_y, v_z \right) $ is ordered
\begin{align} \label{v-order}
\left| \vec{v} \right| \sim v_{\text{t,i}} \text{,}
\end{align}
where $v_{\text{t,i}} = \sqrt{2T_{\text{i}}/m_{\text{i}}} $. % and $T_{\text{i}}$ is the ion temperature.
The system is solved to lowest order under the assumption in (\ref{scale-sep}), which implies that  
%The inequality $ \lambda_D / \rho_{\text{i}} \ll  1$ allows us to assume a quasineutral magnetic presheath, because Debye shielding of the wall potential relative to the quasineutral plasma occurs in a very thin region at $x \sim \lambda_{\text{D}} \ll \rho_{\text{i}}$. 
$x=0$ is the interface between magnetic presheath and Debye sheath, $\lambda_{\text{D}} \ll x \ll \rho_{\text{i}} $,  
%The inequality $\rho_{\text{i}} / \alpha \lambda_{\text{mfp}}  \ll 1$ allows us to assume a collisionless system because the collisional length scale is so much larger than the system length scale that collisional effects can determine nothing more than the boundary condition at the magnetic presheath entrance, $x\rightarrow \infty$. 
%Again, this location is really the interface between the magnetic presheath and the collisional layer, $\rho_{\text{i}} \ll x \ll \alpha \lambda_{\text{mfp}} $.
while $x \rightarrow \infty$ is the interface between the magnetic presheath and the collisional layer, $\rho_{\text{i}} \ll x \ll \alpha \lambda_{\text{mfp}} $.
 Splitting the boundary layer in different scale separated regions and using a matching procedure to join them is common in studies of the plasma-wall boundary, and has been justified in reference \cite{Riemann-2005-matching}. 

The fact that the magnetic field is assumed constant in time implies that the electric field can be expressed in terms of the gradient of an electrostatic potential, $\vec{E} = -\nabla \phi$. We define the electrostatic potential $\phi (x)$ such that $\phi \rightarrow 0$ at $x\rightarrow \infty$ and order it as large as the electron temperature $T_{\text{e}}$ (consistent with \cite{Chodura-1982}),
\begin{align}
\phi \left( x \right)  \sim \frac{T_{\text{e}}}{e} \text{.}
\end{align}
The electric field is $\vec{E} = - \phi' (x) \hat{\vec{x}}$, with % = |\vec{E}|$ 
\begin{align}
\phi' (x) \equiv \frac{d\phi}{dx}(x) \sim \frac{T_{\text{e}}}{e\rho_{\text{i}}}  \sim v_{\text{t,i}} B \text{.}
\end{align}
The second ordering arises because the ion and electron temperatures are ordered of similar sizes, $T_{\text{i}} \sim T_{\text{e}}$.

The angle $\alpha$ is ordered
\begin{align} \label{order-alpha}
\sqrt{\frac{m_{\text{e}}}{m_{\text{i}}}} \simeq 0.02 \ll \alpha \ll 1 \text{,}
\end{align}
where $m_{\text{e}}$ is the electron mass and the estimate for the square root of mass ratio is obtained using a Deuterium ion.
We assume $ \alpha \gg \sqrt{m_{\text{e}} / m_{\text{i}}} $ to ensure that the wall is electron-repelling \cite{Geraldini-2017}, which justifies using a Boltzmann distribution for the electron density,
\begin{align} \label{ne-Boltzmann}
n_{\text{e}} \left( x \right) = n_{\text{e}\infty} \exp\left(\frac{e\phi \left( x \right) }{T_{\text{e}}} \right) \text{.}
\end{align}
Here, $n_{\text{e}\infty}$ is the electron density at $x \rightarrow \infty$. 
In practice, we obtain numerical results for a range of angles that satisfy $\alpha \gtrsim \sqrt{m_{\text{e}}/m_{\text{i}}}$, while assuming for simplicity that (\ref{ne-Boltzmann}) holds even when $\alpha \sim \sqrt{m_{\text{e}}/m_{\text{i}}}$. Provided that the wall remains electron-repelling, square root of mass ratio corrections can be included by using the expression for the electron density derived in \cite{Ingold-1972} instead of equation (\ref{ne-Boltzmann}).

\section{Ion trajectories}

Here, we exploit the smallness of $\alpha$ to asymptotically expand the ion trajectories. The equations of motion for an ion moving in the collisionless magnetic presheath are \cite{Geraldini-2017}
\begin{align}
\label{x-EOM-exact}
\dot{x} = v_x \text{,}
\end{align}
\begin{align}
\label{y-EOM-exact}
\dot{y} = v_y \text{,}
\end{align}
\begin{align}
\label{z-EOM-exact}
\dot{z} = v_z \text{,}
\end{align}
\begin{align}
\label{vx-EOM-exact}
\dot{v}_x = -\frac{\Omega \phi'(x) }{B} + \Omega v_{y}\cos\alpha \text{,}
\end{align}
\begin{align}
\label{vy-EOM-exact}
\dot{v}_y =  - \Omega v_{x}\cos\alpha - \Omega v_{z}\sin\alpha \text{,}
\end{align}
\begin{align}\label{vz-EOM-exact}
\dot{v}_z = \Omega v_{y}\sin\alpha \text{,}
\end{align}
where a dot $\dot{}$ denotes a time derivative, $d / dt$. 

This section is structured as follows. Section \ref{subsec-orbit-parameters} is devoted to obtaining the constants of motion resulting from equations (\ref{x-EOM-exact})-(\ref{vz-EOM-exact}) with $\alpha =0$, which are called orbit parameters. 
We express the ion velocity in terms of the instantaneous position and the orbit parameters, using an ``effective potential''.
In Section \ref{subsec-effpot-types} we introduce two distinct types of effective potential curves. %which lead to qualitatively different ion orbits. 
In Section \ref{subsec-closed-orbits} we study ``closed'' orbits, which are periodic solutions to equations (\ref{x-EOM-exact})-(\ref{vz-EOM-exact}) with $\alpha = 0$. Their characteristic period is $1/\Omega$. The main effect of $\alpha \neq 0$ is to break the exact periodicity by making the orbit parameters vary over a characteristic time $1/ \alpha \Omega \gg 1/\Omega$.
 %both in terms of the instantaneous particle position and in terms of the closed orbit which instantaneously approximates the ion trajectory.
%The ion motion can then be thought of a sequence of a large number of approximately closed orbits that slowly move across the magnetic presheath towards the wall. 
A slow variation of the parameters of periodic motion leads to the existence of an adiabatic invariant $\mu$, a quantity that the ion conserves to lowest order in $\alpha$ over the long timescale $1/\alpha \Omega$ \cite{Geraldini-2017, Cohen-Ryutov-1998}. 
%Total energy $U$ is exactly conserved.  
%therefore ions traversing the magnetic presheath conserve two quantities to (at least) lowest order. 
In Section \ref{subsec-appclosedorbits} we study the real ion trajectories, which consist of a sequence of approximately closed orbits, quantify the variation of the orbit parameters to first order in $\alpha$ and write the adiabatic invariant. 
A time $\sim 1/\Omega$ before the ion reaches the wall, the ion is considered in an ``open'' orbit. 
In Section \ref{subsec-openorbits}, we define an open orbit and obtain the conditions that orbit parameters must satisfy for an ion to be in an open orbit.

\subsection{Orbit parameters} \label{subsec-orbit-parameters}

Setting $\alpha = 0$, equations (\ref{vx-EOM-exact})-(\ref{vz-EOM-exact}) become
\begin{align}
\label{vx-EOM-zero}
\dot{v}_x = -\frac{\Omega \phi'(x)}{B} + \Omega v_{y} \text{,}
\end{align}
\begin{align}
\label{vy-EOM-zero}
\dot{v}_y =  - \Omega v_{x}  \text{,}
\end{align}
\begin{align}\label{vz-EOM-zero}
\dot{v}_z = 0 \text{.}
\end{align}
Using (\ref{x-EOM-exact}), direct integration of (\ref{vy-EOM-zero}) leads to
\begin{align} \label{xbar-def}
 \bar{x} = \frac{v_y}{\Omega} +  x \sim \rho_{\text{i}} \text{,}
\end{align}
where $\bar{x}$ is the constant of integration which represents the position of an ion orbit. 
Multiplying (\ref{vx-EOM-zero}) by $v_x$ and adding it to (\ref{vy-EOM-zero}) multiplied by $v_y$, we obtain $\dot{U}_{\perp} = 0$, where
\begin{align} \label{Uperp-def}
U_{\perp} = \frac{1}{2} v_x^2 + \frac{1}{2} v_y^2 + \frac{\Omega \phi(x)}{B} \sim v_{\text{t,i}}^2
\end{align}
is the perpendicular energy.
From (\ref{vz-EOM-zero}), the parallel velocity $v_z$ of the ion is a constant of the motion. Adding the parallel kinetic energy $v_z^2 / 2 $ to the perpendicular energy we obtain the total energy, 
\begin{align} \label{U-def}
U = \frac{1}{2} v_x^2 + \frac{1}{2} v_y^2 + \frac{1}{2} v_z^2 + \frac{\Omega \phi(x) }{B}  \sim v_{\text{t,i}}^2 \text{.}
\end{align}
The quantities $\bar{x}, ~ U_{\perp} \text{ and } U$ constitute the three orbit parameters of the ion motion. % with $\alpha = 0$.
When $\alpha =0$ they are exactly conserved, and when $\alpha \ll 1$ they change slowly (except for $U$ which remains constant).

The ion velocity components $v_x, ~ v_y \text{ and } v_z$ can be expressed in terms of the orbit parameters and the instantaneous ion position $x$. Inserting (\ref{xbar-def}) into (\ref{Uperp-def}) and rearranging, we get
\begin{align} \label{vx-Uperp-xbar-x}
v_x = \sigma_x V_x \left( x, \bar{x}, U_{\perp} \right)  \text{ with } V_x \left( x, \bar{x}, U_{\perp} \right) = \sqrt{ 2\left( U_{\perp} - \chi\left( x, \bar{x} \right) \right) } \text{,}
\end{align}
where we introduced $\sigma_x = \pm 1$ to account for the two possible signs of $v_x$, and an effective potential function
\begin{align} \label{chi-def}
\chi \left(x, \bar{x} \right) = \frac{1}{2} \Omega^2 \left(x - \bar{x} \right)^2 + \frac{\Omega \phi(x)}{B} \text{.}
\end{align}
The $y$-component of the velocity is obtained by rearranging equation (\ref{xbar-def}),
\begin{align} \label{vy-xbar-x}
v_y = \Omega \left( \bar{x} - x \right) \text{.}
\end{align} 
The $z$-component of the velocity is obtained by subtracting equation (\ref{Uperp-def}) from (\ref{U-def}), multiplying by $2$ and taking a square root,
\begin{align} \label{vz-U-Uperp}
v_z = \sigma_{\parallel} V_{\parallel} \left( U_{\perp}, U \right) \text{ with } V_{\parallel} \left( U_{\perp}, U \right) = \sqrt{2\left( U - U_{\perp} \right) } \text{,}
\end{align}
where $\sigma_{\parallel} = \pm 1$ is the sign of $v_z$.

\subsection{Types of effective potential curves} \label{subsec-effpot-types}

\begin{figure}
\centering
\includegraphics[width= 0.6\textwidth]{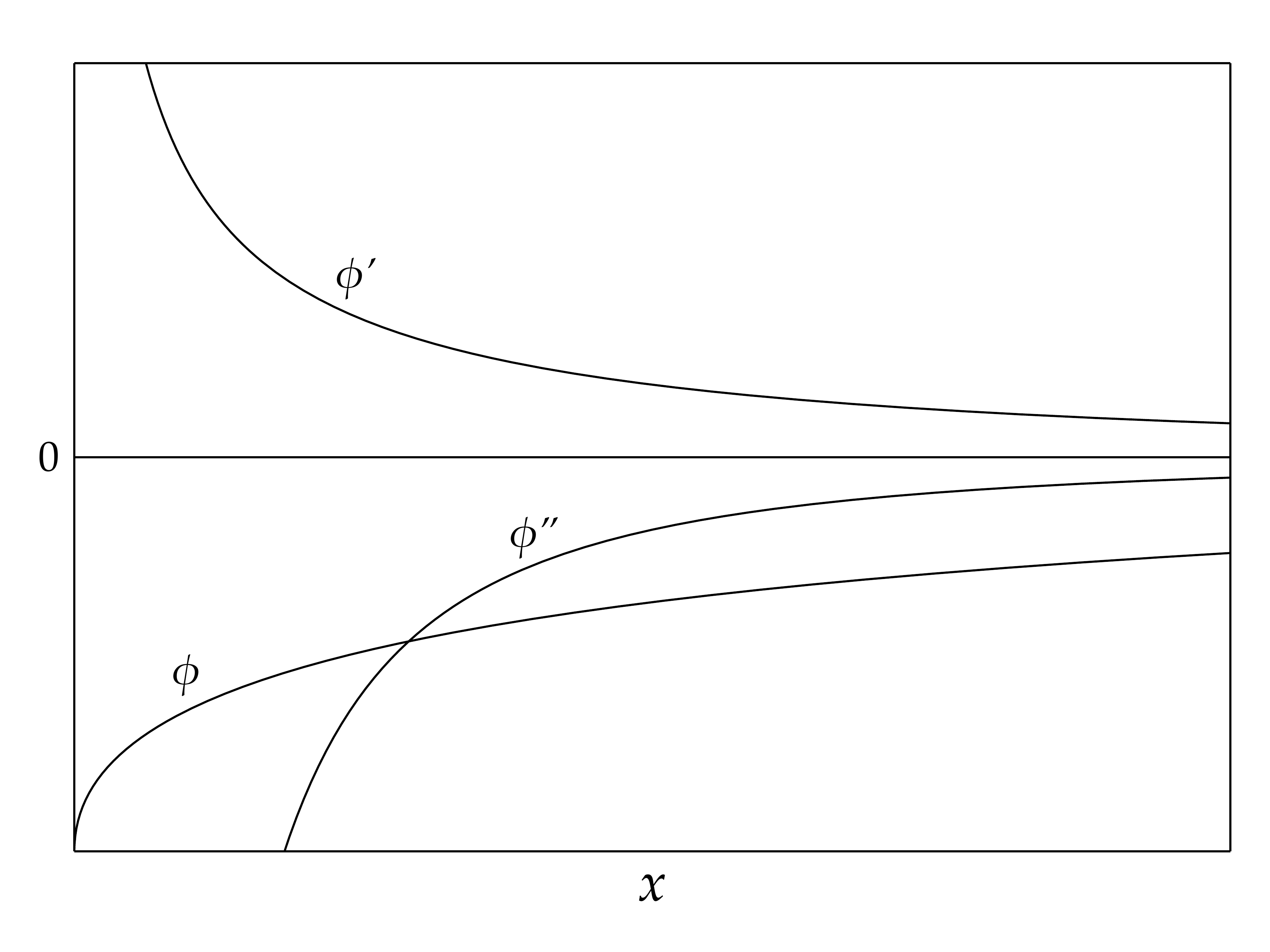}
\caption{An example of a monotonic electrostatic potential profile $\phi(x)$ and its monotonic first and second derivatives $\phi'(x)$ and $\phi''(x)$.
}
\label{fig-phiandderivatives}
\end{figure}

\begin{figure}
\centering
\includegraphics[width= 0.47\textwidth]{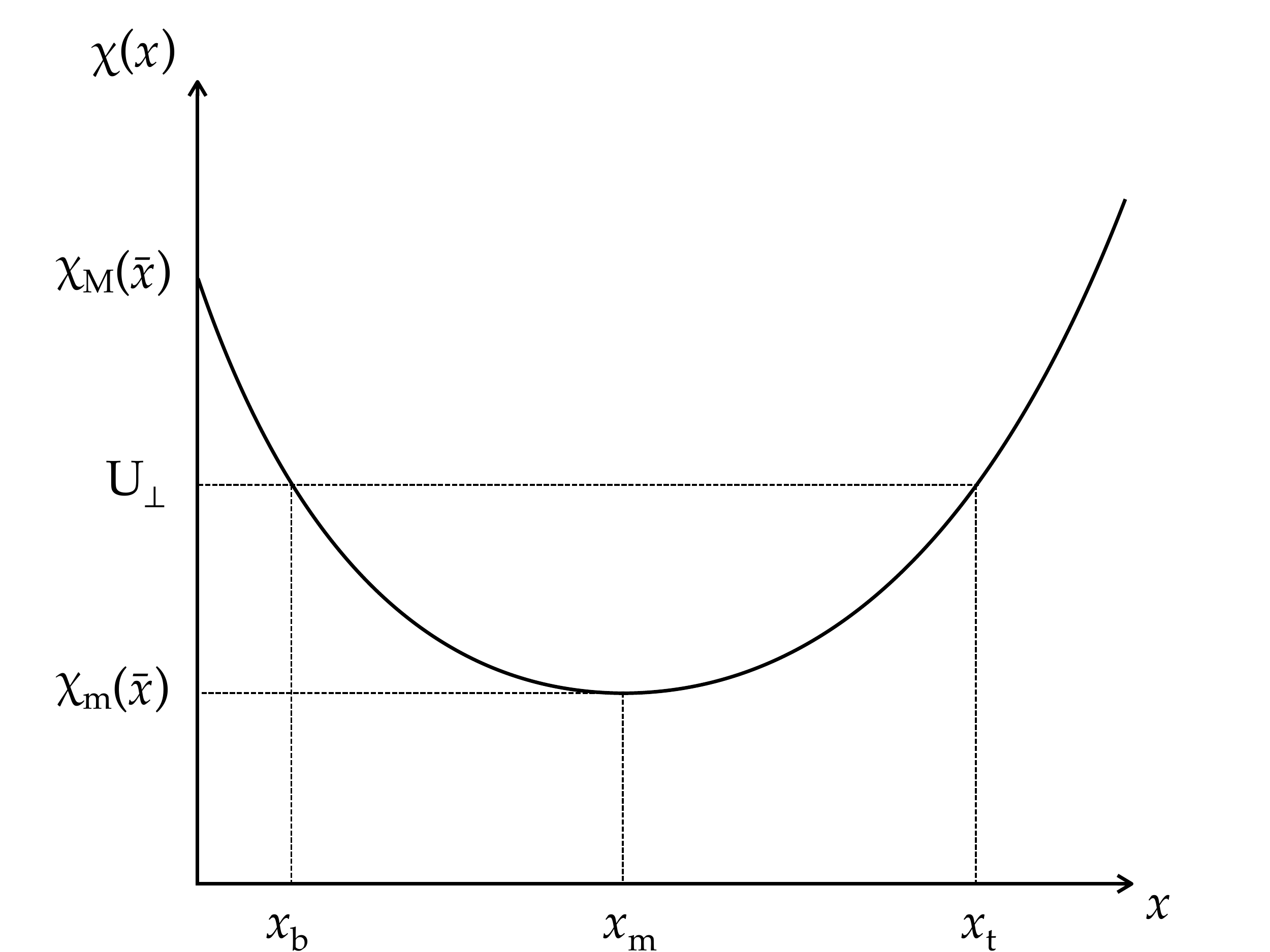}
\includegraphics[width= 0.47\textwidth]{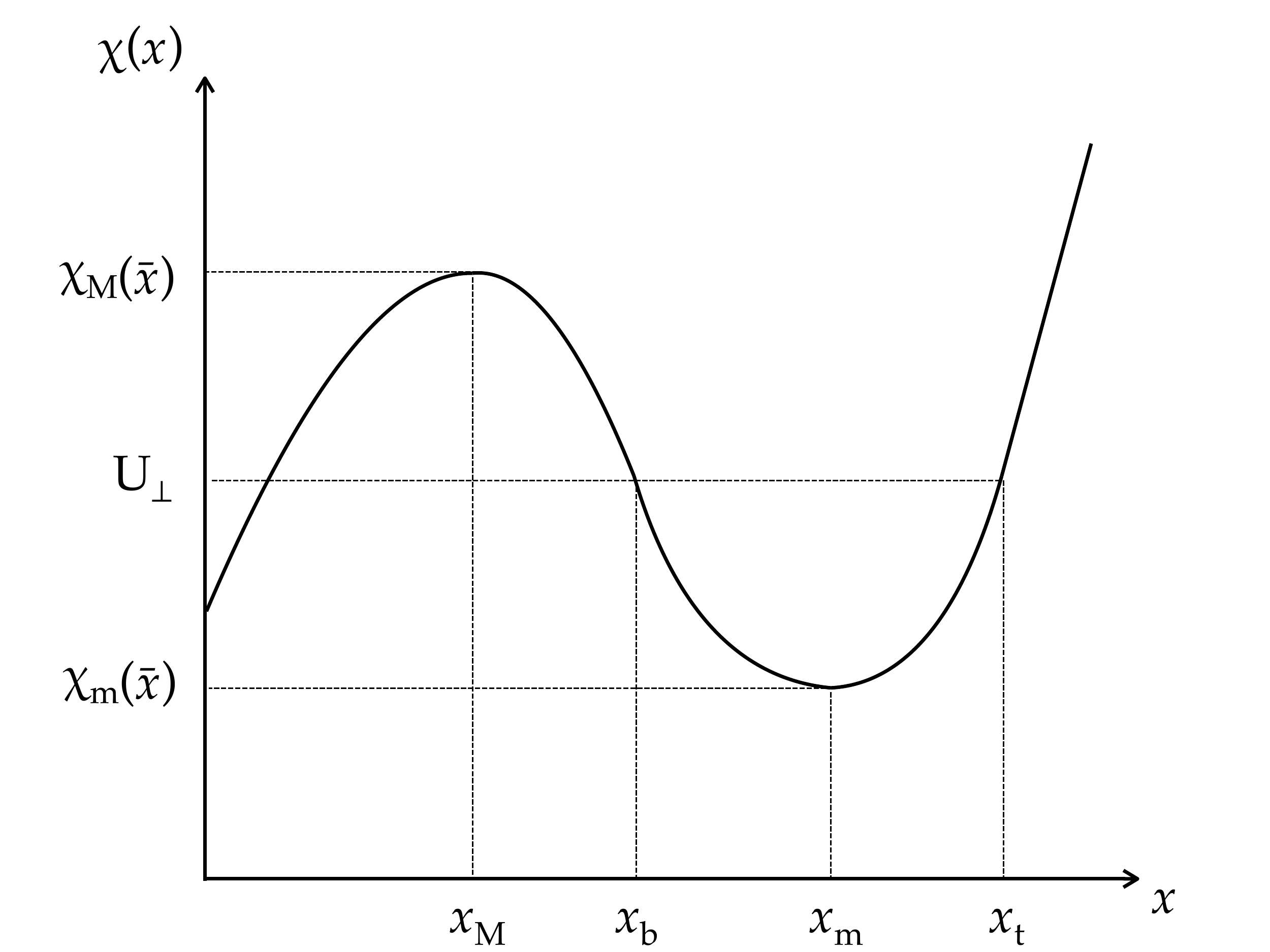}
\caption{Type I (left) and II (right) effective potential curves, both with a stationary minimum at $x=x_{\text{m}}$. A type II curve is characterized by a stationary maximum at $x=x_{\text{M}}$. These curves allow closed orbits for any value of $U_{\perp}$ in the range $\chi_{\text{m}} \left( \bar{x} \right) \leqslant U_{\perp} \leqslant \chi_{\text{M}} \left( \bar{x} \right) $  with bottom and top bounce points at positions $x_{\text{b}}$ and $x_{\text{t}}$.
}
\label{fig-effpotclosed}
\end{figure}

By imposing that $v_x$ be real in equation (\ref{vx-Uperp-xbar-x}), the allowed ion positions must satisfy $U_{\perp} \geqslant \chi \left(x, \bar{x} \right)$.  
%With the aid of Figure \ref{fig-effpotclosed}, one sees that 
A particle moves periodically if, for given values of $U_{\perp}$ and $\bar{x}$, it is trapped around a minimum (with respect to $x$) of the effective potential $\chi(x,\bar{x})$. 
Then, the ion motion is confined between bounce points $x_{\text{b}}$ (bottom) and $x_{\text{t}}$ (top) defined by
\begin{align} \label{closed-orbit-bounce}
U_{\perp} = \chi \left(x_b, \bar{x} \right) =  \chi \left(x_t, \bar{x} \right) \text{ with }  x_{\text{b}} \leqslant x_{\text{t}} \text{.} %0 \leqslant x_{\text{M}} \leqslant  x_b \leqslant x_{\text{m}} \leqslant x_t \text{.}
\end{align}
%Therefore, effective potential minima are a necessary requirement for the presence of closed orbits in the system with $\alpha =0$.
Throughout this work, we assume that the electrostatic potential across the magnetic presheath is such that $\phi (x)$, $\phi'(x) $ and $\phi''(x)$ are all monotonic (our numerical results satisfy these conditions), as shown in Figure \ref{fig-phiandderivatives}. 
Then, for values of $\bar{x}$ for which the effective potential has a stationary minimum, there are two possible types of effective potential $\chi(x, \bar{x})$:
%the curve $\chi \left( x, \bar{x} \right)$ with a stationary minimum for a given $\bar{x}$ can be of two types \cite{Cohen-Ryutov-1998}:
\begin{itemize}
\item a type I effective potential has one stationary minimum at $x_{\text{m}}$, such that $\chi_{\text{m}} \left(\bar{x} \right)  \equiv \chi \left( x_{\text{m}}, \bar{x} \right) $, and \emph{no} stationary maximum --- in this case, it is important to consider the non-stationary local maximum at position $x_{\text{M}} = 0$ with $\chi_{\text{M}} \left( \bar{x} \right) = \chi \left( 0, \bar{x} \right) $;
\item a type II effective potential has two stationary points: one at position $x_{\text{m}}$ which corresponds to a minimum $\chi_{\text{m}} \left(\bar{x} \right)$, and one at position $x_{\text{M}}$ which corresponds to a maximum $\chi_{\text{M}} \left( \bar{x} \right) \equiv \chi \left( x_{\text{M}}, \bar{x} \right)$.
\end{itemize}
These two effective potential types are shown in Figure \ref{fig-effpotclosed}. 
We will refer to the ion trajectories arising due to each curve type as type I and type II orbits \cite{Cohen-Ryutov-1998}.
%From Figure \ref{fig-effpotclosed}, we see that for periodic motion $U_{\perp}$ must lie in the range $\chi_{\text{m}} \left(\bar{x} \right)  \leqslant U_{\perp} \leqslant  \chi_{\text{M}} \left( \bar{x} \right) $.

We proceed to obtain the range of values of $\bar{x}$ for which the effective potential is of either type.
Differentiating equation (\ref{chi-def}) with respect to $x$, we obtain
\begin{align} \label{chi'}
\chi'(x, \bar{x} ) \equiv \frac{\partial \chi}{\partial x} (x, \bar{x}) = \Omega^2 (x - \bar{x} ) + \frac{\Omega \phi'(x)}{B} \text{.}
\end{align} 
%and differentiating again,
%\begin{align} \label{chi''}
%\chi''(x, \bar{x} ) = \Omega^2 + \frac{\Omega \phi''(x)}{B} \text{.}
%\end{align} 
%Note that the second derivative in (\ref{chi''}) is independent of $\bar{x}$.
For type I curves the gradient of the effective potential at $x=0$ must be negative. Hence, from equation (\ref{chi'}), we obtain $-\Omega^2 \bar{x}  + \Omega \phi'(0)/ B < 0$ which leads to the requirement that $\bar{x} > \bar{x}_{\text{m,I}}$ with
\begin{align} \label{xbarmI}
\bar{x}_{\text{m,I}} = \frac{ \phi'(0) }{ \Omega B } \text{.} 
\end{align} 

\begin{figure}
\centering
\includegraphics[width= 0.8\textwidth]{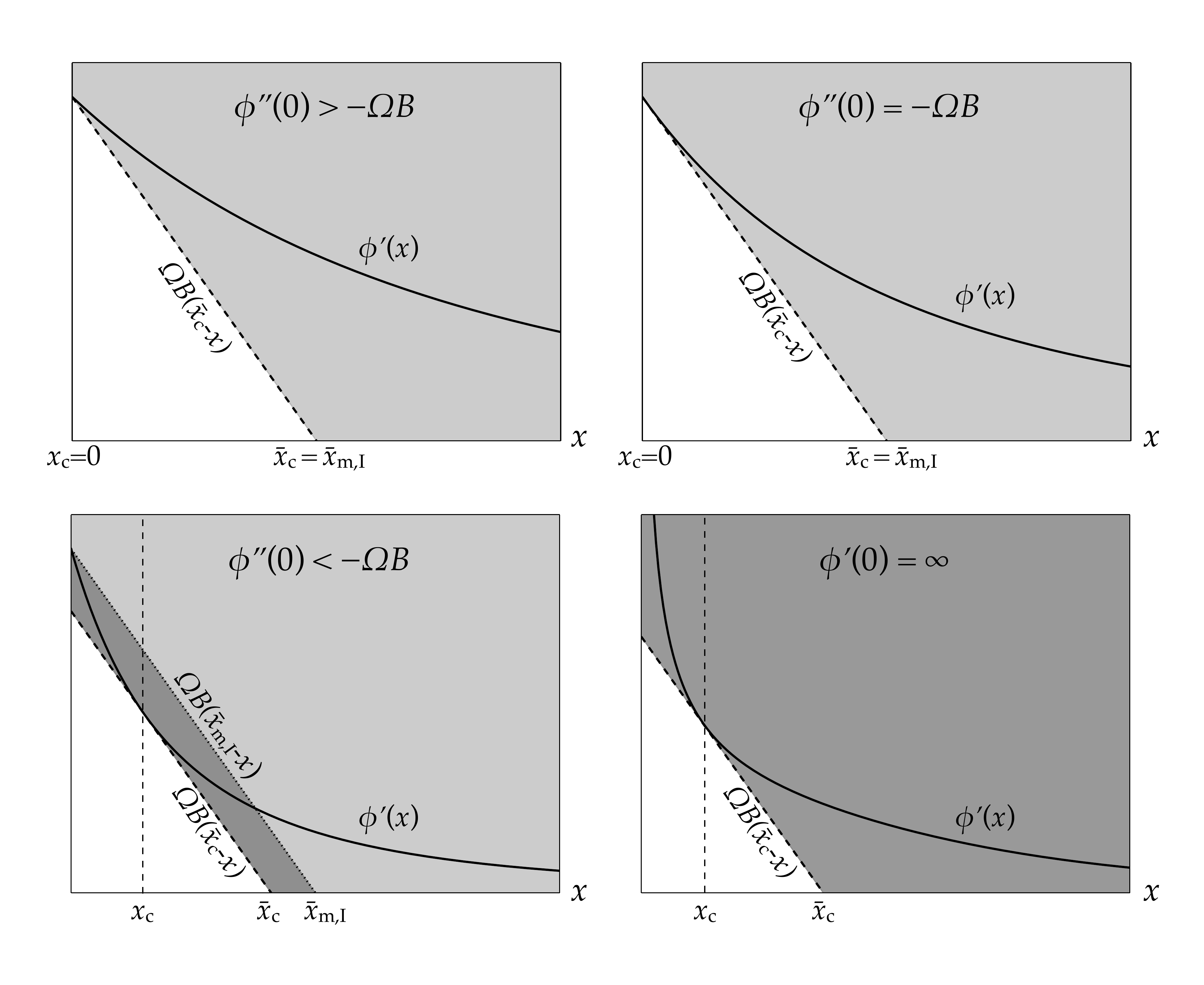}
\caption{%The different types of effective potential curves depend on the shape of the potential. 
The stationary points of the effective potential satisfy equation (\ref{stationary-points}), $ \phi'(x) = \Omega B \left( \bar{x} - x \right)$. 
In each of the four diagrams the solid curves represent $\phi'(x)$, while the function $\Omega B \left( \bar{x} - x \right)$ is the family of straight lines that are parallel to the oblique dashed lines. 
For a given value of $\bar{x}$, equation (\ref{stationary-points}) can have two solutions (dark grey region, $\chi$ is type II), one solution (light grey region, $\chi$ is type I) or no solution (unshaded region, $\chi$ has no minimum).
The smallest value of $\bar{x}$ for which a stationary point exists, at position $x_{\text{c}}$, is $\bar{x}_{\text{c}}$.
The value of $\bar{x}$ which corresponds to a stationary point at $x=0$ is $\bar{x}_{\text{m,I}}$.
%The electrostatic potential $\phi$ can be such that only type I curves are present for $\bar{x} \geqslant \bar{x}_{\text{c}} = \bar{x}_{\text{m,I}}$ (top diagrams, $\phi''(0) \geqslant - \Omega B $), or type I curves are present for $\bar{x} > \bar{x}_{\text{m,I}}$ and type II curves for $\bar{x}_{\text{c}} \leqslant \bar{x} < \bar{x}_{\text{m,I}}$ (bottom diagrams, $\phi''(0) < - \Omega B $).
%Type I curves disappear as $\phi'(0) \rightarrow \infty$ because $\bar{x}_{\text{m,I}} \rightarrow \infty$ (bottom right diagram).
}
\label{fig-typesofcurves}
\end{figure}

Setting equation (\ref{chi'}) to zero gives an equation for the stationary points of $\chi$, which can be rearranged to
\begin{align} \label{stationary-points}
 \phi'( x ) = \Omega B \left( \bar{x}  - x \right)  \text{.}
\end{align}
The stationary points are minima if the second derivative of $\chi$
%\begin{align} \label{chi''}
%\chi''(x, \bar{x} ) = \Omega^2 + \frac{\Omega \phi''(x)}{B} \text{,}
%\end{align} 
is positive.
This condition is equivalent to the gradient of $\phi'(x)$ being larger than the gradient of the line $\Omega B ( \bar{x }- x)$.
%An example of a type I effective potential curve is that arising from a flat electrostatic potential profile, $\phi(x) = 0$, for $\bar{x} > 0$. 
By rearranging equation (\ref{stationary-points}) to an equation for $\bar{x}$ as a function of $x$ and then minimizing it with respect to $x$, we obtain the minimum value of the orbit position $\bar{x}_{\text{c}}$ for which the effective potential has a stationary point,
\begin{align} \label{xbarc}
\bar{x}_{\text{c}} \equiv \min_{x \in [0,\infty]} \left( x  + \frac{\phi'( x )}{\Omega B} \right) \equiv x_{\text{c}} + \frac{\phi'(x_{\text{c}})}{\Omega B}  \text{.}
\end{align}
Note that in equation (\ref{xbarc}) we also defined the position $x_{\text{c}}$ of the stationary point of the effective potential $\chi$ when $\bar{x} = \bar{x}_{\text{c}}$.
%A geometrical interpretation of $\bar{x}_{\text{c}}$ is given in Figure \ref{fig-typesofcurves}: it 
In Figure \ref{fig-typesofcurves}, $\bar{x}_{\text{c}}$ is the smallest value of $\bar{x}$ for which the straight line $\Omega B \left( \bar{x} - x \right) $ touches the curve $\phi'(x)$, and $x_{\text{c}}$ is the value of $x$ at which they intersect. 
% $\chi'(0, \bar{x}) = 0$. the line $\Omega B \left( \bar{x}_{\text{m,I}} - x \right) $ intersecting the curve $\phi'(x)$ at $x=0$,
%If $| \phi''(0) | \leqslant \Omega B$, the gradient of the curve $\phi'(x)$ is everywhere smaller (due to monotonicity) than the gradient of the straight line $\Omega B (\bar{x} - x )$, therefore $x_{\text{c}} = 0$ and $\bar{x}_{\text{c}} = \bar{x}_{\text{m,I}}$, where
From Figure \ref{fig-typesofcurves}, $\bar{x}_{\text{c}}$ and $\bar{x}_{\text{m,I}}$ coincide if $\phi''(0) \geqslant - \Omega B$. 
%Then, any stationary point present at larger values of $\bar{x}$ is a minimum because the gradient of $\phi'(x)$ is clearly everywhere larger than that of the straight line $\Omega B ( \bar{x} - x )$.
Then, all effective potential curves are type I for $\bar{x} > \bar{x}_{\text{c}} = \bar{x}_{\text{m,I}} $.
If $\phi''(0) < - \Omega B$, $\bar{x} = \bar{x}_{\text{c}}$ is the orbit parameter value corresponding to when the straight line $\Omega B (\bar{x} - x )$ touches the curve $\phi'(x)$ tangentially.
Then, for orbit parameter values in the range $\bar{x}_{\text{c}} \leqslant \bar{x} \leqslant \bar{x}_{\text{m,I}}$ there are two stationary points (a minimum in the region $x>x_{\text{c}}$ and a maximum in the region $0\leqslant x < x_{\text{c}}$), corresponding to type II curves, while for $\bar{x} > \bar{x}_{\text{m,I}}$ there is only one stationary minimum, corresponding to type I curves.
Summarizing these observations with the aid of Figure \ref{fig-typesofcurves}:
\begin{itemize}
\item if $  \phi''(0)  \geqslant - \Omega B $, $\chi$ is a type I curve for $\bar{x} > \bar{x}_{\text{c}} = \bar{x}_{\text{m,I}}$;
\item if $ \phi''(0)   < - \Omega B $, $\chi$ is a type II curve for $ \bar{x}_{\text{c}} < \bar{x} < \bar{x}_{\text{m,I}}$ and a type I curve for $ \bar{x} > \bar{x}_{\text{m,I}}$.
\end{itemize}

We will see in Sections \ref{sec-quasi} and \ref{sec-numsol} that our solution to the magnetic presheath electrostatic potential is such that the electric field diverges at $x=0$, $\phi'(0) \rightarrow \infty$. 
Thus, the effective potential curves are type II for all values of $\bar{x}$ larger than  $ \bar{x}_{\text{c}}$ because $\bar{x}_{\text{m,I}} =  \phi'(0) / \Omega B \rightarrow \infty$ (see Figure \ref{fig-typesofcurves}, bottom right diagram). 
It is nonetheless useful to consider also type I curves because we obtain our solution by iterating over possible electrostatic potential profiles starting from the initial guess of a flat potential, $\phi(x) =0$. %, and also because upon numerical discretization with a finite resolution type II curves may look like type I curves when the position of the maximum $x_{\text{M}}$ is close to zero.

\subsection{Closed orbits for $\alpha = 0$} \label{subsec-closed-orbits}

The ion motion for $\alpha =0$ is a periodic (closed) orbit provided that an effective potential minimum exists, $\bar{x}> \bar{x}_{\text{c}}$, and that a pair of bounce points $x_{\text{b}}$ and $x_{\text{t}}$ exist, $U_{\perp} < \chi_{\text{M}} (\bar{x})$ (see Figure \ref{fig-effpotclosed}). 
When the $\alpha = 0$ motion of an ion is a closed orbit, we can write its position as a function of a gyrophase angle which parameterizes the particular point of the orbit in which the particle lies. 
The period of the orbit, $2\pi/ \overline{\Omega}$, where $\overline{\Omega}$ is the generalized gyrofrequency, is the integral of all the time elements $dt = dx/v_x$ over a whole orbit,
\begin{align} \label{Omegabar-def}
\frac{2\pi}{\overline{\Omega}} = 2 \int_{x_{\text{b}}}^{x_{\text{t}}} \frac{dx}{V_x \left( x, \bar{x}, U_{\perp} \right)} \text{.}
\end{align}
The gyrophase angle $\varphi$ of the orbit is defined as $\overline{\Omega} t$, where $t$ is defined in the interval $-\pi / \overline{\Omega} < t < \pi / \overline{\Omega} $ and is (when positive) the time elapsed since the particle last reached the top bounce point,
\begin{align} \label{varphi-def}
\varphi = \sigma_x \overline{\Omega} \int_{x_{\text{t}}}^x \frac{ds}{V_x \left( s, \bar{x}, U_{\perp} \right) } \text{.}
\end{align}

It will be useful to define the gyroaveraging operation as an average over possible values of gyrophase, or equivalently as an average over the period of a closed orbit,
\begin{align} \label{gyroaverage}
\left\langle \ldots \right\rangle_{\varphi} = \frac{1}{2\pi} \int_{-\pi}^{\pi} \left( \ldots \right) d\varphi = \sum_{\sigma_x = \pm 1} \frac{\overline{\Omega}}{2\pi} \int_{x_{\text{b}}}^{x_{\text{t}}} \frac{\left( \ldots \right) dx}{V_x \left( x, \bar{x}, U_{\perp} \right)} \text{.}
\end{align}
The second equality in (\ref{gyroaverage}) is obtained using (\ref{varphi-def}). The closed orbit has an $\vec{E}\times \vec{B}$ drift in the $y$ direction (parallel to the wall), with drift velocity $V_{\vec{E} \times \vec{B} }$ defined as the gyroaverage of $v_y$,
\begin{align} \label{V-ExB}
V_{\vec{E} \times \vec{B} } \left( \bar{x}, U_{\perp} \right) = \frac{ \overline{\Omega} }{\pi} \int_{x_{\text{b}}}^{x_{\text{t}}} \frac{\Omega \left( \bar{x} - x \right)}{V_x \left( x, \bar{x}, U_{\perp} \right)} dx = \frac{ \overline{\Omega} }{\pi} \int_{x_{\text{b}}}^{x_{\text{t}}} \frac{ \phi' ( x) / B }{V_x \left( x, \bar{x}, U_{\perp} \right)} dx \text{.}
\end{align}
The second equality in (\ref{V-ExB}) comes from using equation (\ref{chi'}) and the result
\begin{align} \label{gyroaveragevy-trick}
\int_{x_{\text{b}}}^{x_{\text{t}}} \frac{\chi' \left( x, \bar{x} \right) }{V_x \left( x, \bar{x}, U_{\perp} \right)} dx = - \int_{x_{\text{b}}}^{x_{\text{t}}}   V_x' \left( x, \bar{x}, U_{\perp} \right)  dx = V_x \left( x_{\text{b}}, \bar{x}, U_{\perp} \right)  - V_x \left( x_{\text{t}} , \bar{x}, U_{\perp} \right) = 0 \text{,}
\end{align}
where we used $V_x \left( x_{\text{b}}, \bar{x}, U_{\perp} \right)  = V_x \left( x_{\text{t}} , \bar{x}, U_{\perp} \right) = 0$. The first equality in (\ref{gyroaveragevy-trick}) comes from differentiating equation (\ref{vx-Uperp-xbar-x}).

\subsection{Approximately closed orbits for $\alpha \ll 1$} \label{subsec-appclosedorbits}

\begin{figure}
\centering
\includegraphics[width= 0.8\textwidth]{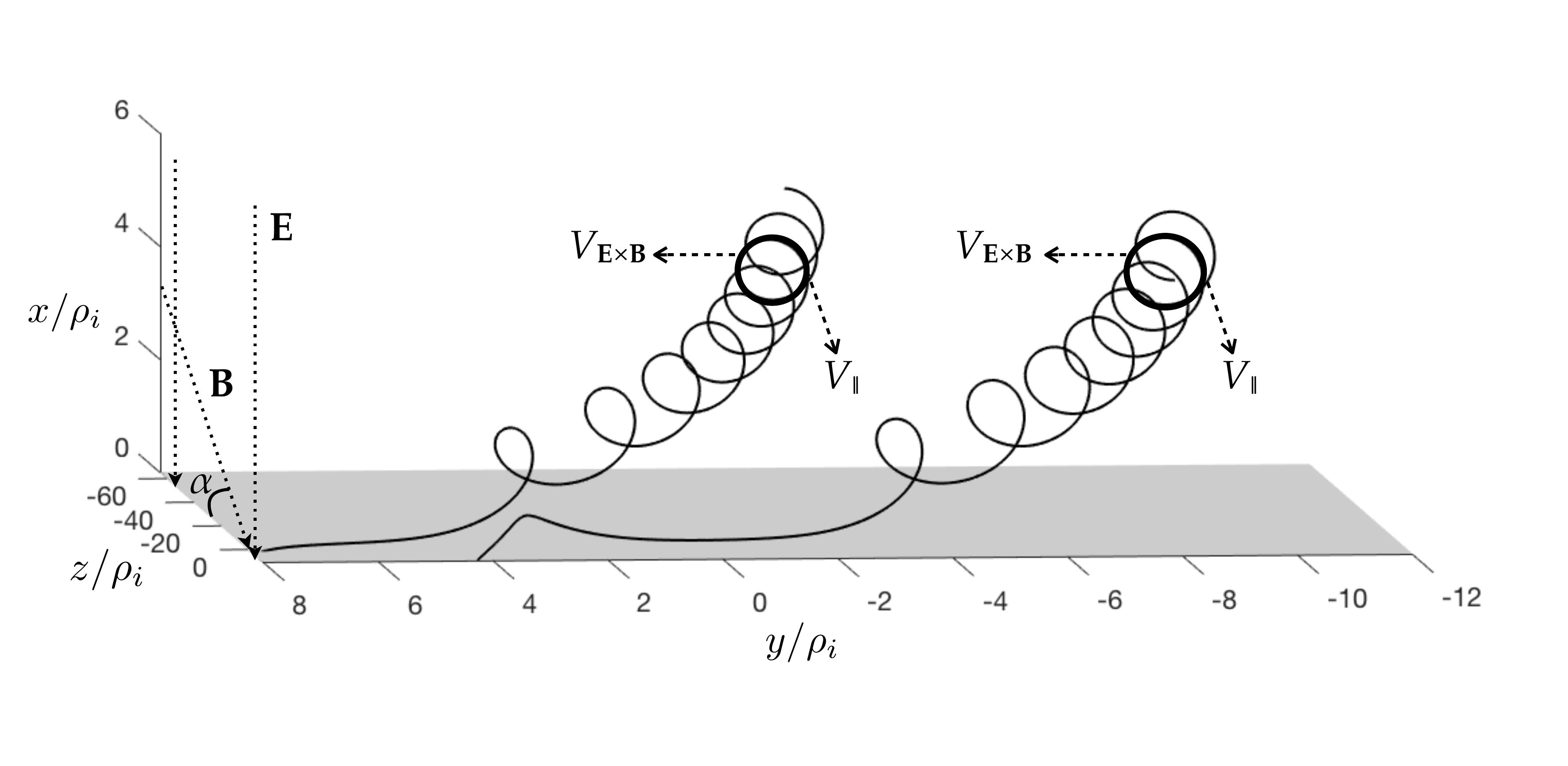}
\caption{Two ion trajectories approaching the wall, represented as a grey surface at $x=0$, are shown as black lines for $\alpha = 0.05 \text{ radians}$. The electric and magnetic fields are marked with dotted arrows, and the angle that the magnetic field makes with the wall is $\alpha = 0.05$. The angle $\alpha$ looks large to the naked eye because the $z$ direction has been squashed in order to draw the 3 dimensional trajectory of the ions. Most of the ion path is locally approximated by closed orbits, represented as superimposed rings. Ions stream along the magnetic field $\vec{B}$ at velocity $V_{\parallel} \left( U_{\perp}, U \right)$, and the strong electric field towards the wall causes the approximately closed orbits to $\vec{E} \times \vec{B}$ drift at velocity $V_{\vec{E} \times \vec{B} } \left( \bar{x}, U_{\perp} \right) $ in the $y$ direction. The increasing electric field as the orbits approach $x=0$ causes the  $\vec{E} \times \vec{B}$ velocity to noticeably increase (see equation (\ref{V-ExB})). 
%Eventually the electric force pushing the ion towards the wall becomes larger than the magnetic force trying to turn it away from the wall.
}
\label{fig-iontraj}
\end{figure}

When $\alpha = 0$ an ion moves in a closed orbit which $\vec{E}\times \vec{B}$ drifts in the $y$ direction (equation (\ref{V-ExB})) and streams parallel to the magnetic field in the $z$ direction (equation (\ref{vz-U-Uperp})). When $\alpha \ll 1$, the motion is \emph{approximately} periodic because the orbit parameters vary
% satisfy $\bar{x} / \dot{\bar{x}} \sim U_{\perp} / \dot{U}_{\perp} \sim 1/ \alpha \Omega $, so they 
over a timescale $1/\alpha \Omega$ that is much longer than the typical gyroperiod $1/ \Omega$. %This means that the gyroaveraged time derivatives of the orbit parameters are equivalent to their non-gyroaveraged counterparts at long timescales $1/\alpha \Omega $. 
Differentiating (\ref{xbar-def}) with respect to time and using (\ref{vy-EOM-exact}), we find
\begin{align} \label{xbardot}
\dot{\bar{x}} = - \sigma_{\parallel} \alpha V_{\parallel} \left( U_{\perp}, U \right)  +  O \left( \alpha^2 v_{\text{t,i}} \right) \text{.}
\end{align}
Physically, this represents the small component of the parallel motion which moves the approximately closed ion orbit in the $x$ direction when $\alpha \neq 0$. Note that
\begin{align} \label{Udot}
\dot{U} = 0 \text{}
\end{align}
is true to every order in $\alpha$ because energy is exactly conserved in the absence of explicit time dependence. 
Differentiating (\ref{Uperp-def}) and using (\ref{vx-EOM-exact}) and (\ref{vy-EOM-exact}) we get
\begin{align} \label{Uperpdot}
\dot{U}_{\perp}  = - \sigma_{\parallel} \alpha \Omega^2 V_{\parallel} \left( U_{\perp}, U \right) \left( \bar{x} - x\right)  + O \left( \alpha^2 \Omega v_{t,i}^2 \right) \text{,}
\end{align}
which depends on the instantaneous particle position $x$ and therefore on the gyrophase $\varphi$. 
Since the orbit parameters are varying over the long timescale $1/\alpha \Omega$, they are approximately constant over a single orbit, and hence the time derivative of $U_{\perp}$ is approximately periodic at small timescales (recall that $x$ is approximately periodic). 
Then, $\dot{U}_{\perp}$ can be split in a gyroaveraged piece, $\langle \dot{U}_{\perp} \rangle_{\varphi}$, which remains approximately constant over a  few gyroperiods, and an oscillatory piece, $\dot{U}_{\perp} -  \langle \dot{U}_{\perp} \rangle_{\varphi}$, whose contribution to $U_{\perp}$ averages to zero after a few gyroperiods.
Thus, the gyroaveraged time derivative of $U_{\perp}$ determines the behaviour of $U_{\perp}$ at long timescales. 
Exploiting (\ref{gyroaverage}) and (\ref{V-ExB}), the gyroaverage of (\ref{Uperpdot}) is
\begin{align}  \label{Uperpdot-gyro}
\left\langle \dot{U}_{\perp} \right\rangle_{\varphi} =  - \sigma_{\parallel} \alpha  \Omega V_{\parallel} \left( U_{\perp}, U \right) \frac{ \overline{\Omega} }{\pi} \int_{x_{\text{b}}}^{x_{\text{t}}} \frac{ \phi' ( x) / B }{V_x \left( x, \bar{x}, U_{\perp} \right)} dx + O \left( \alpha^2 \Omega v_{t,i}^2 \right) \text{.}
\end{align}
%The decrease of the perpendicular energy is due to the small component of the electric field directed parallel to the magnetic field which accelerates the parallel velocity of the ion. 
Two ion trajectories, which were obtained by varying the orbit parameters according to equations (\ref{xbardot})-(\ref{Uperpdot}), are shown in Figure \ref{fig-iontraj}.

In a Hamiltonian system, when the parameters of periodic motion change over a timescale much longer than the period of the motion, an \emph{adiabatic invariant} exists. Here, it is given by \cite{Cohen-Ryutov-1998, Geraldini-2017}
\begin{align} \label{mu-Uperp-xbar}
\mu = \mu_{\text{gk}} \left( \bar{x}, U_{\perp} \right) \equiv \frac{1}{\pi} \int_{x_b}^{x_t} V_x \left( x, \bar{x}, U_{\perp} \right) dx \sim \frac{v_{\text{t,i}}^2}{\Omega} \text{.}
\end{align}
Unlike $\bar{x}$ and $U_{\perp}$, the adiabatic invariant (\ref{mu-Uperp-xbar}) is conserved to lowest order over the much longer timescale $ 1/\alpha\Omega$, %, which is the characteristic time it takes for an orbit drifting at $\dot{\bar{x}} \sim \alpha v_{t,i}$ to cross the magnetic presheath of size $\rho_{\text{i}}$. It satisfies
\begin{align}
\left \langle \dot{\mu} \right \rangle_{\varphi} = O \left( \alpha^2 v_{\text{t,i}}^2 \right) \simeq 0 \text{.}
\end{align}
The picture that emerges of the ion trajectory in a grazing-angle magnetic presheath is that of a sequence of approximately closed orbits whose parallel streaming brings them slowly towards the wall, as shown in Figure \ref{fig-iontraj}. The adiabatic invariant $\mu_{\text{gk}} \left( \bar{x}, U_{\perp} \right) $ and total energy $U$ 
%corresponding to the closed orbit with parameters $\bar{x},~ U_{\perp} \text{ and } U$ (which approximates the ion motion for a short time $1/\Omega$), 
are conserved as the ion traverses the magnetic presheath.
%\begin{align}
%\left\langle \dot{\mu} \right\rangle_{\varphi} = O\left( \alpha^2 \Omega \mu \right) \simeq 0 \text{.}
%\end{align}

In this work, we assume an electron-repelling wall, hence $\phi' \left( x \right) > 0$ in the sheath-presheath system. Since the wall is absorbing, any ion present in the system must be coming from $x \rightarrow \infty$ and moving towards $x=0$, and therefore it has $\dot{\bar{x}} < 0$ and $\sigma_{\parallel} = +1$.
Then, from (\ref{Uperpdot-gyro}), $U_{\perp}$ decreases as the ion moves across the magnetic presheath with $\sigma_{\parallel} = +1$. 
The decrease in $U_{\perp}$ is caused by the small component of the electric field which is parallel to the magnetic field and therefore accelerates ions in the parallel direction, such that $ V_{\parallel} \left( U_{\perp}, U  \right)$ increases as the particle approaches the wall and $\sigma_{\parallel}$ never changes sign. 
Hence, from here on we take $\sigma_{\parallel} = + 1$ for all ions.

\subsection{Open orbits} \label{subsec-openorbits}

%If $U_{\perp} > \chi_{\text{M}} \left(\bar{x} \right) = \chi \left( x_{\text{M}}, \bar{x} \right) $, the ion trajectory is an \emph{open} orbit that reaches $x=0$ and is lost from the system after a time $1/\Omega$.
%Ions travel across the magnetic presheath in approximately closed orbits until they reach the wall. 
The time that it takes for an ion to cross the magnetic presheath is $\sim 1/\alpha \Omega$. 
During this time the ion motion can be approximated by that of a periodic gyro-orbit with period $\sim 1/\Omega$. 
The parameters of the periodic motion change over the slow timescale $1/\alpha \Omega$.
When the ion reaches values of the orbit parameters for which its lowest order motion intersects the wall (and is therefore no longer periodic), it reaches the wall and is lost from the system over the fast timescale $1/\Omega$ (as we will show).
In this short period of time, the ion is in an open orbit.
The number of ions in open orbits is small (higher order in $\alpha$) compared with the number of ions in closed orbits because open orbits exist for a much shorter time. 
%However, there is a narrow region near the wall where the small number of ions in open orbits dominates.
However, the number of ions in closed orbits that cross a point arbitrarily close to the wall is small because it only includes those ions that are near the bottom bounce point of their orbit (and therefore, from equation (\ref{varphi-def}), it only includes ions with a small range of gyrophases around $\varphi = \pm \pi$). 
%\footnote{The orbit position $\bar{x}$ constrains the adiabatic invariant $\mu$ (the size of the orbit). The density of ions in closed orbits at the wall is zero in the same way that the volume of a plane is zero.} 
%Moreover, for the electrostatic potential solution we will find, tangential solutions to the ion trajectories arbitrarily close to the wall are impossible.
Therefore, it is essential to obtain the contribution to the density due to ions in open orbits. %, which are the ones that intersect the wall in their future trajectory instead of bouncing back from a bottom bounce point $x_{\text{b}}$.

\begin{figure}[h]
\centering
\includegraphics[width=0.75\textwidth]{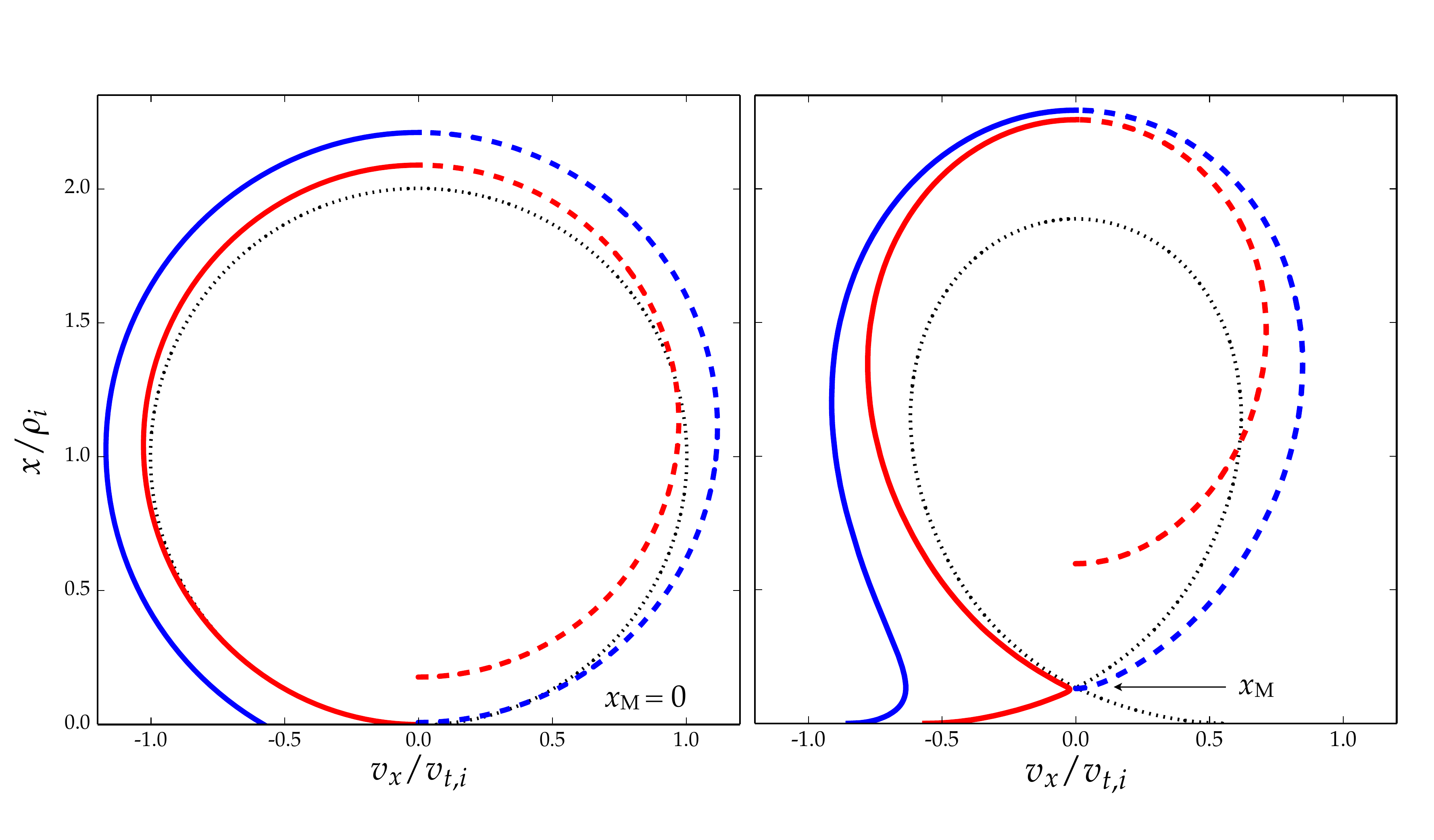}
\caption{Two sets of phase space trajectories corresponding to type I (left diagram) and type II (right diagram) orbits.
The type I trajectories are obtained using $\phi \left(x \right) =0$, while the type II trajectories are evaluated using the electrostatic potential solution of Section \ref{sec-numsol} for $\alpha = 0.02$. 
The dotted lines are trajectories of motion with $\alpha =0$ when $U_{\perp} =\chi_{\text{M}} (\bar{x}) $, with $\bar{x} = \rho_{\text{i}}$ (type I) and $\bar{x} = 1.6 \rho_{\text{i}}$ (type II). 
The solid and dashed lines are trajectories calculated by integrating equations (\ref{x-EOM-exact}), (\ref{vx-Uperp-xbar-x}) and (\ref{xbardot})-(\ref{Uperpdot}) in the past from $x=0$ with $\alpha = 0.02$, starting with the same value of $\bar{x}$ used to obtain the dotted trajectories and with $U - \chi_{\text{M}}(\bar{x}) = v_{\text{t,i}}^2$. 
The solid lines are the open orbit pieces of the trajectories, while the dashed lines are approximately closed orbits according to our definition.
In each diagram, the red trajectory corresponds to the ion crossing $x_{\text{M}}$ with $v_x \simeq 0$, while the blue trajectory corresponds to the ion crossing $x_{\text{M}}$ with the largest possible value of $|v_x|$. 
}
\label{fig-phase}
\end{figure}

It is clear that an ion is in an open orbit when $x \leqslant x_{\text{M}}$, because a closed orbit cannot access this region by definition (see Figure \ref{fig-effpotclosed}).
For the ion to reach $x \leqslant x_{\text{M}}$, it must have crossed the maximum of the effective potential $\chi$ from the region $x>x_{\text{M}}$.
The exact point $x>x_{\text{M}}$ at which we consider its orbit to be open is arbitrary, but this arbitrariness does not matter because the ion density for $x>x_{\text{M}}$ is dominated by closed orbits.
%Given that we consider all ions with $D<0$ at $x>x_{\text{M}}$ to be in approximately closed orbits, the natural criterion for an ion to be in an open orbit at $x>x_{\text{M}}$ is that it have $D>0$. 
We exploit this to generalize the open orbit definition in a way that includes all ions at $x\leqslant x_{\text{M}}$ and smoothly extends the open orbit density to $x> x_{\text{M}}$. We consider an ion to be in an open orbit if:
\begin{enumerate}
\item at future times, its trajectory has no bounce points,
\item at past times, its trajectory has several bounce points (the trajectory becomes an approximately closed orbit).
\end{enumerate}
Note that criterion (ii) is equivalent to the past ion trajectory reaching a bottom bounce point $x_{\text{b}}$. % such that $\sigma_x = v_x / |v_x|$ changes from $-1$ to $+1$.
Examples of pieces of trajectories considered to be open orbits are shown in Figure \ref{fig-phase} by solid lines.
We consider open orbit the part of a trajectory between the wall and the top bounce point.

To study open orbits, it will be useful to consider the difference between the perpendicular energy and the effective potential maximum as a separate quantity $D$, %$U_{\perp} - \chi_{\text{M}} \left(\bar{x} \right)$
\begin{align} \label{D-def}
D = U_{\perp} - \chi_{\text{M}} \left(\bar{x} \right) \text{.}
\end{align}
The velocity component $v_x$, given by equation (\ref{vx-Uperp-xbar-x}), is
\begin{align} \label{vx-open}
v_x = \sigma_x V_x \left( x, \bar{x}, D + \chi_{\text{M}} \left( \bar{x}\right)  \right) = \sigma_x \sqrt{ 2\left( D + \chi_{\text{M}} \left( \bar{x}\right) - \chi \left( x, \bar{x} \right)\right) } \text{.}
\end{align}
When $x=x_{\text{M}}$ is reached from $x> x_{\text{M}}$, the velocity is given by $v_x = -\sqrt{2D}$, hence only ions with $D > 0$ cross the effective potential maximum and reach $x\leqslant x_{\text{M}}$.
To obtain the rate of change of $D$, we calculate the rate of change of $\chi_{\text{M}} \left( \bar{x} \right)$,
\begin{align} \label{chiMdot1}
\dot{\chi}_{\text{M}} \left( \bar{x} \right) = \frac{\partial \chi}{\partial \bar{x}}\left( x_{\text{M}}, \bar{x} \right) \dot{\bar{x}} +  \chi' \left( x_{\text{M}}, \bar{x}  \right) \frac{\partial x_{\text{M}}}{\partial \bar{x}} \dot{\bar{x}} \text{.}
\end{align} 
%Note that type I and type II shapes for the effective potential correspond to qualitatively different trajectories, which we henceforth refer to as type I and type II orbits following the nomenclature of Cohen and Ryutov \cite{Cohen-Ryutov-1998}. 
For both type I and type II orbits, the second term in (\ref{chiMdot1}) vanishes (type I curves have $\partial x_{\text{M}}  / \partial \bar{x} = 0$, while type II curves have $ \chi' \left( x_{\text{M}}, \bar{x} \right)   = 0$) and, using (\ref{xbardot}) with $\sigma_{\parallel} = +1$, we find
\begin{align} \label{chiMdot}
\dot{\chi}_{\text{M}} \left( \bar{x} \right) = \alpha \Omega^2 V_{\parallel} \left( U_{\perp}, U \right) \left( x_{\text{M}} - \bar{x} \right) + O\left( \alpha^2 \Omega v_{\text{t,i}}^2 \right)  \text{.}
\end{align}
Combining (\ref{chiMdot}) with the result for $\dot{U}_{\perp}$ in (\ref{Uperpdot}), we get (using $\sigma_{\parallel} = +1$)
\begin{align} \label{D-dot-general}
\dot{D} = \alpha \Omega^2 V_{\parallel} ( U_{\perp} , U ) \left( x - x_{\text{M}} \right) + O\left( \alpha^2 \Omega v_{\text{t,i}}^2 \right) \text{.}
\end{align}

Consider an ion that reaches $U_{\perp} = \chi_{\text{M}} (\bar{x})$ at a position $x' > x_{\text{M}}$ and is travelling towards the maximum ($\sigma_x = -1$).
We use the relation
\begin{align} \label{dt-open-naive}
dt = \frac{dx}{v_x} \simeq \frac{dx}{ \sigma_x V_x \left( x, \bar{x}, U_{\perp} \right) }
\end{align}
to estimate the time taken for the ion to reach the effective potential maximum,
\begin{align} \label{t-open}
\delta t_{\text{M}} = \int dt \simeq \int^{x'}_{x_{\text{M}}}  \frac{ds}{ V_x \left( s, \bar{x}, U_{\perp} \right) } \text{.}
\end{align}
We assume that the difference between $U_{\perp}$ and $\chi_{\text{M}} (\bar{x})$ stays small and that the change in $\bar{x}$ during the time $\delta t_{\text{M}}$ is small (which we will show to be true), so that $U_{\perp} \simeq \chi_{\text{M}} (\bar{x})$.
If the effective potential curve is of type I, $\delta t_{\text{M}}^{\text{I}} \sim 1/\Omega $, whereas for type II curves $\delta t_{\text{M}}^{\text{II}}$ diverges according to equation (\ref{t-open}).
We show this by expanding $V_x \left( x, \bar{x}, U_{\perp} \right) $ near $x \simeq x_{\text{M}}$ for a type II curve, using $U_{\perp} \simeq \chi_{\text{M}} (\bar{x})$ and defining $\chi''_{\text{M}} \equiv \chi'' \left( x_{\text{M}}, \bar{x} \right)$ to obtain 
\begin{align} \label{VxII}
V_x^{\text{II}} \left( x, \bar{x}, U_{\perp} \right)  \simeq V_x^{\text{II}} \left( x, \bar{x},  \chi_{\text{M}} \left( \bar{x} \right) \right) \simeq \sqrt{\left| \chi''_{\text{M}} \right| } \left| x - x_{\text{M}} \right| \text{.}
\end{align}
The time $\delta t_{\text{M}}^{\text{II}}$ is then
%estimated as the time taken to reach a point $x \simeq x_{\text{M}}$ from a neighbouring point $x' > x$,
\begin{align}
\delta t_{\text{M}}^{\text{II}} \simeq \int^{x'}_{x_{\text{M}}} \frac{ds}{ \sqrt{\left| \chi''_{\text{M}} \right| } \left( s - x_{\text{M}} \right)  } \rightarrow \infty \text{.}
%
%\simeq \frac{1}{\sqrt{ \left| \chi''_{\text{M}} \right| } } \ln \left( \frac{ x' - x_{\text{M}} }{ x_{\text{M}} - x_{\text{M}}  } \right)    \text{,}
\end{align}
Despite this apparent divergence, the variation of $D$ \emph{during} the time $\delta t_{\text{M}}$ can be evaluated using (\ref{dt-open-naive}).
Using $U_{\perp} \simeq \chi_{\text{M}} (\bar{x})$, equation (\ref{D-dot-general}) becomes
\begin{align} \label{D-dot}
\dot{D} = \alpha \Omega^2 V_{\parallel} ( \chi_{\text{M}} (\bar{x}) , U ) \left( x - x_{\text{M}} \right) + O\left( \alpha^2 \Omega v_{\text{t,i}}^2 \right) \text{.}
\end{align} 
Thus, equations (\ref{VxII}) and (\ref{D-dot}) imply that $\dot{D}/ V_x \left( x, \bar{x}, \chi_{\text{M}} (\bar{x}) \right) $ is not divergent at $x = x_{\text{M}}$.
Integrating equation (\ref{D-dot}) in time using (\ref{dt-open-naive}) we have
\begin{equation} \label{D-estimate}
D = \int \dot{D} dt \simeq \alpha \Omega^2 V_{\parallel} (\chi_{\text{M}} (\bar{x}) , U )  \int_{x_{\text{M}}}^{x'}  \frac{ s - x_{\text{M}}  }{V_x \left( s, \bar{x}, \chi_{\text{M}} \left( \bar{x} \right) \right)}  ds \text{,}
\end{equation}
hence we expect $D \sim \alpha v_{\text{t,i}}^2$ for both orbit types, justifying $U_{\perp} \simeq \chi_{\text{M}} (\bar{x})$ a posteriori. %which justifies the ordering $D \sim \alpha v_{\text{t,i}}^2$. 
Using $U_{\perp} = \chi_{\text{M}} (\bar{x}) + D$ with $D \sim \alpha v_{\text{t,i}}^2$, equation (\ref{t-open}) can be used to obtain the more accurate estimate $\delta t_{\text{M}}^{\text{II}} \sim \ln \left( 1/\alpha \right) /\Omega$. Putting together the estimates for both orbit types, we have 
\begin{align} \label{deltat_M}
\Omega \delta t_{\text{M}} \sim \begin{cases}
1 & \text{ for type I orbits,} \\
 \ln \left( \frac{1}{\alpha} \right) & \text{ for type II orbits.}
\end{cases}
\end{align}

We proceed to find the possible values of $D$ which satisfy the open orbit criteria that we have defined.
If $x < x_{\text{M}}$ the particle has already crossed the effective potential maximum and we have to integrate backwards in time to obtain the value of $D$ at the moment $x_{\text{M}}$ was crossed, denoted $D_{\text{X}}$, and further back to obtain the value of $D$ during the last bounce from the bottom bounce point $x_{\text{b}} \simeq x_{\text{M}}$, denoted $D_{\text{B}}$. If $x>x_{\text{M}}$, we must integrate $\dot{D}$ forwards in time to obtain $D_{\text{X}}$ (because by definition the particle trajectory must cross $x_{\text{M}}$ when it next reaches it, otherwise it would not be an open orbit), and backwards in time to obtain $D_{\text{B}}$. 

We first obtain $D_{\text{X}} - D$ in terms of $x$, $\bar{x}$ and $U$. If $x>x_{\text{M}}$ we integrate $\dot{D} > 0$ \emph{forwards} in time (so $dt>0$) and if $x<x_{\text{M}}$ we integrate $\dot{D} < 0$ \emph{backwards} in time (so $dt<0$), hence we expect a positive quantity, denoted $\Delta_+$, in both cases. From equation (\ref{D-estimate}), such quantity is approximately
%\begin{align} \label{Delta+}
% \Delta_+ \left(x,  \bar{x}, U \right)  = \alpha \Omega^2 V_{\parallel} (\chi_{\text{M}} (\bar{x} ) , U )  \left[  \int_{x_{\text{M}}}^{x} \frac{ \left( s - x_{\text{M}} \right)  }{ V_x \left( x, \bar{x}, \chi_{\text{M}} \left( \bar{x} \right) \right) + \dot{x}_{\text{M}}   } ds    \right] \text{,}
%\end{align}
\begin{align} \label{Delta+}
D_{\text{X}} - D \simeq  \Delta_+ \left(x,  \bar{x}, U \right)  = \alpha \Omega^2 V_{\parallel} (\chi_{\text{M}} (\bar{x} ) , U )  \int_{x_{\text{M}}}^{x} \frac{ \left( s - x_{\text{M}} \right)  }{ V_x \left( s, \bar{x}, \chi_{\text{M}} \left( \bar{x} \right) \right)   } ds   \sim \alpha v_{\text{t,i}}^2 \text{,}
\end{align}
therefore $D_{\text{X}}$ is
\begin{align} \label{D-X}
D_{\text{X}} = D + \Delta_+ \left(x,  \bar{x}, U \right)  + O\left( \alpha^{1+p} v_{\text{t,i}}^2 \right)   \text{.}
\end{align}
%The order of the error $E$ depends on the open orbit type and is (see \ref{app-open-orbits})
%\begin{align}
%E \sim \begin{cases} 
%\alpha & \text{ for type I orbits,} \\
%\alpha^{1/2} & \text{ for type II orbits.} 
%\end{cases}
%\end{align}
The power $p$ used to quantify the error is given by 
\begin{align} \label{p-def}
p = \begin{cases}
1 & \text{ for type I orbits,} \\
\frac{1}{2} & \text{ for type II orbits.}
\end{cases}
\end{align}
The larger error from type II orbits comes from the fact that $D \sim \alpha v_{\text{t,i}}^2 $ is neglected when we use $dt \simeq ds / V_x \left(s, \bar{x}, \chi_{\text{M}}(\bar{x})  \right)$. Estimating $|v_x|$ more accurately in the region near the maximum, we have
\begin{align} \label{VxII-modified}
V_x^{\text{II}} \left( x, \bar{x},  U_{\perp} \right)  = V_x^{\text{II}} \left( x, \bar{x},  \chi_{\text{M}}  \left( \bar{x} \right) + D \right) \simeq \sqrt{\left| \chi''_{\text{M}} \right| \left( x - x_{\text{M}} \right)^2 + 2D } \text{.}
\end{align}
Hence, there is a region of size $| x-x_{\text{M}}| \sim \alpha^{1/2} \rho_{\text{i}}$ %in which $V_x^{\text{II}} \left( x, \bar{x},  U_{\perp}  \right) \sim \alpha^{1/2} v_{\text{t,i}}$, 
where the estimate (\ref{VxII}) is incorrect. 
The contribution from this region to the integral (\ref{Delta+}) is therefore incorrect, and the size of this contribution is the size of the error in equation (\ref{D-X}).
Indeed, multiplying the size of the region ($\int_{x_{\text{M}}}^{x} ds \sim \alpha^{1/2} \rho_{\text{i}}$) by the size of the integrand ($| x-x_{\text{M}}|/V_x^{\text{II}} \sim 1/\Omega$) and by the prefactor ($\alpha \Omega^2 v_{\text{t,i}}$), we obtain an error of $\alpha^{3/2} v_{\text{t,i}}^2$, in accordance with equation (\ref{D-X}) with $p=1/2$.

We proceed to obtain $D_{\text{B}} - D_{\text{X}}$ by integrating $\dot{D} > 0$ \emph{backwards} in time (so $dt < 0$) from the point at which the maximum is crossed. 
The result is a negative quantity of magnitude $\Delta_{\text{M}}$, which is an integral from the bottom bounce point $x_{\text{b}} \simeq x_{\text{M}}$ to the top bounce point $x_{\text{t}} \simeq x_{\text{t,M}}$ and back, where $x_{\text{t,M}}$ is the top bounce point corresponding to $U_{\perp} =  \chi_{\text{M}} (\bar{x})$. 
The backward integration is identical to the forward one, hence using equation (\ref{D-estimate}) we obtain,
\begin{align} \label{DeltaM}
D_{\text{X}} - D_{\text{B}} \simeq \Delta_{\text{M}}  \left( \bar{x}, U \right) = 2\alpha \Omega^2 V_{\parallel} (\chi_{\text{M}} (\bar{x} ) , U )  \int_{x_{\text{M}}}^{x_{\text{t,M}}} \frac{ \left( s - x_{\text{M}} \right)  }{ V_x \left( s, \bar{x}, \chi_{\text{M}} \left( \bar{x} \right) \right)  } ds  \sim 2\pi \alpha v_{\text{t,i}}^2 \text{.}
\end{align}
The factor of $2\pi$ in the final scaling of (\ref{DeltaM}) is due to having integrated in time over a gyroperiod $\sim 2\pi / \Omega$.
Then, $D_{\text{B}}$ is
\begin{align} \label{D-turn}
D_{\text{B}} = D_{\text{X}} -  \Delta_{\text{M}}  \left( \bar{x}, U \right)   + O\left( \alpha^{1+p} v_{\text{t,i}}^2 \right)   \text{.}
\end{align}

The criteria used to determine whether an ion is in an open orbit can be re-expressed in terms of $D_{\text{B}}$ and $D_{\text{X}}$:
\begin{align} \label{cond-open-DX}
& \text{(i) at future times, the ion's trajectory has no bounce points } \nonumber \\
& ~~~~ \implies D_{\text{X}} > O  \left( \alpha^{1+p} v_{\text{t,i}}^2 \right) \text{;}  \\
& \text{(ii) at past times, the ion's trajectory has several bounce points } \nonumber   \label{cond-open-DB} \\ 
& ~~~~ \implies D_{\text{B}} < O \left( \alpha^{1+p} v_{\text{t,i}}^2 \right) \text{.}
\end{align}
Note that condition (i) is automatically satisfied if $x<x_{\text{M}}$ (and $\sigma_x = -1$); in this case condition (ii) is directly related to both (\ref{cond-open-DX}) and ({\ref{cond-open-DB}).
The limited accuracy in the evaluation of $D_{\text{X}}$ and $D_{\text{B}}$ leads to the  $O \left( \alpha^{1+p} v_{\text{t,i}}^2 \right) $ error in the inequality.
Using conditions (\ref{cond-open-DX}) and (\ref{cond-open-DB}), and equations (\ref{D-X}) and (\ref{D-turn}), we have the inequality
\begin{align} \label{D-range}
 -  \Delta_+ \left(x,  \bar{x}, U \right)  + O \left( \alpha^{1+p} v_{\text{t,i}}^2 \right)  <  D  <  \Delta_{\text{M}} \left( \bar{x}, U \right)  -  \Delta_+  \left(x,  \bar{x}, U \right) + O \left( \alpha^{1+p} v_{\text{t,i}}^2 \right)   \text{.}
\end{align}
From equations (\ref{vx-open}) and (\ref{D-range}), there is a range of possible particle velocities $v_x$ for open orbits, with maximum given by $-V_{x+} \left( x, \bar{x}, U \right)$, where
\begin{align} \label{Vx+}
V_{x+} \left( x, \bar{x}, U \right) = \sqrt{ 2 \left(  -  \Delta_+ \left(x,  \bar{x}, U \right) + \chi_{\text{M}} \left( \bar{x} \right) - \chi \left( x, \bar{x} \right) \right)  + O \left( \alpha^{1+p} v_{\text{t,i}}^2 \right)  } \text{,}
\end{align}
and with range of values given by
\begin{align} \label{Deltavx}
\Delta v_x = &  \sqrt{ 2 \left(   \Delta_{\text{M}} \left( \bar{x}, U \right) -  \Delta_+ \left(x,  \bar{x}, U \right) + \chi_{\text{M}} \left( \bar{x} \right) - \chi \left( x, \bar{x} \right) \right) + O \left( \alpha^{1+p} v_{\text{t,i}}^2 \right)   } \nonumber \\
 & - \sqrt{ 2 \left(  -  \Delta_+ \left(x,  \bar{x}, U \right) + \chi_{\text{M}} \left( \bar{x} \right) - \chi \left( x, \bar{x} \right) \right)  + O \left( \alpha^{1+p} v_{\text{t,i}}^2 \right)  } \text{,}
\end{align}
such that
\begin{align} \label{vx-range}
- V_{x+} \left( x, \bar{x}, U \right) - \Delta v_x <  v_x <  - V_{x+} \left( x, \bar{x}, U \right)  \text{.}
\end{align}
Note that equations (\ref{Vx+})-(\ref{vx-range}) are defined, for a given $\bar{x}$ and $U$, in the region $0 \leqslant x \leqslant x_{\text{t,M}+}$, where $ x_{\text{t,M}} - x_{\text{t,M}+}  \sim \alpha \rho_{\text{i}}$ and $ x_{\text{t,M}+}$ is obtained by setting $V_{x+}  \left( x_{\text{t,M}+}, \bar{x}, U \right)  $ to zero, $\chi_{\text{M}} (\bar{x} ) - \chi  \left( x_{\text{t,M+}} ,  \bar{x} \right)  - \Delta_+ \left(x_{\text{t,M+}} ,  \bar{x}, U \right) = 0$.
In Section \ref{subsec-open-density}, we will obtain a useful approximation to equations (\ref{Vx+})-(\ref{vx-range}) which eliminates the dependence on $\Delta_+$ and is defined in the region $0 \leqslant x \leqslant x_{\text{t,M}}$ (instead of $0 \leqslant x \leqslant x_{\text{t,M}+}$).

\section{Ion distribution function and density}
 
Suppose that the plasma entering the magnetic presheath, at $x \rightarrow \infty$, has an ion species whose distribution function is $f_{\infty} \left( v_x , v_y, v_z \right)$.
This function is re-expressed in terms of the variables $\mu$ and $U$ by applying the change of variables $\left( v_{x}, v_{y} , v_{z} \right) \rightarrow \left( \varphi, \mu, U \right)$ at $x \rightarrow \infty$. 
The adiabatic invariant and total energy at $x\rightarrow \infty$ are given by $\mu = (v_x^2 + v_y^2)/2\Omega$ and $U = (v_x^2 + v_y^2 + v_z^2)/2$, as shown in \ref{subapp-mu-expansion}.
%To lowest order in $\alpha$, we have
%\begin{align} \label{vx-infty}
%v_{x} = - \sqrt{2\mu \Omega}  \sin \varphi \text{,}
%\end{align}
%\begin{align} \label{vy-infty}
%v_y = - \sqrt{2\mu\Omega} \cos \varphi   \text{,}
%\end{align}
%\begin{align} \label{vz-infty}
%v_{z} = \sqrt{2\left( U - \mu\Omega \right)}   \text{.}
%\end{align}
The distribution function must be independent of gyrophase $\varphi$ to lowest order in $\alpha$ \cite{Geraldini-2017}, hence the result of the change of variables is a function of $\mu$ and $U$ only,
 % because the gyration timescale is faster than any other timescale
\begin{align} \label{F-closed}
F_{\text{cl}} \left( \mu , U \right)  \simeq f_{\infty} \left( v_x , v_y, v_z \right) \text{.} 
\end{align}
The subscript ``cl'' in equation (\ref{F-closed}) is short for ``closed'', because $F_{\text{cl}}$ refers to the distribution function of approximately closed orbits.
%Ions entering the magnetic presheath have $v_{z} > 0$ (corresponding to $\sigma_{\parallel} = 1$ and $\dot{\bar{x}} < 0$) at $x \rightarrow \infty$.
Using conservation of the two invariant quantities $\mu$ and $U$, the distribution function of ions in the magnetic presheath is $F_{\text{cl}} (\mu, U)  $ to lowest order in $\alpha$ \cite{Cohen-Ryutov-1998, Geraldini-2017}. In this section, we obtain expressions for the density of ions in approximately closed and open orbits in terms of this distribution function.

\subsection{Closed orbit ion density}
 
Using equation (\ref{mu-Uperp-xbar}) for $\mu_{\text{gk}} \left( \bar{x}, U_{\perp} \right)$, and equations (\ref{xbar-def})-(\ref{U-def}) for the change of variables $\left( v_x , v_y, v_z \right) \rightarrow \left( \bar{x}, U_{\perp}, U \right) $, we obtain the distribution function of ions in approximately closed orbits,
\begin{align} \label{fclosed}
f_{\text{cl}} (x, v_x, v_y, v_z) \simeq F_{\text{cl}} \left( \mu_{\text{gk}} \left( \bar{x}, U_{\perp}  \right) , U \right)  \Theta \left( \bar{x} - \bar{x}_{\text{m}}(x) \right) \Theta \left( \chi_{\text{M}} (\bar{x}) - U_{\perp} \right)    \text{,}
\end{align}
where $\Theta$ is the Heaviside step function,
\begin{align} \label{Heaviside}
\Theta ( y ) = \begin{cases} 1 & \text{ for } y \geqslant 0 \text{,} \\
0 & \text{ for } y < 0 \text{.}
\end{cases}
\end{align}
The function $\Theta \left( \bar{x} - \bar{x}_{\text{m}}(x) \right)$ is necessary to consider only values of $\bar{x}$ for which closed orbits that cross position $x$ can exist.
An ion in a closed orbit must be in the region enclosed by the largest possible orbit, $x_{\text{M}} \leqslant x \leqslant x_{\text{t,M}}$, which leads to $\bar{x} > \bar{x}_{\text{m}} \left( x \right) $ \cite{Geraldini-2017}, with
\begin{align} \label{xbarm-general}
\bar{x}_{\text{m}} \left( x \right) = \min_{s \in [0,x)} \left\lbrace \frac{1}{2} \left( x + s \right) + \frac{\phi(x) - \phi(s)}{\Omega B \left( x-s \right)}  \right\rbrace \text{.}  
\end{align}
The function $\Theta \left( \chi_{\text{M}} (\bar{x}) - U_{\perp} \right)$ is necessary to consider only values of $U_{\perp}$ for which a pair of bounce points $x_{\text{b}}$ and $x_{\text{t}}$ exist. 
%Finally, $ \Theta \left( x - x_{\text{M}} \right) $ is necessary to avoid including ions in the region $x<x_{\text{M}}$ which is not accessible by closed orbits. %because for type II effective potential curves $x<x_{\text{M}}$ is associated with $U_{\perp} < \chi_{\text{M}}$.
%\begin{align} \label{fclosed}
%F_{\text{cl}} (x, v_x, v_y, v_z) \simeq F\left( \mu_{\text{gk}} \left( x + \frac{v_y}{\Omega}, \frac{v_x^2 + v_y^2}{2}  + \frac{\Omega \phi \left( x \right) }{B}  \right) , \frac{\left| \vec{v} \right|^2}{2} + \frac{\Omega \phi \left( x \right) }{B} \right)  \text{.}
%\end{align}
The density of ions crossing position $x$ in approximately closed orbits is an integral in velocity space of the distribution function (\ref{fclosed}),
\begin{align}
n_{\text{i,cl}} (x) = \int  f_{\text{cl}} (x, \vec{v})  d^3v   \text{.}
\end{align}
Changing to the set of variables $(U_{\perp}, \bar{x}, U)$ \cite{Geraldini-2017}, we obtain
\begin{align} \label{ni-closed}
n_{\text{i,cl}} (x) \simeq \int_{\bar{x}_{\text{m}}(x)}^{\infty} \Omega d\bar{x} \int_{\chi\left( x, \bar{x} \right)}^{\chi_{\text{M}}(\bar{x})} \frac{2dU_{\perp}}{\sqrt{2\left(U_{\perp} - \chi\left(x, \bar{x} \right) \right)}} \int_{U_{\perp}}^{\infty} \frac{F_{\text{cl}} \left( \mu_{\text{gk}} \left( \bar{x}, U_{\perp} \right), U \right) }{\sqrt{2\left( U - U_{\perp} \right)}} dU  \text{.}
\end{align}

 It is worth noting that $n_{\text{i,cl}} (0) = 0$, because for type I orbits $\chi_{\text{M}} (\bar{x}) = \chi (0, \bar{x} )$ while for type II orbits $x=0 < x_{\text{M}} $. The fact that $n_{\text{i,cl}} (0) = 0$ means that we cannot naively impose quasineutrality with only the approximately closed orbit contribution to the ion density. % in order to obtain the lowest order contribution to the ion density at $x=0$ (and sufficiently near it) we must find the density of ions in open orbits. This is because 
An attempt to impose $Zn_{\text{i,cl}} (0) = n_{\text{e}} \left( 0 \right)$ leads to $n_{\text{e}} \left( 0 \right) = n_{e\infty} \exp\left( e\phi ( 0 ) / T_{\text{e}} \right) = 0$ and therefore $\phi ( 0 ) = - \infty$. This is an unphysical result which stems from the fact that we have not kept the dominant contribution to the ion density at (and near) the wall, which comes from ions in open orbits.

\subsection{Open orbit ion density} \label{subsec-open-density}

Consider an ion at position $x$ in an open orbit, when $U_{\perp } = \chi_{\text{M}} \left( \bar{x} \right) + D$ and $D$ lies in the range (\ref{D-range}). 
The ion transitioned from being in a closed orbit to being in an open orbit a time $\sim \delta t_{\text{M}} $ before the instant in time that we consider. 
At this time, the orbit position differed from $\bar{x}$ by $O\left(\alpha \Omega \delta t_{\text{M}} \rho_{\text{i}} \right) $, which is small. To lowest order, the ion conserved its adiabatic invariant up to the point where $U_{\perp} = \chi_{\text{M}} (\bar{x})$. 
Using $U_{\perp} \simeq \chi_{\text{M}} \left( \bar{x} \right)$, the adiabatic invariant of the ion was $\mu_{\text{gk}} \left( \bar{x}, \chi_{\text{M}} \left( \bar{x} \right) \right) + O\left( \alpha \Omega \delta t_{\text{M}} v_{\text{t,i}} \rho_i \right)$. Hence, the distribution function is $F_{\text{cl}} \left( \mu_{\text{gk}} \left( \bar{x}, \chi_{\text{M}} \left( \bar{x} \right) \right) , U \right) $ to lowest order, which is independent of the value of $D$ \cite{Cary-1986, Neishtadt-1987}. 

For an ion in an open orbit to be at position $x$, the range of possible values of $\bar{x}$ (to lowest order) is determined by two constraints. A time $\sim \delta t_{\text{M}}$ before being in an open orbit, the ion must have been in an approximately closed orbit whose existence depends on the presence of an effective potential minimum. Hence, we require a stationary point to exist, which implies that $\bar{x} > \bar{x}_{\text{c}}$ is necessary. Moreover, we require that $x< x_{\text{t,M}}$.
For $x< x_{\text{c}} $, it is impossible for an ion to be in the region $x>x_{\text{t,M}}$ because $x_{\text{c}} \leqslant x_{\text{m}} \leqslant x_{\text{t,M}} $, and therefore $\bar{x} > \bar{x}_{\text{c}}$ is the necessary and sufficient condition for an open orbit crossing position $x$ in this case. 
For $x> x_{\text{c}}$, we use the fact that $x_{\text{M}} < x_{\text{c}}$ to conclude that the ion must be in the region $x_{\text{M}} < x < x_{\text{t,M}}$; the criterion for an open orbit crossing position $x$ is therefore identical to that of a closed orbit crossing position $x$, $\bar{x} > \bar{x}_{\text{m}} (x)$.
%the top bounce point $x_{\text{t}}$ corresponding to $U_{\perp} = \chi_{\text{M}} \left( \bar{x} \right) = \chi \left( x_{\text{t}}, \bar{x} \right)$ is larger than $x$.
Therefore, the condition for an ion in an open orbit to be present at position $x$ is $\bar{x} > \bar{x}_{\text{m,o}} \left( x \right)$, where
\begin{align} \label{xbarm-open}
\bar{x}_{\text{m,o}} (x) = \begin{cases} 
\bar{x}_{\text{c}}  & \text{ for } x < x_{\text{c}}  \\
\bar{x}_{\text{m}} (x) &  \text{ for } x \geqslant  x_{\text{c}} \text{.}  
\end{cases}
\end{align}
Two examples of how the constraint $\bar{x} > \bar{x}_{\text{m,o}} \left( x \right)$ arises are shown in Figure \ref{fig-effpot-open}.
This constraint is valid to lowest order in $\alpha \Omega \delta t_{\text{M}}$. For any $\bar{x}$ larger than $\bar{x}_{\text{m,o}}$, the component $v_y$ of the velocity is given by (\ref{vy-xbar-x}). The ion's total energy has to be larger than the effective potential maximum, $U > \chi_{\text{M}} \left( \bar{x} \right)$, and we can approximate the $z$ component of the velocity as $ V_{\parallel} \left( \chi_{\text{M}} \left( \bar{x} \right) , U \right) $. In order to relate values of $v_y$ and $v_z$ to lowest order values of $\bar{x}$ and $U$ for ions in open orbits, in what follows we will refer extensively to equations (\ref{xbar-def}) and
\begin{align} \label{U-open}
U = \chi_{\text{M}} \left( \bar{x} \right) + \frac{1}{2} v_z^2 + O\left( \alpha v_{\text{t,i}}^2 \right) \text{,}
\end{align} 
where the latter equation is obtained by rearranging the equation $v_z \simeq V_{\parallel} \left( \chi_{\text{M}} \left( \bar{x} \right) , U \right) $.

\begin{figure}
\centering
\includegraphics[width= 0.8\textwidth]{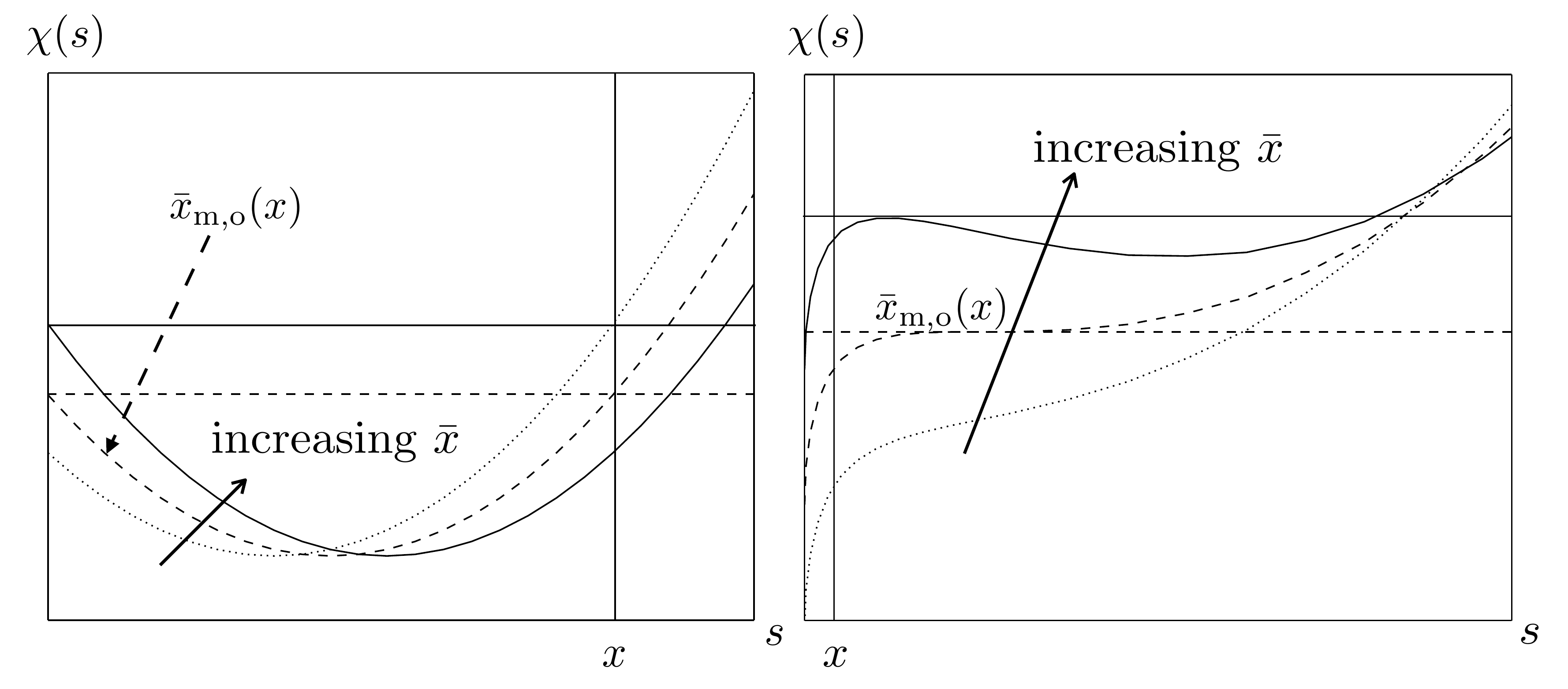}
\caption{Type I and II effective potential curves are shown on the left and right respectively. The dashed curves correspond to an orbit position $\bar{x} = \bar{x}_{\text{m,o}} \left( x \right) $, which is the minimum value of $\bar{x}$ above which open orbits crossing the position $x$ (vertical line) exist. The solid effective potential curves are the ones corresponding to $\bar{x} > \bar{x}_{\text{m,o}} (x)$. The horizontal lines correspond to $U_{\perp} = \chi_{\text{M}} \left( \bar{x} \right)$, which is the lowest order perpendicular energy of an ion in an open orbit. The dotted curves correspond to $\bar{x} < \bar{x}_{\text{m,o}} \left( x\right)$: no open orbits crossing position $x$ exist for such values of $\bar{x}$ because there are no closed orbits at $s\geqslant x$.
}
\label{fig-effpot-open}
\end{figure}

The velocity component $v_x$ lies in the range (\ref{vx-range}), which is obtained from the range of values of $D$ for given values of $x$, $\bar{x}$, and $U$. For the evaluation of the distribution function and density of ions in open orbits, the value of $\Delta v_x$ is crucial because  at a given $x$, $\bar{x}$ and $U$ it gives the \emph{small} range of values of $v_x$ in which the distribution function is non-zero.
The \emph{exact} value of the maximum and minimum $v_x$ only needs to be known to lowest order.
Hence, we can shift $V_{x+} \left( x, \bar{x}, U\right)$ by a small amount provided we preserve the same value of $\Delta v_x$.
With this in mind, we proceed to obtain simpler expressions for $V_{x+} \left( x, \bar{x}, U\right)$ and $\Delta v_x$. 
We need to distinguish two regions: $|x-x_{\text{M}}| \sim \rho_{\text{i}}$ where $\chi_{\text{M}} (\bar{x} ) - \chi (x, \bar{x}) \sim v_{\text{t,i}}^2$, and $|x-x_{\text{M}}| \sim \alpha^p \rho_{\text{i}}$ where $\chi_{\text{M}} (\bar{x} ) - \chi (x, \bar{x})  \sim \alpha v_{\text{t,i}}^2$ (with $p$ defined in equation (\ref{p-def})).

In the region $|x - x_{\text{M}} | \sim \rho_i$, we have
\begin{align} \label{outer-ordering-2}
\alpha^{1+p} v_{\text{t,i}}^2 \ll \Delta_{\text{M}}  \sim \Delta_+  \sim \alpha v_{\text{t,i}}^2 \ll \chi_{\text{M}}  - \chi  \sim v_{\text{t,i}}^2 \text{.}
\end{align}
By using equations (\ref{Vx+}) and (\ref{Deltavx}), the ordering (\ref{outer-ordering-2}) leads to
\begin{align} \label{outer-ordering-1}
 \Delta v_x \sim \alpha  v_{\text{t,i}} \ll  V_{x+}  \sim  v_{\text{t,i}} \text{.} 
 \end{align}
When we neglect the term $\Delta_+ \sim \alpha v_{\text{t,i}}^2$ in the square root of equation (\ref{Vx+}) for $V_{x+}$, we obtain 
\begin{align}  \label{Vx+-simpler}
V_{x+} \left( x, \bar{x}, U \right) & = \sqrt{ 2 \left(   \chi_{\text{M}} \left( \bar{x} \right) - \chi \left( x, \bar{x} \right) \right)  + O \left( \alpha v_{\text{t,i}}^2 \right)  } \nonumber \\  
&  = V_x \left( x, \bar{x}, \chi_{\text{M}} \left( \bar{x} \right) \right) +  O\left( \alpha v_{\text{t,i}}  \right)   \text{.}
\end{align}
%equation (\ref{Vx+-simpler}), which is valid in both regions $|x - x_{\text{M}}| \sim \alpha^p \rho_{\text{i}}$ and $|x - x_{\text{M}}| \sim \rho_{\text{i}}$ with the same error.
%\begin{align}  \label{Vx+-simpler}
%V_{x+} \left( x, \bar{x}, U \right) & = \sqrt{ 2 \left(   \chi_{\text{M}} \left( \bar{x} \right) - \chi \left( x, \bar{x} \right) \right)  + O \left( \alpha v_{\text{t,i}}^2 \right)  } \nonumber \\  
%&  = V_x \left( x, \bar{x}, \chi_{\text{M}} \left( \bar{x} \right) \right) +   O\left( \alpha v_{\text{t,i}} \right)   \text{.}
%\end{align}
%Note that for type I orbits in the region $|x - x_{\text{M}} | \sim \alpha \rho_{\text{i}}$ the error in $V_{x+}$ is, from the earlier discussion, $O(\alpha^{3/2} v_{\text{t,i}})$, but the error arising in the region $|x - x_{\text{M}} | \sim \rho_{\text{i}}$ is $O\left( \alpha v_{\text{t,i}} \right)$; therefore equation (\ref{Vx+-simpler}) and its associated error are correct in both regions. 
%For type II orbits, the error in equation (\ref{Vx+-simpler}) has the same size in both regions.
When we expand the terms $\Delta_{\text{M}}$ and $\Delta_+$ out of the square root in equation (\ref{Deltavx}) for $\Delta v_x$ using the ordering (\ref{outer-ordering-2}), we obtain
\begin{align} \label{Deltavx-simplifying}
\Delta v_x  = & \left[  \sqrt{ 2 \left(  \Delta_{\text{M}} \left( \bar{x}, U \right) - \Delta_+ \left( x, \bar{x}, U \right)  + \chi_{\text{M}} \left( \bar{x} \right) - \chi \left( x, \bar{x} \right) + O\left( \alpha^{1+p} v_{\text{t,i}}^2 \right)  \right)  } \right.  \nonumber \\ & \left.  - \sqrt{ 2 \left(  -\Delta_+ \left( x, \bar{x}, U \right) + \chi_{\text{M}} \left( \bar{x} \right)  - \chi \left( x, \bar{x} \right) + O\left( \alpha^{1+p} v_{\text{t,i}}^2 \right) \right)  } \right]  \nonumber \\
= & \frac{  \Delta_{\text{M}} \left( \bar{x}, U \right)  }{ \sqrt{2\left( \chi_{\text{M}}(\bar{x}) - \chi(x, \bar{x}) \right) } } \left( 1 + O\left( \alpha^{p}  \right) \right) \text{.}
\end{align} 
Note that the terms proportional to $\Delta_+$ have cancelled to first order, and the error in the last line of (\ref{Deltavx-simplifying}) comes from the $O\left( \alpha^{1+p} v_{\text{t,i}}^2 \right)$ error in the range of values of $D$ (see equation (\ref{D-range})). 
For convenience, we re-express (\ref{Deltavx-simplifying}) to the form
\begin{align} \label{Deltavx-simpler}
\Delta v_x  = & \left[  \sqrt{ 2 \left(  \Delta_{\text{M}} \left( \bar{x}, U \right) + \chi_{\text{M}} \left( \bar{x} \right) - \chi \left( x, \bar{x} \right) \right)  } \right.  \nonumber \\ & \left.  - \sqrt{ 2 \left(  \chi_{\text{M}} \left( \bar{x} \right) - \chi \left( x, \bar{x} \right) \right)  } \right] \left( 1 +  O\left( \alpha^{p}  \right) \right) \text{.}
\end{align} 
%The relative error from neglecting terms proportional to $\Delta_+^2$ is $O\left(\alpha\right)$, which is equivalent for type I orbits with $p=1$.
%, which for type I orbits ($p=1$) is larger than $\Delta_+ \sim \alpha^{2p+1/2}$, and for type II orbits ($p=1/2$) is as large as $\Delta_+$.
% recover equation (\ref{Deltavx-simpler}). 
%In principle we could expand $\Delta_{\text{M}}$ out of the square root owe don't because equation (\ref{Deltavx}).
%also holds in the region $|x-x_M| \sim \alpha^p \rho_{\text{i}}$, where $\Delta_+$ could be neglected.
%\begin{align} \label{Deltavx-simpler}
%\Delta v_x  = & \left[  \sqrt{ 2 \left(  \Delta_{\text{M}} \left( \bar{x}, U \right) + \chi_{\text{M}} \left( \bar{x} \right) - \chi \left( x, \bar{x} \right) \right)  } \right.  \nonumber \\ & \left.  - \sqrt{ 2 \left(  \chi_{\text{M}} \left( \bar{x} \right) - \chi \left( x, \bar{x} \right) \right)  } \right] \left( 1 +  O\left( \alpha^{p}  \right) \right) \text{}
%\end{align}
%The error in equations (\ref{}) and (\ref{Vx+-simpler}) is different depending on whether the open orbit is type I or type II. 

We proceed to show that equations (\ref{Vx+-simpler}) and (\ref{Deltavx-simpler}) are also valid in the region $|x-x_{\text{M}}| \sim \alpha^p \rho_{\text{i}}$.
In this region, we have the scalings
%\begin{align}
% \Delta_+ \sim \alpha^{2p+1/2} v_{\text{t,i}}^2 \lesssim \alpha^{1+p} v_{\text{t,i}}^2  \ll \Delta_M \sim \chi_{\text{M}}  - \chi  \sim \alpha v_{\text{t,i}}^2
% \end{align}
\begin{align} \label{inner-ordering}
 \Delta_+  \lesssim \alpha^{1+p} v_{\text{t,i}}^2  \ll \Delta_{\text{M}} \sim \chi_{\text{M}}  - \chi  \sim \alpha v_{\text{t,i}}^2 \text{.}
 \end{align}
From equations (\ref{Vx+}), (\ref{Deltavx}) and (\ref{inner-ordering}) we have
\begin{align} \label{inner-ordering-1}
 \Delta v_x \sim  V_{x+}  \sim  \alpha^{1/2} v_{\text{t,i}} \text{.} 
 \end{align}
The term $\Delta_+$ in the ordering (\ref{inner-ordering}) is small because the range of integration in equation (\ref{Delta+}) is small. 
Importantly, the $O\left( \alpha^{1+p} v_{\text{t,i}}^2 \right) $ error in the range of possible values of $D$ is larger than (or comparable to) $\Delta_+$.
Hence, the term $\Delta_+ $ is negligible in equations (\ref{Vx+}) and (\ref{Deltavx}), and equations (\ref{Vx+-simpler}) and (\ref{Deltavx-simpler}) are valid in the region $|x-x_{\text{M}}| \sim \alpha^p \rho_{\text{i}}$.

%The ion density near $x_{\text{t,M}}$ is expected to be dominated by closed orbits, hence the inaccuracy in the small number of open orbits does not matter.
%Moreover, we will argue after equation (\ref{ni-open-order}) that the contribution to the open orbit density from the region $| x - x_{\text{t,M}} | \sim \alpha \rho_{\text{i}} $ is small.
%The error in equation (\ref{Vx+-simpler}) is obtained by considering the worst case of $p=1/2$ for type II orbits.
%This error comes from the $O\left( \alpha^{1+p} v_{\text{t,i}}^2 \right)$ error in evaluating $D_X$ and $D_B$ (equations (\ref{D-X}) and (\ref{D-turn})), which for type I orbits ($p=1$) is larger than $\Delta_+ \sim \alpha^{2p+1/2}$, and for type II orbits ($p=1/2$) is as large as $\Delta_+$.

\begin{figure}[h]
\centering
\includegraphics[width=0.75\textwidth]{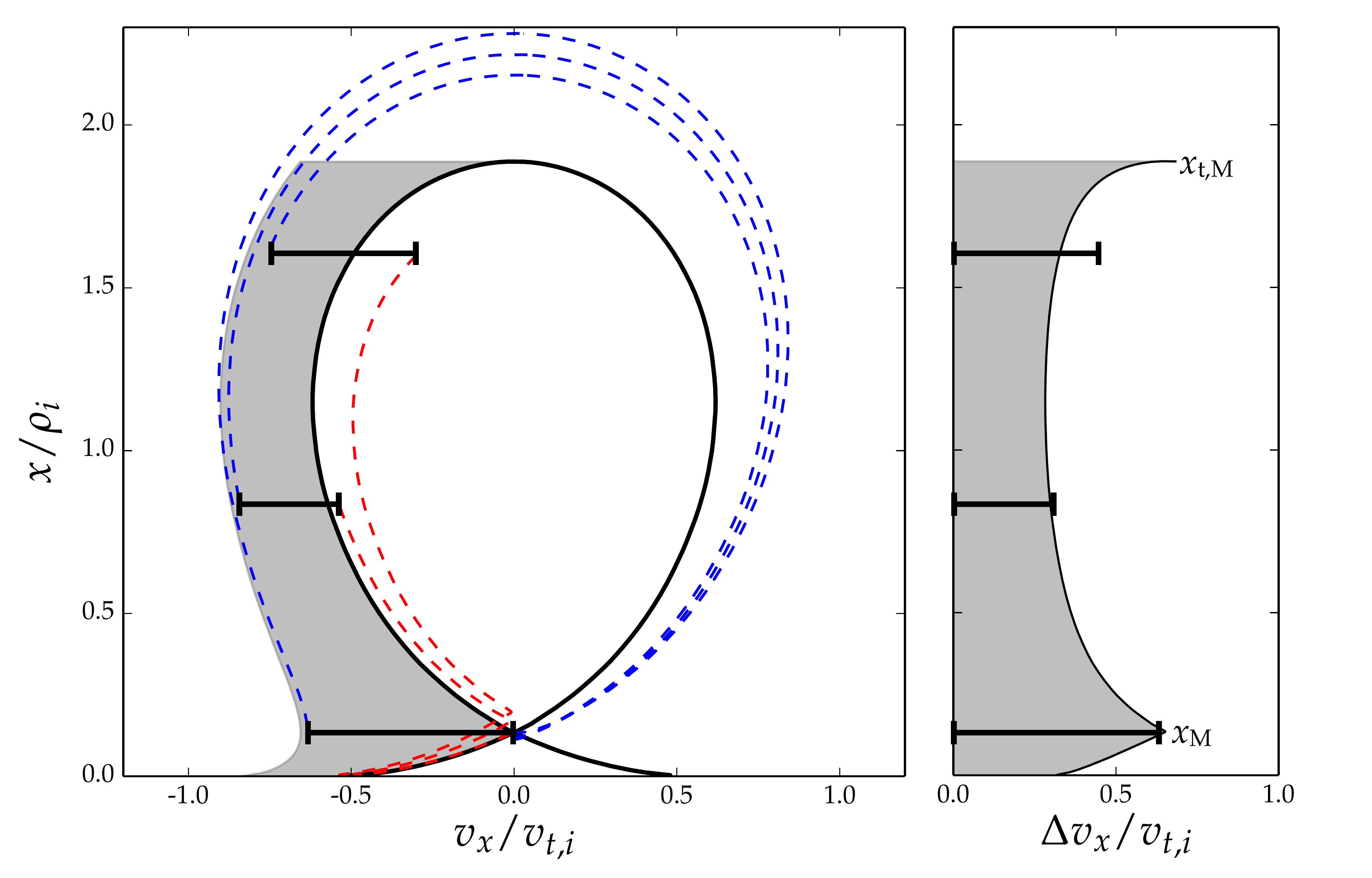}
\caption{The left diagram shows ion trajectories (dashed lines) for $\alpha = 0.02$, obtained using equations (\ref{x-EOM-exact}), (\ref{vx-Uperp-xbar-x}) and (\ref{xbardot})-(\ref{Uperpdot}). At a given time, the trajectories have $\bar{x} = 1.6 \rho_{\text{i}}$ and $U - \chi_{\text{M}} (\bar{x}) =  v_{\text{t,i}}^2$ at three different positions (marked with a thick black line). Blue lines are past ion trajectories chosen to have the largest value of $U_{\perp}$ for which a bottom bounce point exists. Red lines are future ion trajectories chosen to have the smallest value of $U_{\perp}$ for which the ion crosses the effective potential maximum $x_{\text{M}}$ and reaches the wall. The thick black lines connect the red and blue trajectories at the three positions, thus they measure the difference between the maximum and minimum velocities of the open orbits. The shaded region on the left is $- V_x \left( x, \bar{x}, \chi_{\text{M}} \left( \bar{x} \right) \right) - \Delta v_x  <  v_x <  - V_x \left( x, \bar{x}, \chi_{\text{M}} \left( \bar{x} \right) \right)$. On the right diagram, the difference between the maximum and minimum velocities of the open orbits at the three values of $x$ is compared to the width of the shaded region, given by $\Delta v_x$. 
%The approximation is not good at the position which is closest to $x_{\text{t,M}}$.
}
\label{fig-Deltavx}
\end{figure}

In the above discussion we neglected the factor of $2\pi$ in the scaling $\Delta_{\text{M}} \sim 2\pi \alpha v_{\text{t,i}}^2$ of equation (\ref{DeltaM}). 
From equation (\ref{Deltavx-simpler}) we obtain, when we include this factor, the scaling 
\begin{align} \label{Deltavx-size}
2\pi \alpha v_{\text{t,i}} \lesssim \Delta v_x \lesssim \sqrt{2\pi \alpha } v_{\text{t,i}} \text{,}
\end{align} 
where $\Delta v_x \sim \sqrt{2\pi \alpha } v_{\text{t,i}}$ holds in the neighbourhood of the effective potential maximum $x_{\text{M}}$, while $\Delta v_x \sim 2\pi \alpha  v_{\text{t,i}}$ holds almost everywhere else.
The behaviour of $\Delta v_x$ as a function of $x$ is shown in Figure~\ref{fig-Deltavx}.
Note that there is a small region near the top bounce point that satisfies $|x-x_{\text{t,M}}| \sim \alpha \rho_{\text{i}} $ in which $\Delta v_x \sim \sqrt{2\pi \alpha } v_{\text{t,i}}$. 
In this region, equations (\ref{Vx+-simpler}) and (\ref{Deltavx-simpler}) are not valid because $ \Delta_+  \sim \Delta_{\text{M}} \sim \chi_{\text{M}}  - \chi  \sim \alpha v_{\text{t,i}}^2$ and thus $\Delta_+$ cannot be neglected.
However, we will argue after equation (\ref{ni-open-order}) that the contribution to the density of ions in an open orbit due to this region at a given position $x$ is small.
Recall that $\Delta v_x$, calculated from equation (\ref{Deltavx-simpler}), should be equal to the difference between the maximum and minimum velocity that an open orbit with a given $\bar{x}$ and $U$ can have.
Indeed, from Figure \ref{fig-Deltavx} we see that $\Delta v_x$ is a good approximation to the range of allowed velocities at two out of three positions shown, and is a bad approximation only at the position close to $x_{\text{t,M}}$. 
%Equation (\ref{Deltavx-simpler}) has a relative error of $O\left( \alpha^{1/2} \right)$ instead of $O\left( \alpha \right)$ in the region $x - x_{\text{M}} \sim \alpha \rho_{\text{i}}$ for type II orbits, but this does not affect the error in the open orbit density and in the distribution function at $x=0$. 

The range of velocities in (\ref{vx-range}) reduces, using equations (\ref{Vx+-simpler}) and (\ref{Deltavx-simpler}), to 
\begin{align} \label{vx-range-simpler}
- V_x \left( x, \bar{x}, \chi_{\text{M}} \left( \bar{x} \right) \right) - \Delta v_x  <  v_x <  - V_x \left( x, \bar{x}, \chi_{\text{M}} \left( \bar{x} \right) \right)  \text{.}
\end{align}
Note the major simplification: equations (\ref{Vx+-simpler}) and (\ref{Deltavx-simpler}), and therefore the range (\ref{vx-range-simpler}), are independent of $\Delta_+$. 
The ``open orbit integral''
\begin{align} \label{open-orbit-integral}
I( \bar{x} ) = \int_{x_{\text{M}}}^{x_{\text{t,M}}} \frac{ \left( s - x_{\text{M}} \right)  }{ V_x \left( s, \bar{x}, \chi_{\text{M}} \left( \bar{x} \right) \right)  } ds 
\end{align} 
is a function of $\bar{x}$ only. Using $I(\bar{x})$ we can re-express $\Delta_{\text{M}}$, defined in equation (\ref{DeltaM}), as $\Delta_{\text{M}} \left( \bar{x}, U \right) = 2\alpha \Omega^2 V_{\parallel} (\chi_{\text{M}} (\bar{x} ) , U ) I( \bar{x} )$.
Equation (\ref{vx-range-simpler}) gives the range of values of $v_x$ for which the distribution function of open orbits is non-zero. Using this range in $v_x$ and $\bar{x} > \bar{x}_{\text{m,o}}(x)$, we have
%\begin{align} \label{fopen}
%f_{\text{open}} (x, v_x, v_y, v_z) \simeq F\left( \mu_{\text{gk}} \left( x + \frac{v_y}{\Omega}, \chi_{\text{M}} \left( x + \frac{v_y}{\Omega} \right) \right) , \chi_M \left( x + \frac{v_y}{\Omega} \right) + \frac{v_z^2}{2}  \right)  \nonumber  \\
% \times \hat{\Pi} \left(  v_x ,  - V_x \left( x, \bar{x}, \chi_{\text{M}} \right) - \Delta v_x  , - V_x \left( x, \bar{x}, \chi_{\text{M}} \right)  \right) \text{,}
%\end{align}
\begin{align} \label{fopen}
f_{\text{op}} (x, v_x, v_y, v_z) \simeq & F_{\text{cl}} \left( \mu_{\text{gk}} \left( \bar{x}, \chi_{\text{M}} \left( \bar{x} \right) \right) , U \right) \Theta \left( \bar{x} - \bar{x}_{\text{m,o}}(x) \right)   \nonumber  \\
 & \times \hat{\Pi} \left(  v_x ,  - V_x \left( x, \bar{x}, \chi_{\text{M}} (\bar{x}) \right) - \Delta v_x  , - V_x \left( x, \bar{x}, \chi_{\text{M}} (\bar{x}) \right)  \right) \text{,}
\end{align}
where we defined the top-hat function $\hat{\Pi}\left(r, l_1, l_2\right)$ as
\begin{align}
\hat{\Pi} \left( r, l_1, l_2 \right) = 
\begin{cases}
1 & \text{ if } l_1 \leqslant r < l_2 \text{,} \\
0 & \text{ else.} 
\end{cases}
\end{align}
In equation (\ref{fopen}) we can use (\ref{xbar-def}) and (\ref{U-open}) to re-express $\bar{x}$ and $U$ in terms of $x$, $v_y$ and $v_z$. The subscript ``op'' stands for ``open''.
The density of ions in open orbits is an integral of the distribution function in velocity space at fixed $x$, hence
\begin{align}
n_{\text{i,op}}  \left( x \right) = \int f_{\text{op}} \left( x, \vec{v} \right) d^3v
 \text{.}
\end{align}
Changing variables in the integral using equations (\ref{xbar-def}) and (\ref{U-open}) we get
\begin{align} \label{ni-open-Deltavx}
n_{\text{i,op}} \left( x \right) = \int_{\bar{x}_{\text{m,o}}(x)}^{\infty} \Omega d\bar{x} \int_{\chi_{\text{M}} \left( \bar{x}  \right) }^{\infty} \frac{F_{\text{cl}} \left( \mu_{\text{gk}} (\bar{x}, \chi_{\text{M}} (\bar{x}) ), U \right) }{\sqrt{2\left( U - \chi_{\text{M}} \left( \bar{x} \right) \right)}}  \Delta v_x  dU \left[ 1 + O  \left( \alpha^{p} \right) \right]  \text{.}
\end{align}
% Explicitly, the open orbit density is then
%\begin{align} \label{ni-open}
%& n_{i,\text{op}} \left( x \right) = \int_{\bar{x}_{\text{m,o}}}^{\infty} \Omega d\bar{x}
%\int_{\chi_{\text{M}} \left( \bar{x} \right) }^{\infty} 
% \frac{F_{\text{cl}} \left( \mu_{\text{gk}} \left( \bar{x}, \chi_{\text{M}} \left( \bar{x} \right) \right), U \right) }{\sqrt{2\left( U - \chi_{\text{M}} \left( \bar{x} \right) \right)}} dU \nonumber  \\
%& \times  \left( \sqrt{ 2 \left(  \chi_{\text{M}} \left(\bar{x}  \right) - \chi \left( \bar{x} , x \right) + \Delta_M \left( \bar{x}, U \right)  \right) } -  \sqrt{ 2 \left(  \chi_{\text{M}} \left(\bar{x}  \right) - \chi \left( \bar{x}, x \right)  \right) } \right) \left[ 1 + O  \left( \alpha^{p} \right) \right] \text{.}
%\end{align}

From equations (\ref{Deltavx-size}) and (\ref{ni-open-Deltavx}), the characteristic size of the open orbit density is
\begin{align} \label{ni-open-order}
\alpha   n_{\text{e}\infty}  \lesssim n_{i,\text{op}} (x) \lesssim \alpha^{1/2} n_{\text{e}\infty} \text{.}
\end{align}
The ordering $ n_{i,\text{op}} (x) \sim \alpha^{1/2} n_{\text{e}\infty} $ is valid for $x \lesssim \alpha \rho_{\text{i}}$ only if there is a sufficiently large number of type I orbits, that is, $\bar{x}_{\text{m,I}} \sim \rho_{\text{i}}$ (see Figure \ref{fig-typesofcurves}). 
Type I effective potential curves have $x_{\text{M}} = 0$ by definition, so all type I ion orbits must cross the effective potential maximum at the same position $x=0$, with a range of values of $v_x$ given by $\Delta v_x \sim \alpha^{1/2} v_{\text{t,i}}$.
For type II orbits, the open orbit density is always $ n_{i,\text{op}} (x) \sim \alpha n_{\text{e}\infty} $ because ions with different values of $\bar{x}$ cross the effective potential maximum at different locations $x_{\text{M}} $. 
At some position $x$, there is a small range of values of $x_{\text{M}}$ (and therefore of $\bar{x}$), given by $|x-x_{\text{M}}| \sim \alpha^{1/2} \rho_{\text{i}}$, in which $\Delta v_x\sim \alpha^{1/2} v_{\text{t,i}}$. 
Multiplying the factor $\alpha^{1/2}$ from the range of values of $x_{\text{M}}$ by the factor $\alpha^{1/2}$ from the size of $\Delta v_x$ gives a contribution of order $\alpha n_{\text{e}\infty}$ to the ion density from ions in the region $|x-x_{\text{M}}| \sim \alpha^{1/2} \rho_{\text{i}}$. % from the slow ions crossing the maximum. % due to ions crossing the effective potential maximum.
%Consider the motion of the ions in the $x$ direction, normal to the wall.
Physically, the ions approach the wall more slowly near the effective potential maximum (where $v_x$ is smaller), leading to a larger number of ions in this region due to flux conservation.
However, ions in type II orbits slow down at different locations depending on their orbit position $\bar{x}$.
Thus, there is not a \emph{single} location where the ions in type II orbits accumulate.
Therefore, their contribution to the density has the same characteristic size at all values of $x$.
Conversely, ions in type I orbits are all slowly crossing the effective potential maximum at the same position $x=0$, and therefore their contribution to the density at $x=0$ is larger.
Despite the fact that $\Delta v_x \sim \alpha^{1/2} v_{\text{t,i}}$ near $x_{\text{t,M}} $, the contribution to the density from ions in this region is of order $\alpha^{3/2} n_{\text{e}\infty} $ because the ions must be very close to $x_{\text{t,M}}$ for $\Delta v_x$ to be large, that is, $|x-x_{\text{t,M}}| \sim \alpha \rho_{\text{i}}$. 
Consequently, the fact that $\Delta v_x$ is a bad approximation to the range of values of $v_x$ near $x_{\text{t,M}}$ (Figure \ref{fig-Deltavx}) is unimportant. 
%does not affect the open orbit density to lowest order.

 \section{Quasineutrality } \label{sec-quasi}

The previous section provides the equations from which the ion distribution function and density can be obtained across the magnetic presheath \text{if} the electrostatic potential profile and the distribution function at $x \rightarrow \infty$ are known. However, the electrostatic potential is not known a priori, but has to be determined by the quasineutrality equation. With the electron density given by (\ref{ne-Boltzmann}) and the closed and open orbit ion densities given by (\ref{ni-closed}) and (\ref{ni-open-Deltavx}), quasineutrality gives
\begin{align} \label{quasineutrality-compact}
n_{e\infty} \exp \left( \frac{e\phi\left( x \right) }{T_{\text{e}}} \right) = Zn_{\text{i}} \left( x \right) \equiv Z\left( n_{\text{i,cl}}(x) + n_{\text{i,op}}(x) \right) \text{.}
\end{align}

In this section, we expand the quasineutrality equation (\ref{quasineutrality-compact}) near the magnetic presheath entrance $x\rightarrow \infty$ and then near the Debye sheath entrance $x=0$. These expansions are useful to gain analytical insight into the system, and from a more practical point of view, they make the task of finding the numerical solution easier. 
From the expansion near $x\rightarrow \infty$, we deduce:
\begin{itemize}
\item a solvability condition for the distribution function at the magnetic presheath entrance, with which we choose a realistic boundary condition for the ion distribution function at $x\rightarrow \infty$;
\item the form of the electrostatic potential near $x\rightarrow \infty$, which is needed to determine the potential above a certain value of $x$ in our numerical scheme.
\end{itemize}
From the expansion near $x=0$, we deduce:
\begin{itemize}
\item that the self-consistent solution of the system requires the ion distribution function at $x=0$ to marginally satisfy the kinetic Bohm condition, with which we can check the numerically calculated distribution function;
\item the self-consistent form of the potential near $x=0$, with which we choose a suitable numerical discretization for the system.
\end{itemize}

\subsection{Expansion of quasineutrality near $x \rightarrow \infty$ } \label{subsec-quasi-expansion}

At sufficiently large values of $x$, the electrostatic potential must be small, such that
\begin{align} \label{hatphismall}
\hat{\phi} = \frac{e\left| \phi\left(x\right) \right|}{T_{\text{e}}} \ll 1 \text{.}
\end{align}
In this subsection we also assume that the length scale of changes in the electrostatic potential is very large at sufficiently large $x$, such that
\begin{align} \label{epsilonsmall}
\epsilon = \frac{\rho_{\text{i}} \phi'\left( x\right)}{\phi\left( x \right)} \ll 1 \text{.}
\end{align}
Our assumption (\ref{epsilonsmall}) is not the most general one, as $\epsilon$ can be of order unity, but it is useful because it is correct for the boundary condition at $x\rightarrow \infty$ that we choose in Section \ref{sec-numsol}. 
In general, $\hat{\phi } \lesssim \epsilon^2 \lesssim 1 $, but in this paper we take the more constrained limit
\begin{align} \label{epssmallerdelta}
\hat{\phi } \lesssim \epsilon^2 \ll 1 \text{.}
 \end{align}

Near $x \rightarrow \infty$, the open orbit density is higher order in $\alpha$ than the closed orbit density. Moreover, if the distribution function is exponentially decaying with energy, like the one we use, the open orbit density near $x \rightarrow \infty$ is exponentially small because only very large orbits with very large energies can extend all the way from the wall, $x=0$, to points near $x\rightarrow \infty$. 
%$U > U_{\perp} \simeq \chi_{\text{M}} (\bar{x}) \simeq  \Omega^2 \bar{x}^2 / 2 $. 
Using that $ n_{i,\text{open}} (x) \simeq 0$ for large $x$, the closed orbit density is obtained by expanding the near-circular ion orbits about circular orbits, as shown in \ref{app-quasi-expansion}, to obtain
\begin{align} \label{niclosedinfty}
n_{\text{i,cl}} \left( x \right)   = & \left( 1 +  \frac{\phi''(x)}{\Omega B}  \right) \int_{-\pi}^{\pi} d\varphi \int_0^{\infty} \Omega d\mu \left\lbrace \int_{\Omega \mu}^{\infty} \frac{F_{\text{cl}}(\mu, U') }{\sqrt{2\left(U- \Omega \mu \right)}} dU  \right. \nonumber \\
& \left.  - \sqrt{2\delta U_{\perp}} F_{\text{cl}}(\mu, \Omega \mu ) - \delta U_{\perp}  \int_{\Omega \mu}^{\infty} \frac{\partial F_{\text{cl}} \left(\mu, U \right) / \partial U  }{\sqrt{2\left( U -  \Omega  \mu \right)}} dU   \right.  \nonumber \\ 
&  \left.  +  \frac{1}{3} \left( 2\delta U_{\perp} \right)^{3/2} \frac{ \partial F_{\text{cl}} }{\partial U } (\mu, \Omega \mu ) + \frac{1}{2} \delta U_{\perp}^2 \int_{\Omega \mu}^{\infty}  \frac{\partial^2 F_{\text{cl}}(\mu, U)/ \partial U^2 }{\sqrt{2\left( U - \Omega \mu \right)}} dU  \right\rbrace  \nonumber \\  
& + O\left( \hat{\phi} \epsilon^3 n_{e\infty}, \hat{\phi}^2 \epsilon^2 n_{e\infty}, \hat{\phi}^{5/2} n_{e\infty} \right)  \text{,}
\end{align}
where
\begin{align} \label{deltaUperp}
\delta U_{\perp} = -\frac{\Omega \phi\left( x \right) }{B} + \frac{\Omega \phi'\left( x \right) }{B} \sqrt{\frac{2\mu}{\Omega} } \cos \varphi -  \frac{\mu \phi'' \left( x \right) }{2B} \left( 1 + 2 \cos^2 \varphi \right) \nonumber  \\
+ O \left( \hat{\phi} \epsilon^3 v_{t,i}^2, \hat{\phi}^2 \epsilon^2 v_{t,i}^2 \right) \text{.}
\end{align}
Note that equations (\ref{niclosedinfty}) and (\ref{deltaUperp}) are derived to lowest order in $\alpha \ll 1$. The quantity $\delta U_{\perp}$ is defined such that $U_{\perp} = \Omega \mu - \delta U_{\perp}$, and therefore captures the difference between $U_{\perp}$ and $\Omega \mu$ as the ion travels into the magnetic presheath. Outside of the magnetic presheath, at $x \rightarrow \infty$, ion orbits are circular and $U_{\perp} = \Omega \mu $ (using $\phi(\infty) = \phi'(\infty) = \phi''(\infty) = 0$). 

The electron density in (\ref{ne-Boltzmann}) is expanded in $\hat{\phi} \ll 1$ near $x\rightarrow \infty$,
\begin{align} \label{neinfty}
n_{\text{e}} \left( x \right) = n_{\text{e}\infty} + n_{\text{e}\infty} \frac{e\phi(x)}{T_{\text{e}}} + \frac{1}{2}n_{\text{e}\infty}\left( \frac{e\phi(x)}{T_{\text{e}}} \right)^2 + O\left( \hat{\phi}^3 n_{e\infty} \right) \text{.}
\end{align}
Substituting (\ref{niclosedinfty}) and (\ref{neinfty}) in (\ref{quasineutrality-compact}), and using that $n_{\text{i,open}}\left( x \right) = 0$, we obtain the quasineutrality equation expanded in $\hat{\phi}$ and $\epsilon$,
\begin{align} \label{quasineutrality-expanded-full}
& n_{e\infty} + n_{e\infty} \frac{e\phi(x)}{T_{\text{e}}} + \frac{1}{2}n_{e\infty}\left( \frac{e\phi(x)}{T_{\text{e}}} \right)^2 = Z \left( 1 +  \frac{\phi''(x)}{ \Omega B }  \right) \int_{-\pi}^{\pi} d\varphi \int_0^{\infty} \Omega d\mu \nonumber \\
& \times \left\lbrace \int_{\Omega \mu}^{\infty} \frac{F_{\text{cl}}(\mu, U) }{\sqrt{2\left(U-\mu \Omega \right)}} dU - \sqrt{2\delta U_{\perp}} F_{\text{cl}}(\mu, \mu \Omega) - \delta U_{\perp}  \int_{\Omega \mu}^{\infty} \frac{\partial F_{\text{cl}}(\mu, U)/ \partial U }{\sqrt{2\left( U - \mu \Omega \right)}} dU  \right. \nonumber \\ & \left. +  \frac{1}{3} \left( 2\delta U_{\perp} \right)^{3/2} \frac{\partial F_{\text{cl}}}{\partial U} (\mu, \mu \Omega)   + \frac{1}{2} \delta U_{\perp}^2 \int_{\Omega \mu}^{\infty}  \frac{\partial^2 F_{\text{cl}} (\mu, U)/ \partial U^2 }{\sqrt{2\left( U - \mu \Omega \right)}} dU    \right\rbrace 
 \\ & + O\left( \hat{\phi} \epsilon^3 n_{e\infty}, \hat{\phi}^2 \epsilon^2 n_{e\infty}, \hat{\phi}^{5/2} n_{e\infty}  \right)    \text{.}
\end{align}
To zeroth order in $\hat{\phi}$, equation (\ref{quasineutrality-expanded-full}) gives
\begin{align} \label{quasineutrality-expanded-eps0}
Z \int_{-\pi}^{\pi} d\varphi \int_0^{\infty} \Omega d\mu   \int_{\Omega \mu}^{\infty} \frac{F_{\text{cl}}(\mu, U)}{\sqrt{2\left(U - \Omega \mu \right)}} dU  = n_{\text{e}\infty} \text{.}
\end{align}
This is the quasineutrality equation evaluated exactly at $x\rightarrow \infty$, where we have $v_{z} = \sqrt{2\left( U -  \Omega \mu \right)}$.
%In going to the next orders, we will relax the maximal ordering assumption $\epsilon \sim \delta$ and will include the lowest order terms including $\epsilon \ll 1$ and $\delta \ll 1$. Because there is no term of order $\delta$ 
The next order correction to (\ref{quasineutrality-expanded-eps0}) is a term of order $\hat{\phi}^{1/2}$, giving
\begin{align} \label{quasineutrality-eps12}
- Z \int_{-\pi}^{\pi} d\varphi \int_0^{\infty}  \Omega d\mu \sqrt{2\delta U_{\perp}}  F_{\text{cl}}(\mu, \Omega \mu) = 0 \text{.}
\end{align}
The distribution function $F_{\text{cl}}\left(\mu, U \right)$ is non-negative, and hence the integral in (\ref{quasineutrality-eps12}) is zero only if $F_{\text{cl}}(\mu, \Omega \mu ) = 0$ for all possible values of $\mu$. We expect this for an electron-repelling sheath where no ions come back from the magnetic presheath, so $f_{\infty} \left( v_{x}, v_{y}, v_{z} \right) =0$ at $v_{z} < 0$ and therefore $F_{\text{cl}} \left( \mu, \Omega \mu \right) = f_{\infty} \left( v_{x}, v_{y},  0 \right) = 0$.

To next order, $\hat{\phi}$, we collect all terms in (\ref{quasineutrality-expanded-full}) which are proportional to $\phi\left( x\right)$ or one of its derivatives. Integrating by parts 
%\footnote{Re-expressing integrals of the form on the left hand side of (\ref{re-express-Chodura}) to the form on the right hand side using integration by parts is common in the community. However, the form on the left hand side is more general, as has been pointed out by Riemann in the context of the Debye sheath \cite{Riemann-review}.}  
and using $F_{\text{cl}}\left( \mu, \Omega \mu \right) =0$, we have the result
%The term proportional to $\delta U_{\perp}$ is evaluated by substituting (?), taking the integral in gyrophase and integrating by parts in $U$,
%\begin{align}
% & - \Omega \int_{-\pi}^{\pi} d\varphi \int_0^{\infty} \delta U_{\perp} d\mu  \int_{\mu \Omega}^{\infty} \frac{F'(\mu, U) dU}{\sqrt{2\left( U - \mu \Omega \right)}} \\ & = 2\pi \Omega \int_0^{\infty} d\mu \left( \frac{\Omega \phi\left( x \right) }{B} + \mu \frac{\phi'' \left( \bar{x} \right) }{B} \right)   \int_{\mu \Omega}^{\infty} \frac{F(\mu, U) dU}{\left( 2\left( U - \mu \Omega \right) \right)^{3/2}} \text{.}
%\end{align}
\begin{align} \label{re-express-Chodura}
\int_{\mu \Omega}^{\infty} \frac{\partial F_{\text{cl}} (\mu, U) / \partial U }{\sqrt{2\left( U - \mu \Omega \right)}} dU  = \int_{\mu \Omega}^{\infty} \frac{F_{\text{cl}}(\mu, U) }{\left( 2\left( U - \mu \Omega \right) \right)^{3/2}} dU \text{.}
\end{align}
With this result, the order $\hat{\phi}$ piece of (\ref{quasineutrality-expanded-full}) is, keeping terms up to $O\left( \hat{\phi} \epsilon^2 \right)$,
\begin{align} \label{phiinftyk1}
\phi''\left( x \right) = k_1 \phi \left( x \right) + O\left( \hat{\phi} \epsilon^3 \right) \text{,}
\end{align}
where we define $k_1$, a quantity with dimensions of $\left(1/ \text{length}\right)^2$, as 
\begin{align} \label{k1}
k_1 =  \frac{\Omega^2 m_{\text{i}}}{ZT_{\text{e}}} \frac{ n_{e\infty} -  2\pi\frac{Z^2T_{\text{e}}}{m_{\text{i}}}\int_0^{\infty} \Omega d\mu  \int_{\mu \Omega }^{\infty}  \frac{F_{\text{cl}}(\mu, U) dU}{\left(2\left( U - \mu \Omega \right) \right)^{3/2}}  }{ n_{e\infty} + 2\pi Z\int_0^{\infty} \Omega^2  \mu d\mu \int_{\mu \Omega}^{\infty}  \frac{F_{\text{cl}}(\mu, U) dU}{\left(2\left( U - \mu \Omega \right) \right)^{3/2}}} \text{.}
\end{align}
From equation (\ref{phiinftyk1}) and using the boundary condition $\phi =0$ at $x\rightarrow \infty$, we find $\phi \propto \exp\left( - \sqrt{k_1} x \right)$. Consequently, $\sqrt{\left| k_1 \right|} \rho_{\text{i}} \sim \epsilon$ and assumption (\ref{epsilonsmall}) is true only if $k_1$, defined in equation (\ref{k1}), is sufficiently small. 
If this is not the case, we expect $\phi \propto \exp\left( - \lambda x \right)$, but the value of $\lambda$ would have to be determined by carrying out the more general expansion of the quasineutrality equation in $\hat{\phi} \ll 1$ with $\epsilon \sim 1$.

%With the knowledge that equations (\ref{phiinftyk1}) and (\ref{k1}) are valid in the neighbourhood of $k_1 =0$ (which is where $k_1$ changes sign), we can obtain conditions for $k_1$ to be positive or negative. 
In order to impose that $\phi \left( \infty \right) = 0$ we require a non-oscillating potential profile at $x \rightarrow \infty$, which gives $k_1 \geqslant 0$ as a solvability condition. The numerator of $k_1$ determines the sign of $k_1$ because the denominator is always positive.
Hence, we obtain the condition
\begin{align} \label{solvability}
   2\pi Z v_{\text{B}}^2 \int_0^{\infty} \Omega d\mu \int_{\mu \Omega }^{\infty}  \frac{F_{\text{cl}} (\mu, U)  dU}{\left(2\left( U - \mu \Omega \right) \right)^{3/2}}  \leqslant  n_{e\infty}\text{,}
\end{align}
where the Bohm velocity $v_{\text{B}}$ is defined in equation (\ref{Bohm-speed}). 
The equation
\begin{align} \label{changevar-infty}
  2\pi  \int_0^{\infty} \Omega d\mu \int_{\mu \Omega }^{\infty}  \frac{F_{\text{cl}} (\mu, U)  h\left( 2\Omega \mu, 2\left( U - \Omega \mu \right) \right)  }{ \sqrt{ 2\left( U - \mu \Omega \right)  } }   dU   = \int  f_{\infty} \left( \vec{v} \right) h \left( v_x^2 + v_y^2, v_z^2 \right) d^3v \text{,}
\end{align}
is valid for any function $h$ and is obtained using the fact that $\mu = (v_x^2 + v_y^2 )/2\Omega$ and $U=(v_x^2 + v_y^2 +v_z^2)/2$ at $x\rightarrow \infty$ (shown in \ref{subapp-mu-expansion})
We can use equation (\ref{changevar-infty}) to re-express the solvability condition as
\begin{align} \label{solvability-vz}
 Zv_{\text{B}}^2 \int  \frac{f_{\infty} \left( \vec{v} \right)}{v_z^2} d^3v \leqslant n_{\text{e}\infty}  \text{.}
\end{align}
The solvability condition (\ref{solvability-vz}) generalizes Chodura's condition for the magnetic presheath entrance \cite{Chodura-1982} to include the effect of kinetic ions at small $\alpha$. 
%Note that equation (\ref{solvability}) is valid to lowest order in $\alpha$.  %, because the variation of the electrostatic potential that ions experience during their gyration is $O\left( \epsilon T_{\text{e}} / e \right)$. 
%In \ref{app-Chodura-noalpha} we generalize the solvability condition (\ref{solvability}) to a form which is valid for all $\alpha$. 
In \ref{app-coldion}, we show that the cold ion limit of our generalized condition recovers the cold ion limit of Chodura's original condition to lowest order in $\alpha$. % in the same way the condition in \cite{Riemann-review} generalizes the original formulation of Bohm's condition. 

It is believed that solvability conditions such as (\ref{solvability}) are usually satisfied marginally \cite{Riemann-review}. This means that equation (\ref{solvability}) is expected to hold in the equality form, which justifies considering $k_1 =0$ and hence justifies our initial assumption that $\epsilon \ll 1$. 
When $k_1 =0$, terms of size $\hat{\phi}^{3/2}$ in the expansion of quasineutrality become important. From considering terms of this order in (\ref{quasineutrality-expanded-full}), we obtain
\begin{align} \label{phi''phi32}
\phi''\left( x \right) = - k_{3/2} \left[ - \phi \left( x \right)\right]^{3/2} \text{,}
\end{align}
where $k_{3/2}$ 
%has dimensions $ \left( \text{charge} \right)^{1/2} / \left( \text{energy} \right)^{1/2} \left( \text{length} \right)^2  $ and
is given by
\begin{align} \label{k32}
 k_{3/2} =  \sqrt{ \frac{e}{T_{\text{e}}} } \left( \frac{\Omega}{v_{\text{B}}} \right)^2  \frac{ \frac{2\sqrt{2}}{3} 2\pi \int_0^{\infty} \Omega v_{\text{B}}^3   \frac{  \partial F_{\text{cl}} }{ \partial U  } (\mu, \Omega \mu)   d\mu }{ n_{e\infty} + 2\pi Z\int_0^{\infty} \Omega^2  \mu d\mu \int_{\Omega \mu}^{\infty}  \frac{F_{\text{cl}}(\mu, U)}{\left(2\left( U - \Omega \mu \right)\right)^{3/2}}  dU } \geqslant 0 \text{.}
\end{align}
The numerator of (\ref{k32}) is positive because $ F_{\text{cl}}(\mu, U ) = 0 $ for $U \leqslant \Omega \mu$ and hence $\partial F_{\text{cl}} (\mu, \Omega \mu ) / \partial U \geqslant 0$ for all values of $\mu$. Moreover, both terms in the denominator of (\ref{k32}) are explicitly positive, so the inequality in (\ref{k32}) follows. The case $ k_{3/2}  =0 $ only arises if $ \partial F_{\text{cl}} (\mu, \Omega \mu) / \partial U = 0 $ for all $\mu$. Note that this condition implies $(1/v_z)\partial f_{\infty}\left( v_{x}, v_{y}, 0 \right) / \partial v_{z} =0$ for all values of $v_{x}$ and $v_{y}$, which corresponds to a very flat ion distribution function near $v_{z} = 0$. One example of such a flat ion distribution function is a Dirac delta function, which is used to model cold ions in \ref{app-coldion}.

Equation (\ref{phi''phi32}) is solved by multiplying by $\phi' \left( x \right)$ then integrating once and using the boundary condition $\phi' \left( x \right) =0 $ when $\phi \left( x \right) = 0$ to get
\begin{align}
 \phi' \left( x \right)^2  =  \frac{4k_{3/2}}{5}  \left[-\phi \left( x \right) \right]^{5/2}  \text{.}
\end{align}
Taking the square root and integrating again, the potential profile is
\begin{align} \label{phisol1}
\phi  \left( x \right) = - \frac{400}{k_{3/2}^2} \frac{1}{(x+C_{3/2})^4} \text{,}
\end{align}
where $C_{3/2}$ is an integration constant. Equation (\ref{phisol1}) implies that $\epsilon \sim \hat{\phi}^{1/4} \gg \hat{\phi}$. The boundary condition that we use to obtain our numerical results (see Section \ref{subsec-finfty}) has $k_{3/2} \neq 0$, so equation (\ref{phisol1}) is the form of the electrostatic potential to which we must match our numerical solution at large $x$.

If $\partial F \left( \mu, \Omega \mu \right) / \partial U =0$, then $k_{3/2}  = 0$ and we must go to higher order in $\hat{\phi}$ to solve for the electrostatic potential at large $x$. 
Note that $k_{3/2} = 0$ is a case that we do not numerically study in this paper, but we carry out the following analysis because it is necessary to obtain the correct potential profile at large $x$ for cold ions, studied in \ref{app-coldion}.
For $\partial F \left( \mu, \Omega \mu \right) / \partial U =0$, we can integrate by parts twice the term with $\partial^2 F\left( \mu, \Omega \mu \right) / \partial U^2 $ to get
\begin{align}
\int_{\Omega\mu}^{\infty} \frac{\partial^2 F_{\text{cl}}\left( \mu, U \right) / \partial U^2 }{\sqrt{2\left( U -\Omega \mu \right)}} dU = 3\int_{\Omega\mu}^{\infty} \frac{F_{\text{cl}}\left( \mu, U \right) }{\left( 2\left( U -\Omega \mu \right) \right)^{5/2}} dU  \text{.}
\end{align}
Balancing the term of order $\hat{\phi} \epsilon^2$ with terms of order $\hat{\phi}^2$ in (\ref{quasineutrality-expanded-full}), we get
\begin{align} \label{phi''phi2}
\phi''\left( x \right) = - k_2 \left[ \phi \left( x \right) \right]^2 \text{,}
\end{align}
where $k_2$ is given by
\begin{align} \label{k2}
 k_2 =   \frac{\Omega^2 e}{2v_{\text{B}}^2 T_{\text{e}}}  \frac{  6\pi Z v_{\text{B}}^4 \int_0^{\infty}  \Omega d\mu \int_{\Omega \mu}^{\infty}  \frac{F_{\text{cl}}(\mu, U) }{\left( 2\left( U - \Omega \mu \right) \right)^{5/2} } dU  -  n_{e\infty} }{  n_{e\infty}  +  2\pi Z \int_0^{\infty} \Omega^2 \mu d\mu \int_{\Omega \mu}^{\infty}  \frac{F_{\text{cl}}(\mu, U)}{\left(2\left(U-\Omega \mu \right) \right)^{3/2}} dU   } > 0 \text{.}
\end{align}
Both terms in the denominator of (\ref{k2}) are positive; therefore the inequality on the right hand side of (\ref{k2}) is the result of the numerator being positive, which is demonstrated in \ref{app-k2>0} if condition (\ref{solvability-vz}) is satisfied with the equality sign. Equation (\ref{phi''phi2}) is solved in the same way as equation (\ref{phi''phi32}), and the result is
\begin{align} \label{phisol2}
\phi \left( x \right) = - \frac{6}{k_2} \frac{1}{(x+C_2)^2} \text{,}
\end{align}
where $C_2$ is an integration constant.
The fact that $k_2$ is positive and $k_2 \rho_{\text{i}}^2 T_{e}/e \sim 1$ implies that we do not need to carry out the expansion of (\ref{quasineutrality-expanded-full}) any further, because the order $\hat{\phi}^2$ term is guaranteed to be non-zero if the solvability condition (\ref{solvability-vz}) is marginally satisfied. Hence, $\epsilon \gtrsim \hat{\phi}^{1/2}$ as stated in equation (\ref{epssmallerdelta}).

%\subsection{Electrostatic potential form near $x\rightarrow\infty$ when the solvability condition is weakly oversatisfied}
%
%More generally, the solvability condition need not be satisfied in its marginal form. If it is weakly oversatisfied, such that $k_1 \rho_{\text{i}}^2 \sim \delta^2 \ll 1$, the equation for the potential near $x \rightarrow \infty$ is
%\begin{align}
%\phi''\left( x \right) = k_1 \phi\left( x\right) -k_{3/2} \left( -\phi\left( x \right) \right)^{3/2} \text{.}
%\end{align}
%Multiplying this by $\phi'\left( x\right)$ and then integrating it we obtain
%\begin{align}
%\phi'\left(x\right)^2 = k_1 \phi\left( x\right)^2 + \frac{4}{5} k_{3/2}\left( -\phi \left( x \right) \right)^{5/2} \text{.}
%\end{align} 
%Taking the square root and integrating again, we get
%\begin{align}
%\phi\left(x\right) = - \frac{25k_1^2/16k_{3/2}^2}{\sinh^4\left(\frac{1}{4}\sqrt{k_1}\left( x + C_{3/2} \right) \right)} \text{.}
%\end{align}
%
%Now we consider the case in which the distribution function is very flat near $v_z =0$, so that $k_{3/2} =0$. Then, the equation governing the behaviour of the elecrostatic potential near $x \rightarrow \infty$ is
%\begin{align}
%\phi''\left( x \right) = k_1 \phi\left( x\right) - k_{2}  \phi\left( x \right)^{2} \text{.}
%\end{align}
%which has solution 
%\begin{align}
%\phi  = - \frac{3k_1/2k_2}{\sinh^2 \left( \frac{1}{2}\sqrt{k_1}\left( x+C_2 \right) \right)} \text{.}
%\end{align}
 
\subsection{Expansion of quasineutrality near $x=0$ }  \label{subsec-expansion-near0}

Here we expand the quasineutrality equation near the Debye sheath entrance, $x=0$. We define the normalized electrostatic potential relative to $x=0$,
\begin{align}
\delta \hat{\phi} = \frac{ e \delta \phi }{T_{\text{e}} }  = \frac{ e }{T_{\text{e}} } \left( \phi ( x ) - \phi ( 0) \right) \ll 1 \text{.}
 \end{align}
Each term of the quasineutrality equation (\ref{quasineutrality-compact}) can be expanded in $\delta \hat{\phi} \ll 1$ separately, order by order.
Denoting the electron density at $x=0$ as $n_{\text{e}0}$, such that
\begin{align}
n_{\text{e}0} = n_{\text{e}\infty} \exp \left( \frac{e \phi(0) }{T_{\text{e}}} \right) \text{,}
\end{align}
the electron density near $x=0$ is
\begin{align} \label{ne-near0}
n_{\text{e}} (x) = n_{\text{e}0} \exp \left( \frac{e\delta \phi }{T_{\text{e}}} \right) \text{.}
\end{align}
%The lowest order of the quasineutrality equation expanded near $x=0$ is the quasineutrality equation at $x=0$. 
Using the result $n_{\text{i,cl}} (0) = 0$ and equations (\ref{ni-open-Deltavx}) and (\ref{quasineutrality-compact}), we have
\begin{align} \label{quasi-0}
n_{\text{e}0} = Zn_{\text{i,op}} (0) =  \int_{\bar{x}_{\text{c}}}^{\infty}  \Omega d\bar{x}  \int_{\chi_{\text{M}} (\bar{x} )}^{\infty}  \frac{ F_{\text{cl}} ( \mu_{\text{gk}} (\bar{x}, \chi_{\text{M}}) , U ) }{\sqrt{2\left( U - \chi_{\text{M}} (\bar{x}) \right)}} \Delta v_{x0} dU     \text{,}
\end{align}
where we used $\bar{x}_{\text{m,o}}(0) = \bar{x}_{\text{c}}$ (from equation (\ref{xbarm-open})) and we introduced
\begin{align} \label{Deltavx-0}
\Delta v_{x0} = \Delta v_x \rvert_{x=0} = \sqrt{ 2\left( \Delta_{\text{M}} (\bar{x}, U ) + \chi_{\text{M}} (\bar{x} ) - \chi (0, \bar{x} ) \right) } - \sqrt{2\left( \chi_{\text{M}} (\bar{x} ) - \chi (0, \bar{x} )  \right)} \text{.}
\end{align}
Subtracting equation (\ref{quasi-0}) from equation (\ref{quasineutrality-compact}), we obtain the perturbed quasineutrality equation near $x=0$,
\begin{align} \label{quasi-perturbed}
n_{\text{e}} (x) - n_{\text{e}0} = Z \left( n_{\text{i,cl}} (x) +  n_{\text{i,op}}(x) - n_{\text{i,op}}(0) \right) \text{.}
\end{align} 
We will show that $\bar{x}_{\text{m,I}} \rightarrow \infty$ in our system and therefore type I orbits are absent. 
However, we first assume the more general scenario in which both type I and type II orbits are present, with $\phi'(0)$ being finite, and calculate the dominant contribution to equation (\ref{quasi-perturbed}).

We proceed to obtain the term $n_{\text{i,cl}}(x)$ in equation (\ref{quasi-perturbed}) to leading order.
Firstly, we observe that a \emph{closed} orbit near $x=0$ must lie at a position $x$ such that $0 \leqslant x_{\text{M}} \leqslant x$, with $\chi ( x, \bar{x} ) \simeq \chi_{\text{M}} ( \bar{x} ) $. 
Remembering that for a closed orbit the perpendicular energy lies in the range $\chi ( x, \bar{x} ) \leqslant U_{\perp} \leqslant \chi_{\text{M}} ( \bar{x} ) $, we can take the integral over $U_{\perp}$ in (\ref{ni-closed}) by approximating
\begin{align}
F_{\text{cl}}  \left( \mu_{\text{gk}} ( \bar{x}, U_{\perp} ), U \right) \simeq F_{\text{cl}}  \left( \mu_{\text{gk}} ( \bar{x}, \chi_{\text{M}} (\bar{x}) ), U \right)  \text{}
\end{align}
and $ \sqrt{2\left( U- U_{\perp} \right)} \simeq  \sqrt{2\left( U- \chi_{\text{M}} (\bar{x} ) \right)}$.
With these approximations, the integral (\ref{ni-closed}) becomes
\begin{align} \label{ni-closed-expanded}
n_{\text{i,cl}} (x) \simeq &  2\int_{\bar{x}_{\text{m}} (x) }^{\infty} \Omega \sqrt{2 \left(\chi_{\text{M}} (\bar{x}) - \chi (x, \bar{x}) \right) } d\bar{x} \nonumber \\
& \times \int_{\chi_{\text{M}} (\bar{x})}^{\infty} \frac{F_{\text{cl}} \left(\mu_{\text{gk}} \left(\bar{x}, \chi_{\text{M}} \left(\bar{x}\right) \right), U \right) }{\sqrt{2\left( U - \chi_{\text{M}} (\bar{x}) \right) }}  dU  \text{.}
\end{align}
The contributions to $n_{\text{i,cl}}(x)$ of type I and type II closed orbits have different sizes. 
Introducing the small quantity
\begin{align} \label{deltachi}
\delta \chi = \chi(0, \bar{x} ) - \chi(x, \bar{x}) \simeq - \frac{\Omega \delta \phi }{B} + \Omega^2 \bar{x} x  \text{,}
\end{align}
where we neglected the term proportional to $x^2$, the closed orbit density (\ref{ni-closed-expanded}) is dominated by type I closed orbits (which have $\chi_{\text{M}} (\bar{x}) = \chi (0, \bar{x})$), whose leading order density is given by
\begin{align} \label{ni-closed-expanded-lowest}
n_{\text{i,cl}} (x) \simeq  2\int_{\bar{x}_{\text{m,I}}}^{\infty} \Omega \sqrt{2 \delta \chi } d\bar{x} \int_{\chi_{\text{M}}}^{\infty} \frac{F_{\text{cl}} \left(\mu_{\text{gk}} \left(\bar{x}, \chi_{\text{M}} \left(\bar{x}\right) \right), U \right) }{\sqrt{2\left( U - \chi_{\text{M}} (\bar{x}) \right) }} dU  \text{.}
\end{align}
The reason for neglecting the contribution to the density of type II closed orbits is that the contribution from ions with $x_{\text{M}} > 0$ is smaller, as shown explicitly in \ref{app-notypeIIclosed}.

We now obtain the term $n_{\text{i,op}} (x) - n_{\text{i,op}} (0)$ to leading order. 
We can re-express this as
\begin{align} \label{ni-open-perturbed-1}
n_{\text{i,op}} (x) - n_{\text{i,op}} (0) & \simeq \int_{\bar{x}_{\text{c}}}^{\infty} \Omega d\bar{x} \int_{\chi_{\text{M}}(\bar{x})}^{\infty} \frac{ F_{\text{cl}} ( \mu, U ) }{\sqrt{2\left( U - \chi_{\text{M}} (\bar{x}) \right)}} \left[ \Delta v_x - \Delta v_{x0} \right] dU \nonumber \\
& - \int_{\bar{x}_{\text{c}}}^{\bar{x}_{\text{m,o}}(x)} \Omega d\bar{x} \int_{\chi_{\text{M}}(\bar{x})}^{\infty} \frac{ F_{\text{cl}} ( \mu, U ) }{\sqrt{2\left( U - \chi_{\text{M}} (\bar{x}) \right)}} \Delta v_{x0}  dU
\end{align}
where 
\begin{align} \label{Deltavx-Deltavx0}
\Delta v_x - \Delta v_{x0}  & =  \sqrt{ 2\left( \Delta_{\text{M}} (\bar{x}, U ) + \chi_{\text{M}} (\bar{x} ) - \chi (0, \bar{x} ) + \delta \chi \right) } \nonumber \\
& - \sqrt{2\left( \chi_{\text{M}} (\bar{x} ) - \chi (0, \bar{x} ) + \delta \chi \right)}  -  \sqrt{ 2\left( \Delta_{\text{M}} (\bar{x}, U ) + \chi_{\text{M}} (\bar{x} ) - \chi (0, \bar{x} ) \right) } \nonumber  \\
& + \sqrt{2\left( \chi_{\text{M}} (\bar{x} ) - \chi (0, \bar{x} ) \right)}  \text{.}
\end{align}
The second term in (\ref{ni-open-perturbed-1}) is zero if type II orbits are present ($x_{\text{c}} > 0$) because, from equation (\ref{xbarm-open}), $\bar{x}_{\text{m,o}}(x) = \bar{x}_{\text{c}}$ for $ x < x_{\text{c}} \neq 0$.
If no type II orbits are present ($x_{\text{c}} = 0$), equation (\ref{xbarm-open}) gives $\bar{x}_{\text{m,o}}(x) = \bar{x}_{\text{m}} (x)$ and, from equation (\ref{xbarm-general}), we expect the variation in $\bar{x}_{\text{m}} (x)$ to be linear in $x$ and $\delta \phi$.
For type I orbits, $\chi_{\text{M}} (\bar{x} ) = \chi (0, \bar{x} )$;
then, the second term in equation (\ref{Deltavx-Deltavx0}) is of order $\sqrt{\delta \hat{\phi}}$, the fourth term is zero, and the first and third terms together cancel to lowest order leaving a piece of order $\delta \hat{\phi}$. 
Hence, $ \Delta v_x - \Delta v_{x0}  \simeq  - \sqrt{2\delta \chi} \sim v_{\text{t,i}} \sqrt{\delta \hat{\phi}} $ for type I orbits.
Type II open orbits have $\chi_{\text{M}}(\bar{x}) > \chi(0, \bar{x})$, and hence they contribute at most an order $\delta \hat{\phi}$ piece to (\ref{ni-open-perturbed-1})%
\footnote{Some type II open orbits have $\chi_{\text{M}}(\bar{x}) - \chi(0, \bar{x}) \sim \delta \chi$, such that the second term in (\ref{Deltavx-Deltavx0}) is $\sqrt{ \chi''_{\text{M}}  x_{\text{M}}^2 + 2\delta \chi } \sim  \sqrt{\delta\hat{\phi}} v_{\text{t,i}}$. 
However, the values of $x_{\text{M}}$ for which type II orbits satisfy $\Delta v_x \sim \sqrt{\delta \hat{\phi}} v_{\text{t,i}}$ are small, $x_{\text{M}} \sim \sqrt{ \delta \hat{\phi } } \rho_{\text{i}} $.
Hence, the contribution to (\ref{ni-open-perturbed-1}) of such type II orbits is of order $\delta \hat{\phi} n_{\text{e} \infty}$.}. %
Thus the dominant contribution to (\ref{ni-open-perturbed-1}) is of order $\sqrt{\delta \hat{\phi}}$, from type I orbits.
%When $\chi_{\text{M}} (\bar{x} ) = \chi (0, \bar{x} )$, the second term in equation (\ref{Deltavx-Deltavx0}) is of order $\sqrt{\delta \hat{\phi}}$, the fourth term is zero, and the first and third terms together cancel to lowest order leaving a piece of order $\delta \hat{\phi}$. 
%Therefore, the dominant contribution to $\Delta v_x - \Delta v_{x0}$ is of order $\sqrt{\delta \hat{\phi}}$ and comes from type I open orbits, which by definition have $\chi_{\text{M}} (\bar{x} ) = \chi (0, \bar{x} )$.
%Type II open orbits have $\chi_{\text{M}}(\bar{x}) \neq \chi(0, \bar{x})$, and hence they contribute at most an order $\delta \hat{\phi}$ piece to $\Delta v_x - \Delta v_{x0}$.
The minimum value of $\bar{x}$ for which type I open orbits are present near $x=0$ is approximately $\bar{x}_{\text{m,I}}$, giving
\begin{align} \label{ni-open-near0-1/2}
n_{\text{i,op}} (x) - n_{\text{i,op}} (0) \simeq - \int_{\bar{x}_{\text{m,I}}}^{\infty} \sqrt{2 \delta \chi}  \Omega d\bar{x}  \int_{\chi_{\text{M}} (\bar{x} )}^{\infty}  \frac{ F_{\text{cl}}  \left( \mu_{\text{gk}} \left( \bar{x}, \chi_{\text{M}}(\bar{x}) \right), U \right)  }{\sqrt{2\left( U - \chi_{\text{M}} (\bar{x}) \right)}} dU      \text{.}
\end{align}

From equation (\ref{ne-near0}), we see that there is no term in the expansion of the electron density that has a size $\sqrt{\delta \hat{\phi}}$.
Hence, the dominant terms in the perturbed quasineutrality equation (\ref{quasi-perturbed}) for small $x$ are obtained by adding equations (\ref{ni-closed-expanded-lowest}) and (\ref{ni-open-near0-1/2}) and setting the sum to zero,
\begin{align} \label{quasi-near0-leading}
0 =  Z\int_{\bar{x}_{\text{m,I}}}^{\infty} \sqrt{2 \delta \chi}  \Omega d\bar{x}  \int_{\chi_{\text{M}} (\bar{x} )}^{\infty}  \frac{ F_{\text{cl}} \left( \mu_{\text{gk}} \left( \bar{x}, \chi_{\text{M}}(\bar{x}) \right), U \right) }{\sqrt{2\left( U - \chi_{\text{M}}(\bar{x}) \right)}} dU  \text{.} 
\end{align}
The right hand side of equation (\ref{quasi-near0-leading}) vanishes only if $\bar{x}_{\text{m,I}} \rightarrow \infty$, which from equation (\ref{xbarmI}) implies a divergent electric field at $x=0$, $\phi'(0) \rightarrow \infty$.
The fact that $\bar{x}_{\text{m,I}} \rightarrow \infty$ means that only type II orbits are present in the magnetic presheath and $n_{\text{i,cl}}(x)$ is exponentially small, as argued in \ref{app-notypeIIclosed}. 
Therefore, we consider $n_{\text{i,cl}} (x) \simeq 0$ in equation (\ref{quasi-perturbed}) and focus on the perturbed open orbit density $n_{\text{i,op}}(x) - n_{\text{i,op}}(0)$.

With type I orbits absent, %$\bar{x}_{\text{m,I}} \rightarrow \infty$, 
the effective potential maximum lies at $x_{\text{M}} \neq 0$, hence $\chi_{\text{M}}(\bar{x}) \neq \chi(0, \bar{x})$. Taking $\bar{x} \rightarrow \infty$ corresponds to $x_{\text{M}} \rightarrow 0$, and therefore
\begin{align} \label{xbar-infty-limit}
\lim_{\bar{x} \rightarrow \infty} \chi_{\text{M}}(\bar{x}) = \chi(0, \bar{x}) \simeq \frac{1}{2} \Omega^2 \bar{x}^2  \text{.}
\end{align}
If the distribution function $F_{\text{cl}}$ decays exponentially at large energies, it is exponentially small in the region of the integral where $\chi_{\text{M}}(\bar{x}) - \chi(0, \bar{x}) \sim \delta \chi$ (which corresponds to $\bar{x}$ being large). This is because, according to equation (\ref{xbar-infty-limit}), $U_{\perp} \simeq \chi_{\text{M}}(\bar{x})$ is very large in that region. %, which is where (\ref{Deltavx-expanded}) fails. 
As a consequence, $\delta \chi \ll \chi_{\text{M}}(\bar{x}) - \chi(0, \bar{x}) $ for values of $\bar{x}$ where the distribution function is not exponentially small.
When $\delta \chi \ll \chi_{\text{M}} (\bar{x}) - \chi(0, \bar{x} ) $ we can Taylor expand both terms in equation (\ref{Deltavx-Deltavx0}) to obtain
\begin{align} \label{Deltavx-expanded}
\Delta v_x - \Delta v_{x0} = - \Delta \left[\frac{1}{ v_{x0}} \right]  \delta \chi  + \frac{1}{2} \Delta \left[ \frac{1}{v_{x0}^3} \right]  \delta \chi^2 \text{,}
\end{align}
where we introduced the positive quantities
\begin{align}
\Delta \left[ \frac{1}{v_{x0}} \right]  =  \frac{1}{ \sqrt{2\left( \chi_{\text{M}} (\bar{x} ) - \chi (0, \bar{x} )  \right) } }  - \frac{1}{  \sqrt{ 2\left( \Delta_{\text{M}} (\bar{x}, U ) + \chi_{\text{M}} (\bar{x} ) - \chi (0, \bar{x} ) \right) }} \text{,}
\end{align}
and 
\begin{align}
\Delta \left[ \frac{1}{v_{x0}^3} \right]  =  \frac{1}{ \left[ 2\left( \chi_{\text{M}} (\bar{x} ) - \chi (0, \bar{x} ) \right) \right]^{3/2} } - \frac{1}{ \left[ 2\left( \Delta_{\text{M}} (\bar{x}, U ) + \chi_{\text{M}} (\bar{x} ) - \chi (0, \bar{x} ) \right) \right]^{3/2} }\text{.}
\end{align} 
We expand the open orbit density (\ref{ni-open-Deltavx}) using equation (\ref{Deltavx-expanded}) for the expansion of $\Delta v_x$, obtaining
\begin{align} \label{ni-open-perturbed}
n_{\text{i,op}} (x) - n_{\text{i,op}} (0) \simeq  & - \int_{\bar{x}_{\text{c}}}^{\infty}  \delta \chi  \Omega d\bar{x}  \int_{\chi_{\text{M}} (\bar{x} )}^{\infty}  \frac{ F_{\text{cl}} ( \mu_{\text{gk}}(\bar{x}, \chi_{\text{M}}), U ) }{\sqrt{2\left( U - \chi_{\text{M}} (\bar{x}) \right)}} \Delta \left[ \frac{1}{v_{x0}} \right]  dU \nonumber \\ 
& + \frac{1}{2}\int_{\bar{x}_{\text{c}}}^{\infty}  \delta \chi^2  \Omega d\bar{x}  \int_{\chi_{\text{M}} (\bar{x} )}^{\infty}  \frac{ F_{\text{cl}} ( \mu_{\text{gk}}(\bar{x}, \chi_{\text{M}}), U ) }{\sqrt{2\left( U - \chi_{\text{M}} (\bar{x}) \right)}} \Delta \left[ \frac{1}{v_{x0}^3} \right]  dU
\text{.}  
\end{align}

Expanding the electron density (\ref{ne-near0}), we get 
\begin{align} \label{ne-perturbed}
n_{\text{e}} (x) - n_{\text{e}0} \simeq \frac{e\delta \phi}{T_{\text{e}}} + \frac{1}{2} \left( \frac{e\delta \phi}{T_{\text{e}}}  \right)^2 \text{.}
\end{align}
The perturbed quasineutrality equation (\ref{quasi-perturbed}), to order $\delta \hat{\phi}$, then implies that
\begin{align} \label{quasi-near0-1}
n_{\text{e}0} \frac{e\delta \phi }{T_{\text{e}}} =   \frac{\Omega \delta \phi }{B}Z\int_{\bar{x}_{\text{c}}}^{\infty}  \Omega d\bar{x}  \int_{\chi_{\text{M}} (\bar{x} )}^{\infty}  \frac{ F_{\text{cl}} \left( \mu_{\text{gk}}\left(\bar{x}, \chi_{\text{M}}(\bar{x}) \right), U \right)  }{\sqrt{2\left( U - \chi_{\text{M}}(\bar{x})\right)}} \Delta \left[ \frac{1}{v_{x0}} \right]  dU \nonumber \\
 -  x  \Omega Z\int_{\bar{x}_{\text{c}}}^{\infty} \Omega^2 \bar{x}   d\bar{x}  \int_{\chi_{\text{M}} (\bar{x} )}^{\infty}  \frac{ F_{\text{cl}} \left( \mu_{\text{gk}}\left(\bar{x}, \chi_{\text{M}}(\bar{x}) \right), U \right) }{\sqrt{2\left( U - \chi_{\text{M}}(\bar{x})\right)}} \Delta \left[ \frac{1}{v_{x0}} \right]  dU  \text{.} 
\end{align}
This can be rearranged to obtain
\begin{align} \label{deltaphi-1}
 \delta \phi = \phi(x) - \phi(0) = \frac{x}{q_1} \text{,}
\end{align}
where $q_1$ is given by
%has dimensions $\left[ \text{charge} \times \text{length} / \text{energy} \right] $ and 
\begin{align} 
q_1 = \frac{e}{\Omega T_{\text{e}}} \frac{  Zv_{\text{B}}^2\int_{\bar{x}_{\text{c}}}^{\infty}   \Omega d\bar{x}  \int_{\chi_{\text{M}} (\bar{x} )}^{\infty}  \frac{ F_{\text{cl}} \left( \mu_{\text{gk}}\left(\bar{x}, \chi_{\text{M}}(\bar{x}) \right), U \right)  }{\sqrt{2\left( U - \chi_{\text{M}}(\bar{x})\right)}} \Delta \left[ \frac{1}{v_{x0}} \right]  dU -  n_{e0}  }{   Z\int_{\bar{x}_{\text{c}} }^{\infty}  \Omega^2 \bar{x} d\bar{x}  \int_{\chi_{\text{M}} (\bar{x} )}^{\infty}  \frac{ F_{\text{cl}} \left( \mu_{\text{gk}}\left(\bar{x}, \chi_{\text{M}}(\bar{x}) \right), U \right)  }{\sqrt{2\left( U - \chi_{\text{M}}(\bar{x})\right)}} \Delta \left[ \frac{1}{v_{x0}} \right] dU  } \text{.}
\end{align}
Equation (\ref{deltaphi-1}) implies that $\phi'(0) = q_1^{-1}$. The magnetic presheath \emph{is driven} towards $q_1 = 0$ because $\phi'(0) \rightarrow \infty$ is required from equation (\ref{quasi-near0-leading}) and the discussion following it. 
Hence, the numerator of $q_1$ must be zero,
\begin{equation} \label{Bohm-not-obvious}
 Zv_{\text{B}}^2\int_{\bar{x}_{\text{c}}}^{\infty}   \Omega d\bar{x}  \int_{\chi_{\text{M}} (\bar{x} )}^{\infty}  \frac{ F_{\text{cl}} ( \mu_{\text{gk}} (\bar{x}, \chi_{\text{M}} ), U ) }{\sqrt{2\left( U - \chi_{\text{M}}\right)}} \Delta \left[ \frac{1}{v_{x0}} \right]  dU = n_{e0}  \text{.} 
\end{equation} 

We proceed to show that equation (\ref{Bohm-not-obvious}) is equivalent to the marginal form of the kinetic Bohm condition \cite{Harrison-Thompson-1959, Riemann-review},
\begin{align} \label{Bohm-kinetic-marginal}
Zv_{\text{B}}^2 \int \frac{f_0 ( \vec{v} ) }{v_{x}^2}  d^3v  =  n_{e0}  \text{.}
\end{align}
From (\ref{fopen}), the distribution function at $x=0$ is
\begin{align} \label{f0-def}
f_{0} (\vec{v})  & = f_{\text{op}} (0, \vec{v})   \nonumber  \\
& \simeq  F_{\text{cl}} \left( \mu_{\text{gk}} \left( \bar{x}, \chi_{\text{M}} \left( \bar{x} \right) \right) , U \right)   \hat{\Pi} \left(  v_x ,  - V_x \left( 0, \bar{x}, \chi_{\text{M}} \right) - \Delta v_{x0}  , - V_x \left( 0, \bar{x}, \chi_{\text{M}} \right)  \right) \text{.}
\end{align}
Using the definition (\ref{f0-def}) and the change of variables $(\bar{x}, U) \rightarrow (v_y, v_z)$ (equations (\ref{xbar-def}) and (\ref{U-open})) at $x=0$, we can re-express the integral in (\ref{Bohm-not-obvious}) to obtain 
\begin{align} 
& \int_{\bar{x}_{\text{c}}}^{\infty}   \Omega d\bar{x}  \int_{\chi_{\text{M}} (\bar{x} )}^{\infty}  \frac{  F_{\text{cl}} \left( \mu_{\text{gk}} \left( \bar{x}, \chi_{\text{M}} \left( \bar{x} \right) \right) , U \right)   }{\sqrt{2\left( U - \chi_{\text{M}} (\bar{x}) \right)}} \Delta \left[ \frac{1}{v_{x0}} \right]  dU \nonumber \\
& = \int_{\bar{x}_{\text{c}}}^{\infty}   \Omega d\bar{x}  \int_{\chi_{\text{M}} (\bar{x} )}^{\infty}  \frac{ F_{\text{cl}} \left( \mu_{\text{gk}} \left( \bar{x}, \chi_{\text{M}} \left( \bar{x} \right) \right) , U \right) }{\sqrt{2\left( U - \chi_{\text{M}} (\bar{x}) \right)}} dU \nonumber \\
& ~ \times \int_{-\infty}^{0}  \frac{1}{v_x^2} \hat{\Pi} \left(  v_x ,  - V_x \left( 0, \bar{x}, \chi_{\text{M}} \right) - \Delta v_{x0}  , - V_x \left( 0, \bar{x}, \chi_{\text{M}} \right)  \right) dv_x \nonumber \\
 & = \int \frac{f_0 ( \vec{v} ) }{v_{x}^2}  d^3v  \text{.}
\end{align}
This shows that equations (\ref{Bohm-not-obvious}) and (\ref{Bohm-kinetic-marginal}) are equivalent. Hence, our system is driven to marginally satisfying the kinetic Bohm condition (\ref{Bohm-kinetic-marginal}).

Because $q_1=0$, we must consider terms of size $ \sim \delta \hat{ \phi}^2$ in equation (\ref{quasi-perturbed}) in order to balance the left hand side of equation (\ref{deltaphi-1}). % term proportional to $x$ in equation (\ref{quasi-near0-1}).
%If the distribution function decays to zero at large $\bar{x}$ according to equation (\ref{Fop-smallK}), there are $ \delta \hat{ \phi}^{3/2}$ terms one needs to consider as shown in \ref{app-flatness}. 
Using equations (\ref{ni-open-perturbed}) and (\ref{ne-perturbed}), we obtain
\begin{align} \label{quasi-near0-2}
\frac{1}{2} n_{\text{e}0} \left( \frac{e\delta \phi }{T_{\text{e}}} \right)^2 & =  - Zx\int_{\bar{x}_{\text{c}}}^{\infty}  \Omega^3 \bar{x} d\bar{x}  \int_{\chi_{\text{M}} (\bar{x} )}^{\infty}  \frac{ F_{\text{cl}} \left( \mu_{\text{gk}} \left( \bar{x}, \chi_{\text{M}}(\bar{x}) \right), U \right) }{\sqrt{2\left( U - \chi_{\text{M}}(\bar{x})\right)}} \Delta \left[ \frac{1}{v_{x0}} \right] dU \nonumber \\
& +   \frac{Z\Omega^2 \delta \phi^2 }{2B^2}  \int_{\bar{x}_{\text{c}}}^{\infty}   \Omega d\bar{x}  \int_{\chi_{\text{M}} (\bar{x} )}^{\infty}  \frac{ F_{\text{cl}} \left( \mu_{\text{gk}} \left( \bar{x}, \chi_{\text{M}}(\bar{x}) \right), U \right) }{\sqrt{2\left( U - \chi_{\text{M}} (\bar{x}) \right)}} \Delta \left[ \frac{1}{v_{x0}^{3}} \right] dU  \text{.} 
\end{align}
This leads to
\begin{align} \label{phi-near0-2}
 \delta \phi = \phi ( x ) - \phi ( 0 )  = q_{2}^{-1/2}  x^{1/2} \text{,}
\end{align}
where 
\begin{align} \label{q2-def}
q_{2} = \frac{1}{2} \left( \frac{e}{T_{\text{e}}} \right)^2  \frac{ Z v_{\text{B}}^4 \int_{\bar{x}_{\text{c}}}^{\infty}  \Omega d\bar{x} \int_{\chi_{\text{M}}(\bar{x})}^{\infty} \frac{F_{\text{cl}} \left( \mu_{\text{gk}} \left( \bar{x}, \chi_{\text{M}}(\bar{x}) \right), U \right) }{\sqrt{ 2\left( U - \chi_{\text{M}} ( \bar{x} ) \right)  } } dU \Delta \left[ \frac{1}{v_{x0}^{3}}\right] - n_{\text{e}0}  }{ Z\int_{\bar{x}_{\text{c}}}^{\infty}  \Omega^3 \bar{x} d\bar{x}  \int_{\chi_{\text{M}} (\bar{x} )}^{\infty}  \frac{ F_{\text{cl}} \left( \mu_{\text{gk}} \left( \bar{x}, \chi_{\text{M}}(\bar{x}) \right), U \right) }{\sqrt{2\left( U - \chi_{\text{M}}(\bar{x})\right)}} \Delta \left[ \frac{1}{v_{x0}} \right] dU   } > 0 \text{.}
\end{align}
%has dimensions $\left[ \text{length} \times \left( \text{charge}/\text{energy} \right)^2 \right]$.
In \ref{app-k2>0} we show that $q_{2}$ is always positive and never small. Therefore, equation (\ref{phi-near0-2}) is the scaling of the electrostatic potential we expect to observe in our numerical results.

\section{Numerical solution} \label{sec-numsol}

In this section, we present the numerical method we used to solve the system of equations (\ref{ni-closed}), (\ref{ni-open-Deltavx}) and (\ref{quasineutrality-compact}), and our results.
We assume a singly charged ion species, $Z = 1$, hence the ion and electron densities at $x\rightarrow \infty$ are equal and denoted $n_{\infty}$, $n_{\text{i}}(\infty) = n_{\text{e}\infty} = n_{\infty} $.
We first introduce, in Section \ref{subsec-finfty}, the ion distribution function that we assume as a boundary condition at $x \rightarrow \infty$. We then explain, in Section \ref{subsec-iteration}, the iteration scheme that was used to find the self-consistent solution of the potential $\phi(x)$. 
In Section \ref{subsec-numresults}, we present the numerical results.

  \subsection{Incoming ion distribution function} \label{subsec-finfty}
  
\begin{figure}[h] 
\centering
\includegraphics[width=0.7\textwidth]{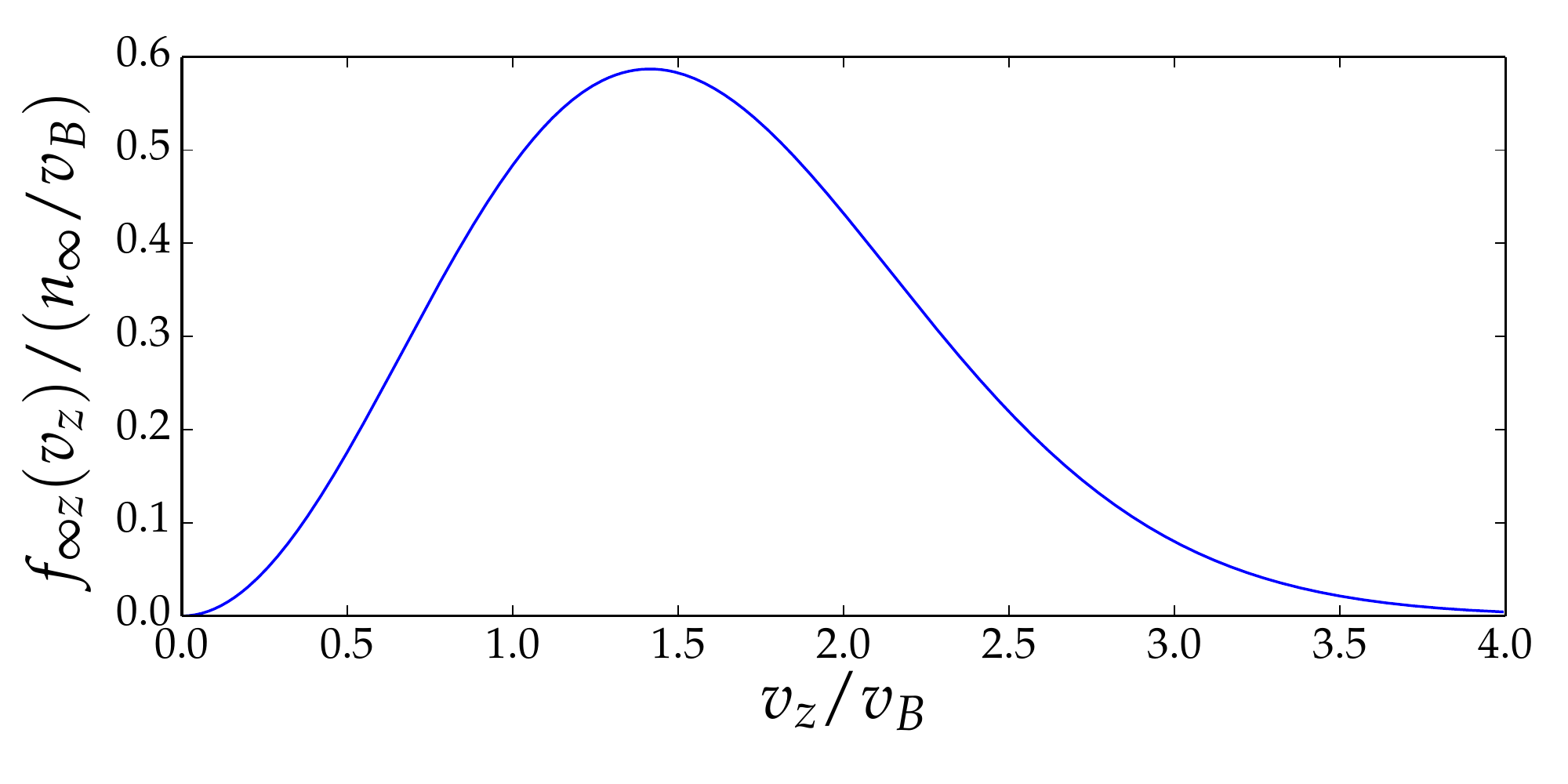}
\caption{The distribution function in (\ref{f-infty}) is shown as a function of the parallel velocity $v_z$ only, $f_{\infty z} \left(v_{z} \right) = \int  \int f_{\infty} \left(\vec{v} \right) dv_x dv_y $. This distribution function marginally satisfies (\ref{solvability}), hence $\int dv_z  f_{\infty z} \left( v_z \right) / v_z^2 = n_{\infty} / v_{\text{B}}^2 $. Its first moment is $u_{z\infty} = \left( 1/n_{\infty} \right) \int dv_z  v_z f_{\infty z} \left( v_z \right) \simeq 1.60 v_{\text{B}} $. }
\label{fig-finfinity}
\end{figure}

\begin{figure}[h] 
\centering
\includegraphics[width=0.7\textwidth]{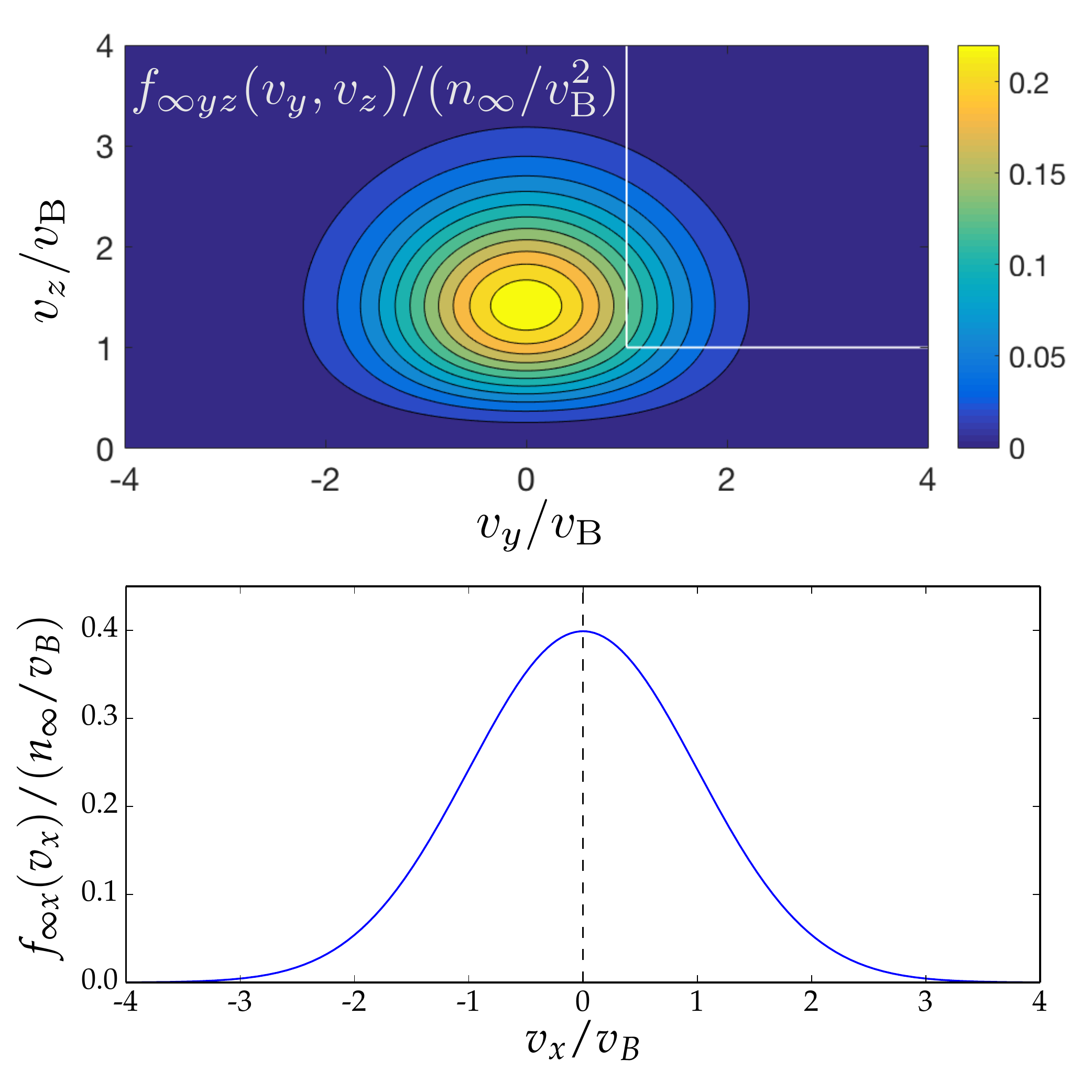}
\caption{The distribution function entering the magnetic presheath is shown as a function of the co-ordinates $\left( v_x, v_y, v_z \right)$. We define $f_{\infty x}(v_x) = \int_{-\infty}^{\infty} dv_y \int_0^{\infty} f_{\infty} (\vec{v} )  dv_z $ and $f_{\infty yz}(v_y, v_z) = \int_{-\infty}^{\infty} f_{\infty} (\vec{v} ) dv_x $. To compare with the distribution function $f_0 \left( \vec{v} \right) $ leaving the magnetic presheath, the box delimited by the white lines and the top right corner in the top diagram has the same size as Figure \ref{fig-f0yz}, and the region to the left of the dashed line in the bottom diagram is the domain of Figure \ref{fig-f0}. }
\label{fig-finfinitycomparison}
\end{figure}
  
Our ordering (\ref{scale-sep}) has allowed us to assume that the collisional layer only affects boundary conditions at $x \rightarrow \infty$.
A solution of the collisional layer would be required to obtain the correct form of $f_{\infty} \left( \vec{v} \right)$.
Alternatively, a drift-kinetic or gyrokinetic code \cite{Shi-Hammett-2017} of the scrape-off layer could be used to obtain such a distribution function. 
In this study, we take $T_{\text{i}} = T_{\text{e}} = T$, studying the dependence of the magnetic presheath on variation of the angle $\alpha$.
The dependence on variation of the ion temperature will be studied in a future publication.
We assume the following form for the lowest order ion distribution function at the magnetic presheath entrance,
\begin{align} \label{f-infty}
f_{\infty} \left( \vec{v} \right) = \frac{4}{\pi^{3/2}} n_{\infty} \left( \frac{m_{\text{i}}}{2T} \right)^{5/2} v_z^2 \exp\left( - \frac{m_{\text{i}} | \vec{v} |^2 }{2T}  \right) \text{.}
\end{align}
Changing to variables $\mu$ and $U$, the distribution function (\ref{f-infty}) is
\begin{align} \label{F-numerical-mu-U}
  F_{\text{cl}} \left(\mu, U \right) =  \frac{8}{\pi^{3/2}} n_{\infty} \left( \frac{m_{\text{i}}}{2T} \right)^{5/2}\left( U - \Omega \mu \right) \exp\left( - \frac{m_{\text{i}} U}{T} \right) \Theta \left(v_z\right) \text{,}
\end{align} 
which is constant throughout the magnetic presheath to lowest order in $\alpha$.
This form was used in other studies, for example \cite{Coulette-Manfredi-2016}, and it is plotted in Figures \ref{fig-finfinity} and \ref{fig-finfinitycomparison}. 
We define the ion thermal velocity $v_{\text{t,i}} = \sqrt{2T/m_{\text{i}}}$ and the ion gyroradius $\rho_{\text{i}} = v_{\text{t,i}}/ \Omega$. 
The sound speed is $v_{\text{B}} = \sqrt{T/m_{\text{i}}} = v_{\text{t,i}}/ \sqrt{2}$. 
The distribution function (\ref{f-infty}) marginally satisfies the solvability condition (\ref{solvability-vz}),
and the coefficient $k_{3/2}$ can be computed from (\ref{k32}), obtaining
\begin{align} \label{k32-Finfty}
\sqrt{ \frac{T}{e} } \left( \frac{v_{\text{B}}}{ \Omega } \right)^2  k_{3/2}  = \frac{8}{3\sqrt{\pi}}  \simeq 1.50  \text{.}
\end{align}
The average ion velocity in the $z$ direction at the magnetic presheath entrance is 
\begin{align} \label{finfty-flow}
u_{z\infty}  = \frac{1}{n_{\infty}} \int f_{\infty}\left( \vec{v} \right) v_{z}  d^3v = 2\sqrt{\frac{2}{\pi}}v_{\text{B}} \simeq 1.60v_{\text{B}} \text{.}
\end{align} 
%  \begin{align}
%  F\left(\mu, U \right) = n_{e\infty} \left( U - \Omega \mu \right) \exp\left( - \frac{m_{\text{i}}}{T_{\text{i}}} \left( U + u\sqrt{2\left( U - \Omega \mu \right)} + \frac{1}{2} u^2 \right) \right)
%  \end{align}
% The distribution function as expressed in equation (\ref{F-numerical-mu-U}) 

\subsection{Numerical method} \label{subsec-iteration}

We discretize the potential on a grid $x_{\eta}$ (labelled by the index $\eta$)
\begin{align} \label{x-grid}
\frac{x_{\eta}}{\rho_{\text{i}}} = \begin{cases}
\left( 0.05\eta \right)^2 &  \text{ for } 0 \leqslant \eta < 10 \text{,} \\
 0.25 +  0.1\left( \eta - 10 \right) &  \text{ for } 10 \leqslant \eta < \eta_2 = 129 \text{.} 
\end{cases}
\end{align}
We numerically calculate the ion density profile $n_{\text{i}}\left( x_{\eta} \right)$ in the region $0 \leqslant x_{\eta} \leqslant x_{\eta_1} = 6.15 \rho_{\text{i}} $ ($\eta_1 = 69$). 
The domain in $x$ is larger than $[0, x_{\eta_1}]$ because the potential profile in the region $x_{\eta_1} < x \leqslant x_{\eta_2} = 12.15 \rho_{\text{i}}$ is necessary to correctly evaluate the ion density at $ x_{\eta_1} $ and in its neighbourhood. 
The electron density profile $n_{\text{e}} \left( x_{\eta} \right)$ is evaluated by inserting $\phi \left( x_{\eta} \right)$ into equation (\ref{ne-Boltzmann}). 
We iterate over electrostatic potential functions $\phi_{\nu} \left( x_{\eta} \right)$, where $\nu$ is an index labelling the iteration number. 
The problem of solving (\ref{quasineutrality-compact}) is equivalent to finding, after $N$ iterations, a $\phi_N \left( x_{\eta}\right)$ for which $n_{\text{e,}N}\left( x_{\eta} \right) \simeq n_{\text{i,}N} \left( x_{\eta} \right)$ in the region $0 \leqslant x \leqslant x_{\eta_1}$. 

\begin{figure}[h] 
\centering
\includegraphics[width=0.47\textwidth]{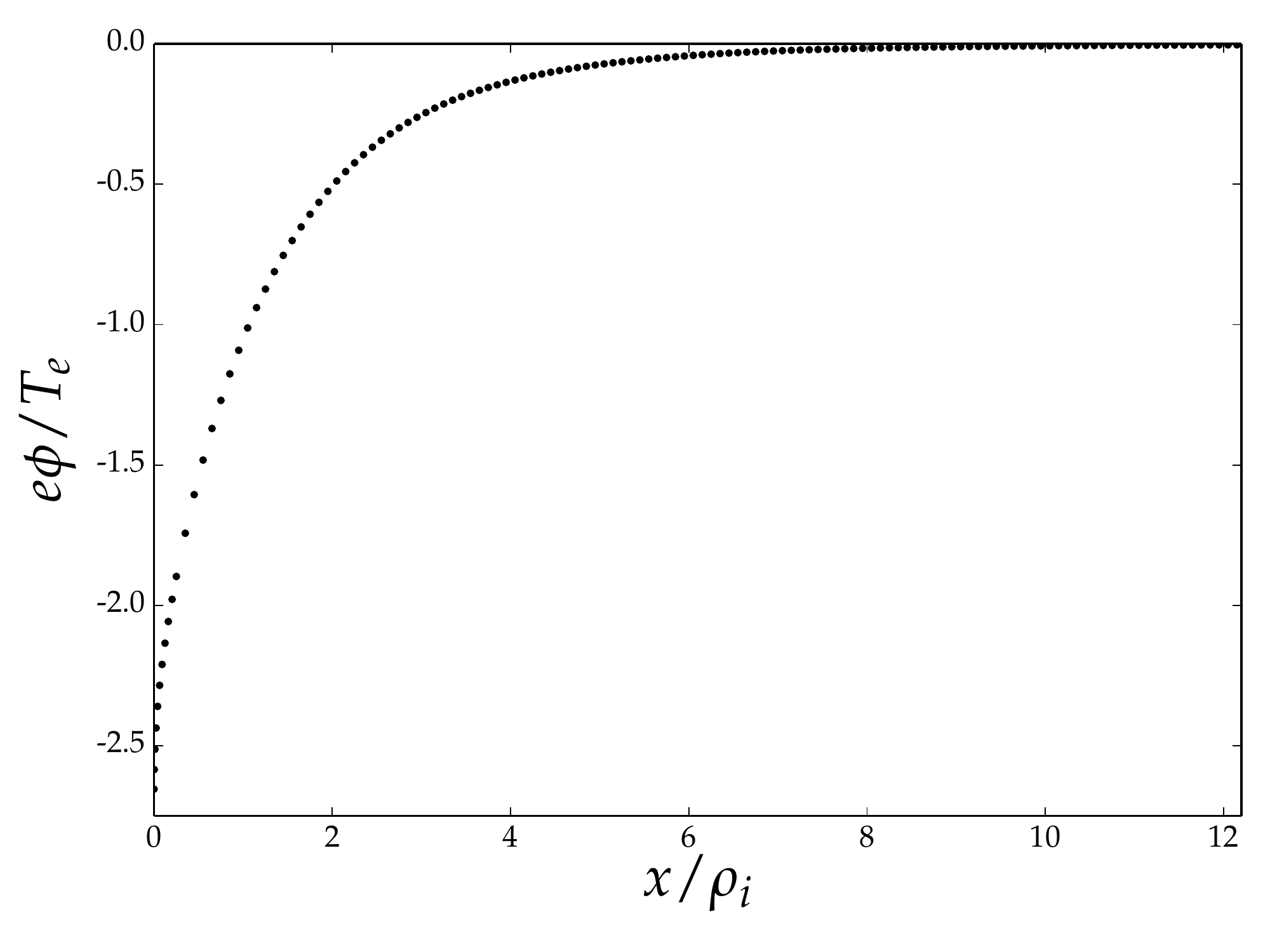}
\includegraphics[width=0.47\textwidth]{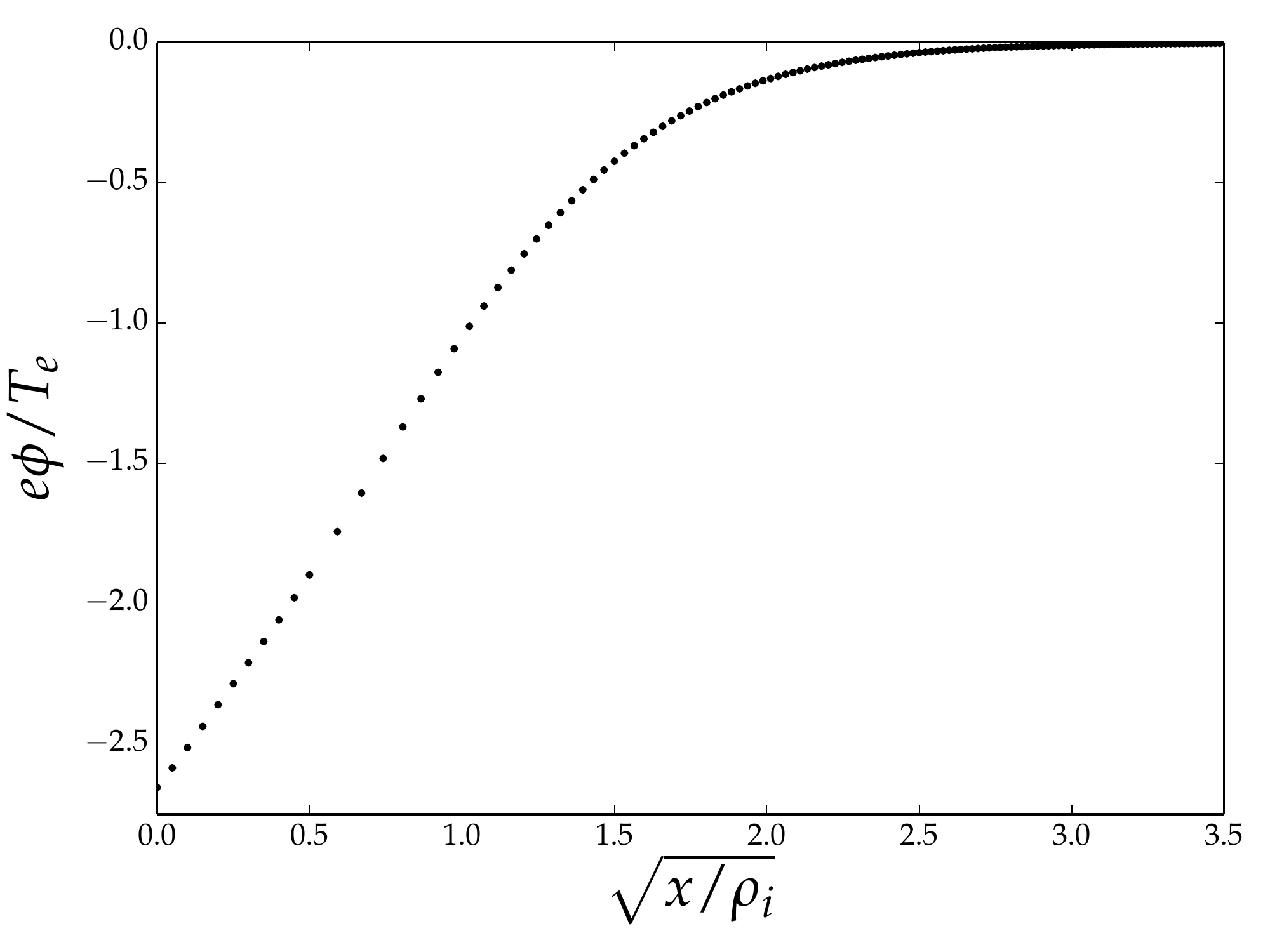}
\caption{An example solution for the electrostatic potential profile (for $\alpha = 0.05$) is plotted on the grid of equation (\ref{x-grid}). Initially $\phi$ increases linearly with $\sqrt{x}$, which justifies our choice of grid. }
\label{fig-phires}
\end{figure}

Near $x=0$, the grid (\ref{x-grid}) that we use to discretize all functions of $x$ has evenly spaced values of $\sqrt{x/\rho_{\text{i}}}$ ranging from $0$ to $0.5$ in intervals of $0.05$. The reason for this is that the self-consistent solution of the electrostatic potential is expected to be proportional to $\sqrt{x}$ near $x=0$, as in equation (\ref{phi-near0-2}). This behaviour of the electrostatic potential is captured by our grid as shown in Figure \ref{fig-phires}. For $\sqrt{x/\rho_{\text{i}}}>0.5$, corresponding to $x/\rho_{\text{i}}>0.25$, our grid has evenly spaced values of $x/\rho_{\text{i}}$, ranging from $0.25$ to $12.15$ in intervals of $0.1$.

The density integrals in equations (\ref{ni-closed}) and (\ref{ni-open-Deltavx}) are evaluated numerically at every point $x_{\eta}$ by employing the trapezoidal rule. In order to evaluate those integrals, we first evaluate the integrands. We introduce a grid of positions $\bar{x}_{\gamma}$ (labelled with the index $\gamma$),
\begin{align} \label{xbar-grid}
\frac{\bar{x}_{\gamma}}{\rho_{\text{i}}} = 0.01\gamma  \text{ for } 0 \leqslant \gamma < 1200 \text{.} 
\end{align}
Then, we evaluate the function $\chi \left( x_{\eta}, \bar{x}_{\gamma} \right)$ at all possible values of $x_{\eta}$ and $\bar{x}_{\gamma}$. We find the location of the effective potential maximum $x_{\text{M}}$ corresponding to the index $\eta_{\text{M}} \left(\gamma \right)$ that satisfies either
%\begin{align} \label{xmax-numerical}
%\chi \left( x_{\eta_{\text{M}} \left(\gamma \right)} , \bar{x}_{\gamma} \right) > \chi \left( x_{\eta_{\text{M}} \left(\gamma \right) + 1} , \bar{x}_{\gamma} \right) & \text{ for }\eta_{\text{M}} \left(\gamma \right) = 0 \text{ (type I),}\nonumber
%\\
%\chi \left( x_{\eta_{\text{M}} \left(\gamma \right)} , \bar{x}_{\gamma} \right) > \chi \left( x_{\eta_{\text{M}} \left(\gamma \right) + 1} , \bar{x}_{\gamma} \right) \text{ AND }  \chi \left( x_{\eta_{\text{M}} \left(\gamma \right)} , \bar{x}_{\gamma} \right) > \chi \left( x_{\eta_{\text{M}} \left(\gamma \right) - 1} , \bar{x}_{\gamma} \right) & \text{ for } \eta_{\text{M}} \left(\gamma \right) \geqslant 1 \text{ (type II),}
%\end{align}
\begin{align} \label{xmax-typeI-numerical}
\chi \left( x_{\eta_{\text{M}} \left(\gamma \right)} , \bar{x}_{\gamma} \right) > \chi \left( x_{\eta_{\text{M}} \left(\gamma \right) + 1} , \bar{x}_{\gamma} \right) & \text{ for } \eta_{\text{M}} \left(\gamma \right) = 0  \text{ (type I) }
\end{align}
or
\begin{align}  \label{xmax-typeII-numerical}
\chi \left( x_{\eta_{\text{M}} \left(\gamma \right)} , \bar{x}_{\gamma} \right) > \chi \left( x_{\eta_{\text{M}} \left(\gamma \right) - 1} , \bar{x}_{\gamma} \right) \nonumber  \\ \text{ and }
\chi \left( x_{\eta_{\text{M}} \left(\gamma \right)} , \bar{x}_{\gamma} \right) > \chi \left( x_{\eta_{\text{M}} \left(\gamma \right) + 1} , \bar{x}_{\gamma} \right) & \text{ for } \eta_{\text{M}} \left(\gamma \right) \geqslant 1 \text{ (type II).}
\end{align}
We also find the location of the effective potential minimum $x_{\text{m}}$ corresponding to the index $\eta_{\text{m}} \left(\gamma \right)$ that satisfies
\begin{align} \label{xmin-numerical}
\chi \left( x_{\eta_{\text{m}} \left(\gamma \right)} , \bar{x}_{\gamma} \right) < \chi \left( x_{\eta_{\text{M}} \left(\gamma \right) - 1} , \bar{x}_{\gamma} \right) \nonumber \\ \text{ and }
\chi \left( x_{\eta_{\text{m}} \left(\gamma \right)} , \bar{x}_{\gamma} \right) < \chi \left( x_{\eta_{\text{M}} \left(\gamma \right) + 1} , \bar{x}_{\gamma} \right) & \text{ for } \eta_{\text{m}} \left(\gamma \right) \geqslant 1 \text{.}
\end{align}
%\begin{align} 
%\chi \left( x_{\eta_{\text{m}} \left(\gamma \right) - 1} , \bar{x}_{\gamma} \right) > \chi \left( x_{\eta_{\text{m}} \left(\gamma \right)} , \bar{x}_{\gamma} \right) < \chi \left( x_{\eta_{\text{m}} \left(\gamma \right) + 1} , \bar{x}_{\gamma} \right) & \text{ for } \eta_{\text{m}} \left(\gamma \right) \geqslant 1 \text{.}
%\end{align}

\begin{figure}[h] 
\centering
\includegraphics[width=0.6\textwidth]{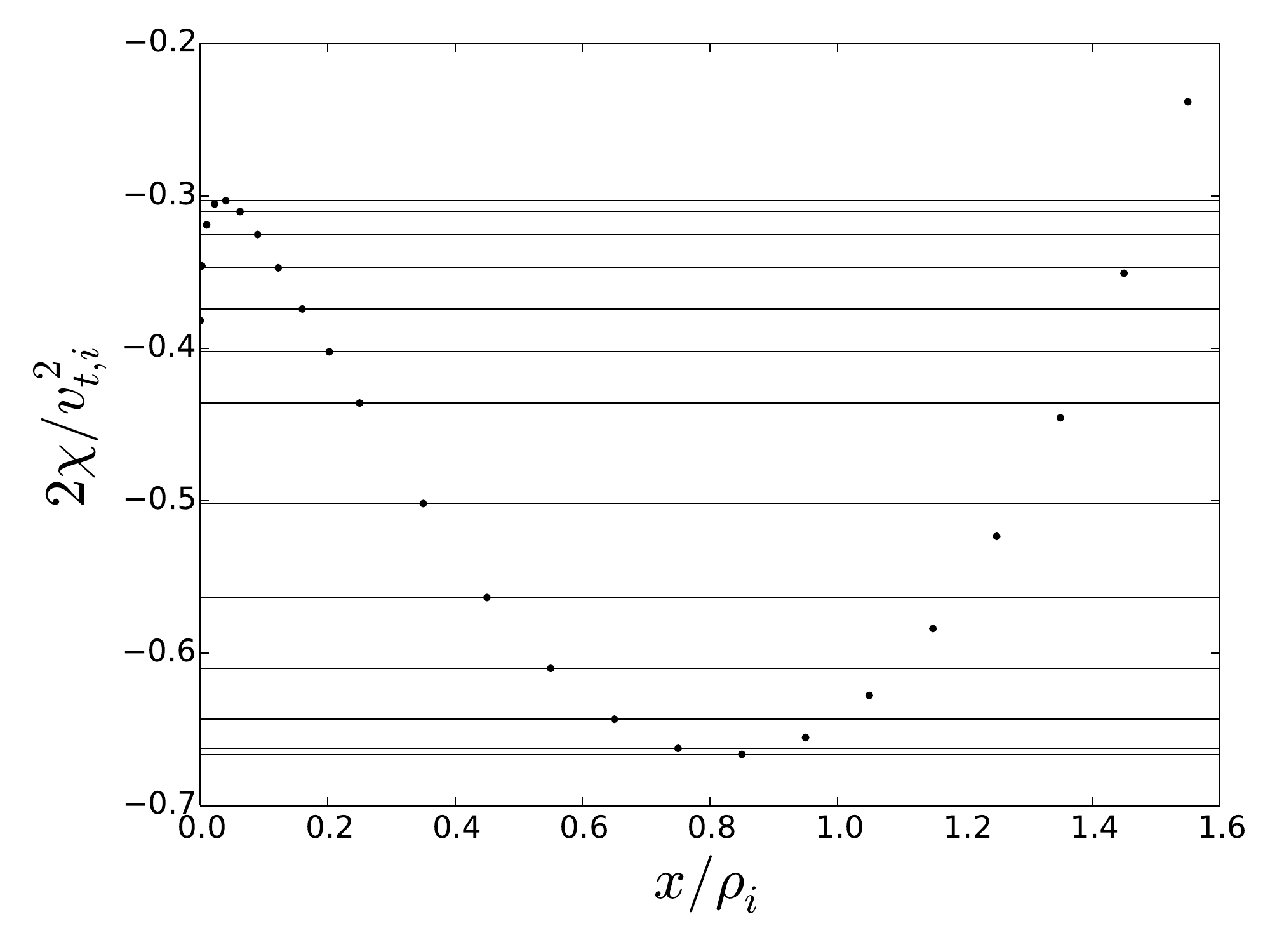}
\caption{ The values of $U_{\perp, \gamma \kappa}$ corresponding to different values of $\kappa$ are shown with horizontal lines on top of the effective potential curve $\chi \left( x_{\eta}, \bar{x}_{\gamma} \right)$, for a particular value of $\gamma$. Here, $\kappa$ ranges from $\kappa = 0$ (top line) to $\kappa = 12$ (bottom line). }
\label{fig-Uperpgrid}
\end{figure}

At every value of the orbit parameter $\bar{x}_{\gamma}$, we obtain a grid of possible values of perpendicular energy $U_{\perp, \gamma \kappa}$, indexed with $\gamma$ and $\kappa$, 
\begin{align} \label{Uperp-grid}
U_{\perp, \gamma \kappa} = \chi \left( x_{\kappa + \eta_{\text{M}} \left(\gamma \right) }, \bar{x}_{\gamma} \right) \text{ for } 0 \leqslant \kappa \leqslant   \eta_{\text{m}} \left(\gamma \right) -  \eta_{\text{M}} \left(\gamma \right) \text{.}
\end{align}
This grid is shown in Figure \ref{fig-Uperpgrid}. For all possible $\bar{x}_{\gamma}$ and $U_{\perp,\gamma \kappa}$, we evaluate the adiabatic invariant by performing the integral (\ref{mu-Uperp-xbar}) using the trapezoidal rule, to obtain the function $\mu_{\text{gk}} \left( \bar{x}_{\gamma}, U_{\perp, \gamma \kappa} \right) $. Similarly, for all possible values of $\bar{x}_{\gamma}$ we evaluate the integral $I \left( \bar{x}_{\gamma} \right)$ in (\ref{open-orbit-integral}) using the trapezoidal rule. For all values of $\gamma$ and $\kappa$, the total energy is labelled by the index $\iota$,
\begin{align} \label{U-grid}
\frac{2U_{\gamma\kappa\iota}}{v_{\text{t,i}}^2} = \frac{2U_{\perp, \gamma \kappa}}{v_{\text{t,i}}^2} + (0.2\iota)^2 \text{ for } 0 \leqslant \iota < \iota_{\text{max}} \text{,}
\end{align}
where $\iota_{\text{max}}$ is such that $2U/v_{\text{t,i}}^2 < 15.0$ and $7.5 v_{\text{t,i}}^2$ is a cutoff energy above which the distribution function is essentially zero. The distribution function $F_{\text{cl}} (\mu, U )$ of equation (\ref{F-numerical-mu-U}) is defined on a square grid of values of $2\Omega \mu /v_{\text{t,i}}^2$ and $2U/v_{\text{t,i}}^2$ which lie between $0$ and $15.0$ in intervals of $0.05$, and bilinearly interpolated at every integration point. The integrals over $U$ and over $U_{\perp}$ in equations (\ref{ni-closed}) and (\ref{ni-open-Deltavx}) are, for numerical convenience, evaluated over $v_z = \sqrt{2\left( U_{\gamma\kappa\iota} - U_{\perp, \gamma \kappa} \right)} $ and $|v_x | = \sqrt{2\left( U_{\perp, \gamma \kappa} - \chi (x_{\eta}, \bar{x}_{\gamma} )\right) }$ respectively (for this reason $U_{\gamma\kappa\iota}$ is defined such that linear increments in $\iota$ correspond to linear increments in $v_z$). Where necessary, the values of the integrands and of the integration limits of equations (\ref{mu-Uperp-xbar}), (\ref{ni-closed}), (\ref{open-orbit-integral}) and (\ref{ni-open-Deltavx}) are found by linear interpolation. 
%The closed orbit density integral (\ref{ni-closed}) is numerically calculated by summing over allowed values of the index $\gamma$ that are multiples of $4$ (using a coarser grid in $\bar{x}$), while the open orbit density integral (\ref{ni-open}) is calculated by summing over all allowed values of $\gamma$.

The iteration scheme we used hinges on imposing
 \begin{align} \label{iterationscheme}
 n_{\text{e,}\nu+1} \left( x_{\eta} \right) = wZn_{\text{i,}\nu} \left( x_{\eta} \right) + \left(1-w\right)n_{\text{e,}\nu} \left( x_{\eta} \right) \text{}
 \end{align}
 at every $(\nu+1)th$ iteration. Here, $w$ is a weight whose value lies in the range $0 < w \leqslant 1$. From (\ref{iterationscheme}), $\phi_{\nu+1} \left( x_{\eta} \right)$ is obtained by inverting the Boltzmann relation for $n_{e,\nu+1} \left( x_{\eta} \right)$, and the new guess for the potential profile is thus obtained for $0 \leqslant \eta \leqslant \eta_1$. For values of $\eta$ in the interval $\eta_1 + 1 \leqslant \eta \leqslant \eta_2$, the electrostatic potential $\phi_{\nu+1}\left(x_{\eta} \right)$ is completed by matching to the appropriate functional form for $\phi \left( x \right)$ near $x \rightarrow \infty$. 
With our choice of distribution function marginally satisfying the Chodura condition (\ref{solvability}), $\phi(x)$ satisfies equation (\ref{phisol1}) for large $x$. %$k_1 = 0$ (numerically $k_1 \approx 0$) and 
The value of $k_{3/2}$ is calculated numerically and coincides (to within a numerical error of $2\%$) with equation (\ref{k32-Finfty}). The value of $C_{3/2}$ is obtained by imposing $\phi_{\nu+1} \left( x_{\eta_1} \right) = - 400 k_{3/2}^{-2} (x_{\eta_1}+C_{3/2})^{-4}  $  to get
\begin{align}
C_{3/2} = \sqrt{\frac{20}{k_{3/2}}}  \left[ - \phi_{\nu+1} \left( x_{\eta_1} \right) \right]^{-1/4} - x_{\eta_1} \text{.}
\end{align} 
The new guess for the electrostatic potential is then
\begin{align}
\phi_{\nu+1} \left( x_{\eta} \right) = \begin{cases} \frac{T_{\text{e}}}{  e} \ln \left( w \frac{ Zn_{i\nu} \left( x_{\eta} \right) }{  n_{\infty} } + \left(1-w\right) \frac{ n_{e\nu} \left( x_{\eta} \right) }{ n_{\infty} } \right) & \text{for } 0 \leqslant \eta \leqslant \eta_1 \text{,} \\ - \frac{ 400 }{  k_{3/2}^2  (x_{\eta}+C_{3/2})^4 }  & \text{for } \eta_1 + 1 \leqslant \eta \leqslant \eta_2 \text{.}
\end{cases} 
\end{align} 
This can be used to evaluate $n_{\text{i,}\nu+1}(x_{\eta})$ in the region $0 \leqslant \eta \leqslant \eta_1$ and continue the iteration. The first potential guess we use is a flat potential profile ($\phi_0 (x_{\eta}) = 0$ for all $\eta$). After $N$ iterations, a numerical solution $\phi_N \left(x_{\eta}\right)$ which satisfies $n_{\text{e,}N} (x_{\eta} ) \simeq n_{\text{i,}N} ( x_{\eta} )$ for all $\eta$ is found. The deviation of $\phi_{\nu} \left( x_{\eta} \right)$ from the exact solution (which satisfies $n_{\text{i}}\left( x_{\eta} \right) = n_{\text{e}}\left( x_{\eta} \right)$) is measured by calculating the quantity
\begin{align}
\tilde{n}_{\nu} \left( x_{\eta} \right) = 1 - \frac{n_{\text{i,}\nu} \left( x_{\eta} \right)}{n_{\text{e,}\nu} \left( x_{\eta} \right)} \text{.}
\end{align} 
Convergence to an acceptable solution is given by the criterion that the root mean square value of $\tilde{n}_{\nu} \left( x_{\eta} \right)$ be less than some number $E$,
\begin{align}\label{convergence-criterion}
\left[  \sum_{\eta=0}^{\eta_1} \frac{1}{\eta_1 + 1} \tilde{n}_{\nu}^2  \left( x_{\eta} \right)  \right]^{1/2} < E \text{.}
\end{align} 
In obtaining the numerical results in the next subsection, we used $E = 0.007$ for all values of $\alpha$. %, except $\alpha = 0.01$, for which we used $E = 0.01$.
%Because the solution of the problem we have posed is a lowest order solution in $\alpha$, it is important that $\tilde{n} _{N}\left( x\right) \lesssim \alpha$ everywhere, which is checked. % always be less than or roughly equal to $\alpha$. %In fact, near $x=0$, $\tilde{Q}\left( x\right)$ only has to be less than $\sqrt{\alpha}$.

\begin{figure}[h] 
\centering
\includegraphics[width=0.6\textwidth]{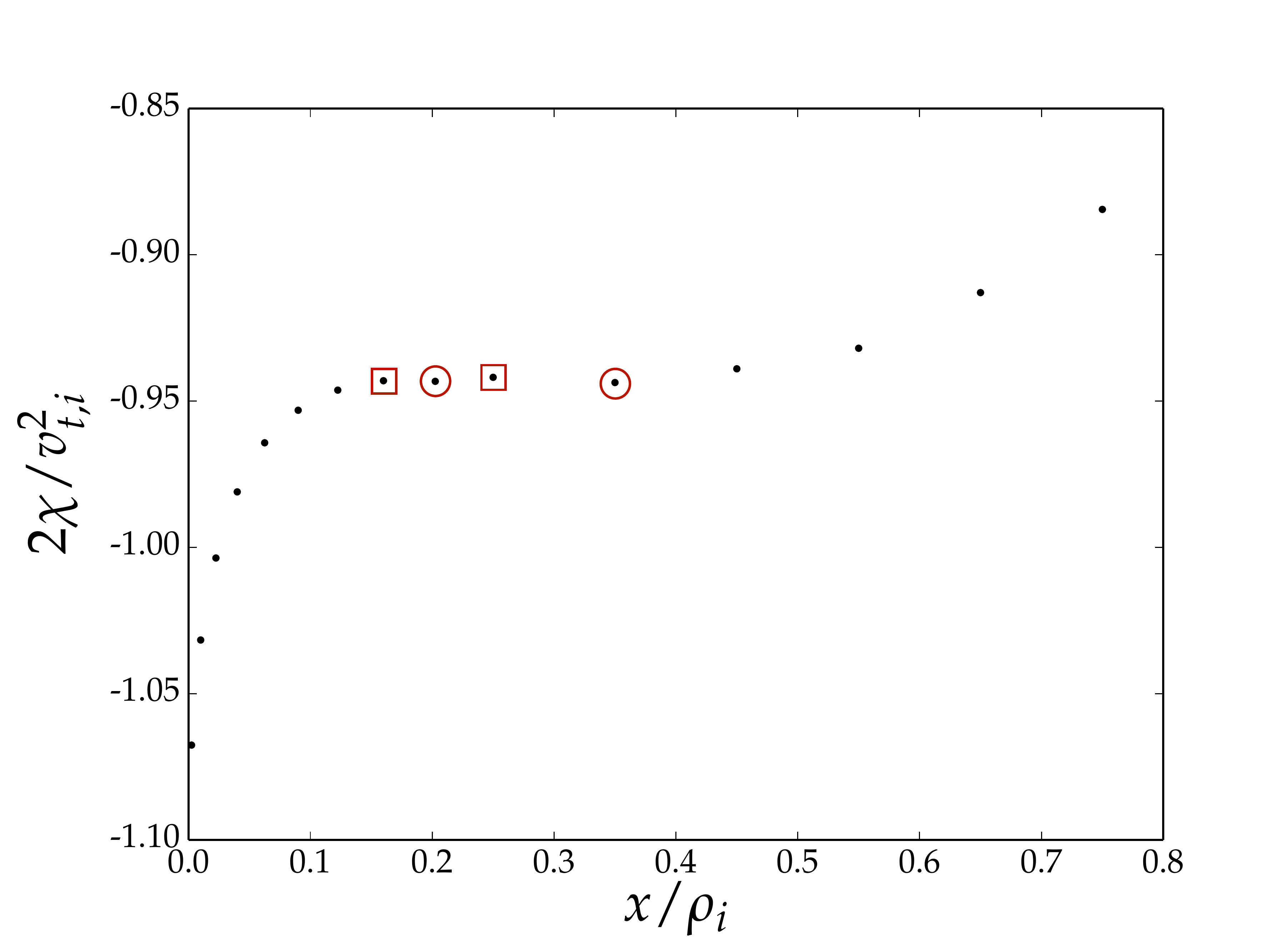}
\caption{An example of an effective potential $\chi \left( x_{\eta}, \bar{x}_{\gamma} \right) $ in which our algorithm for generating the grid $U_{\perp, \gamma \kappa}$ fails, because it does not take into account the possibility of finding multiple effective potential minima (marked with circles) and maxima (marked with squares) for a given $\gamma$. %This curve was obtained using $\bar{x} = 0.98 \rho_{\text{i}}$ ($\gamma = 98$) from an electrostatic potential which resulted from an iteration performed without smoothing, with $w = 0.6$.  
}
\label{fig-effpot-prob}
\end{figure}

The method that we use can give a non-smooth numerical second derivative of the potential $\phi_{\nu} \left( x_{\eta} \right)$.
The numerical noise in the second derivative is problematic because the algorithm fails to take into account the possibility of more than one maximum or minimum of the effective potential existing for some value of $\bar{x}$. 
If at some point during the iteration the function $\phi_{\nu} \left( x_{\eta} \right)$ is such that, for some value of $\gamma$, the function $\chi ( x_{\eta}, \bar{x}_{\gamma} )$ has more than one index $\eta_{\text{M}} (\gamma)$ that satisfies either (\ref{xmax-typeI-numerical}) or (\ref{xmax-typeII-numerical}) (and more than one index $\eta_{\text{m}} (\gamma)$ that satisfies (\ref{xmin-numerical})), a more sophisticated analysis than the one we presented is necessary to obtain the grid of values of $U_{\perp}$.
The appearance of multiple maxima and minima, shown in Figure \ref{fig-effpot-prob}, can be due to the numerical second derivative of $\phi \left( x_{\eta} \right)$ having pronounced oscillations, even when $\phi \left( x_{\eta} \right)$ looks smooth to the naked eye.
To avoid the appearance of multiple maxima and minima, in this work we perform a smoothing operation on the second derivative of $\phi_{\nu} \left( x \right)$ (with respect to $\sqrt{x}$) before iteration number $\nu+1$, for a certain number of iterations until the densities obtained using $\phi_{\nu} (x)$ are close to satisfying criterion (\ref{convergence-criterion}). After that, we carry out the last few iterations without smoothing. 
%The smoothing procedure does not allow a good convergence to the solution when the potential guess is already quite close to it, but it allows a faster initial convergence to the solution because a greater value of $w$ can be used without worrying that the resulting next guess for the potential will have a non-smooth second derivative. 
%Moreover, smoothing is used again if, in the later stages of the iteration, the second derivative becomes so non-smooth that the density calculation fails (in these cases, smoothing usually modifies the potential in such a way that density evaluation can be performed again without failing, although it makes the potential guess worse than the unsmoothed one). 
In our iterations, $w=0.5$ when the smoothing operation is performed, while $w=0.2$ when it is not.

The computing time necessary to obtain the numerical solutions is small.
The number of iterations required for convergence is typically less than 20, and each iteration runs in approximately 3 seconds on a laptop. 
Consequently, the total run time of the code on a laptop is typically less than one minute.
The computing time can be further reduced by using a better initial guess, improving the integration schemes and reducing the number of integration points.

From here on, we omit all indices associated with quantities and functions evaluated numerically.

\subsection{Results and discussion} \label{subsec-numresults}

\begin{figure}[h] 
\centering
\includegraphics[width=0.6\textwidth]{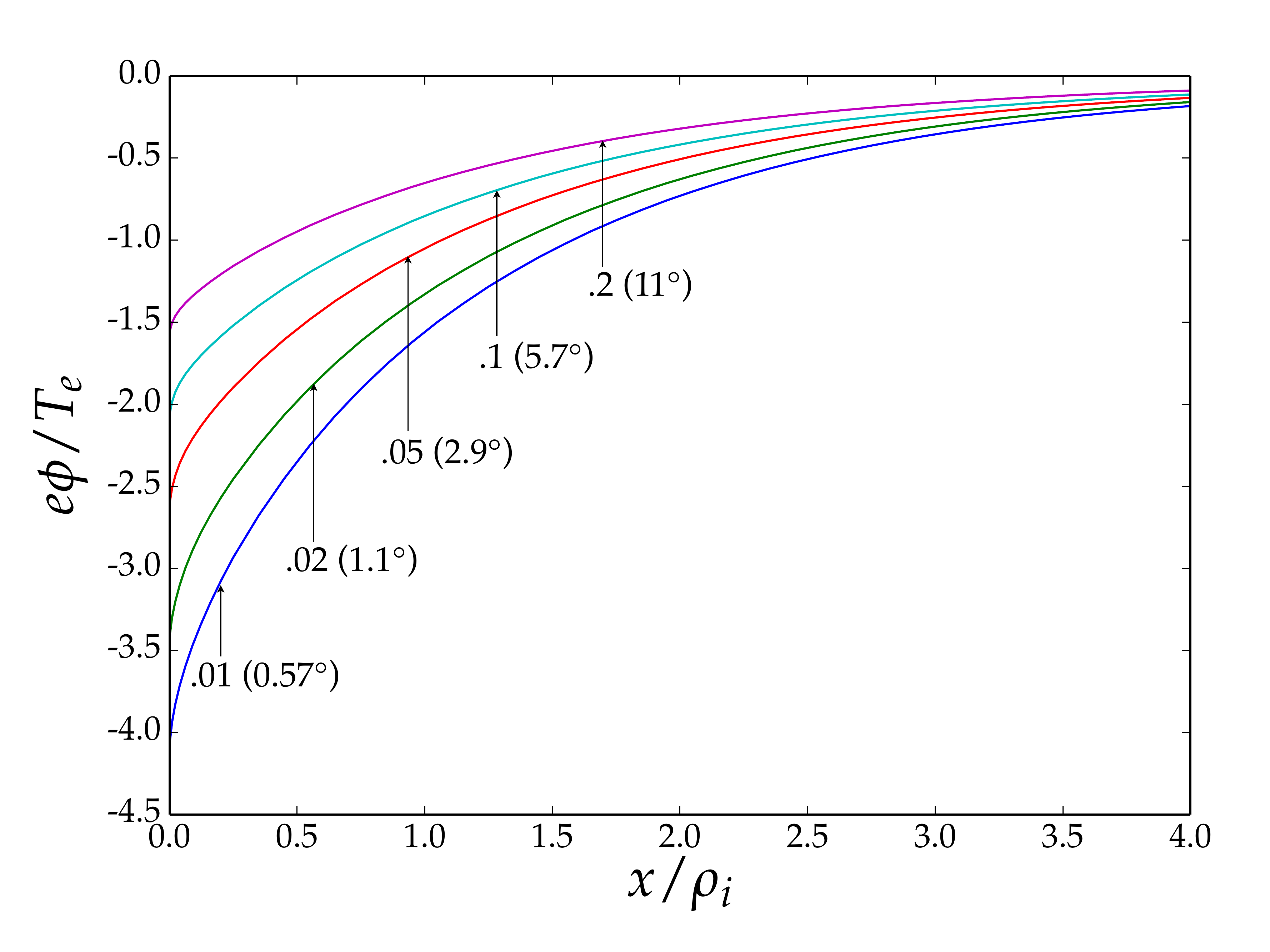}
\caption{The electrostatic potential profile is plotted for a range of angles $\alpha$ labelled in radians (degrees). Near $x=0$, $\phi(x) - \phi (0) \propto \sqrt{x} $. }
\label{fig-phiprofile}
\end{figure}
\begin{figure}[h]
\centering
\includegraphics[width=0.8\textwidth]{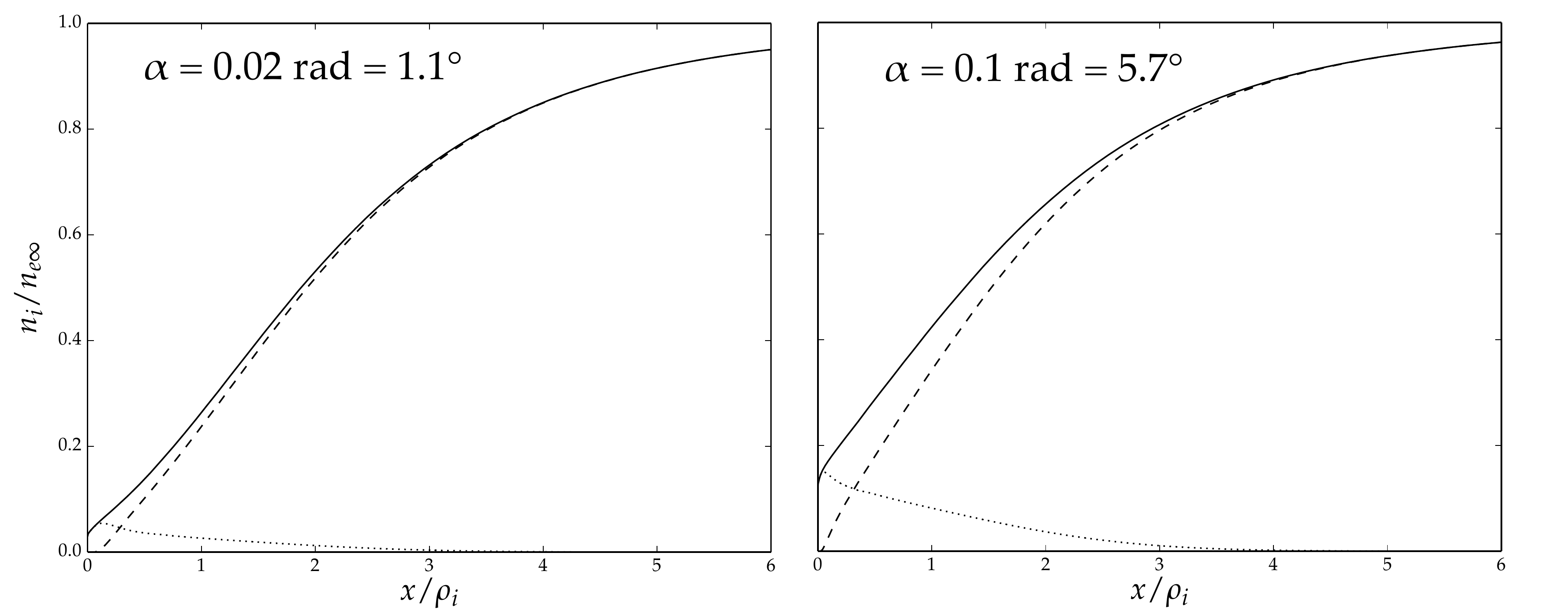}
\caption{The ion density (solid line) for $\alpha = 0.02$ and $\alpha = 0.1$ is shown with the contributions from the closed ion orbits (dashed line) and the open orbits (dotted line) clearly marked. The open orbits clearly dominate in a very small region near $x=0$, then there is an overlap region in which the open orbit contribution and the closed orbit contribution have a similar size, while at larger values of $x$ the closed orbit density dominates.}
\label{fig-ni}
\end{figure}
\begin{figure}[h]
\centering
\includegraphics[width=0.6\textwidth]{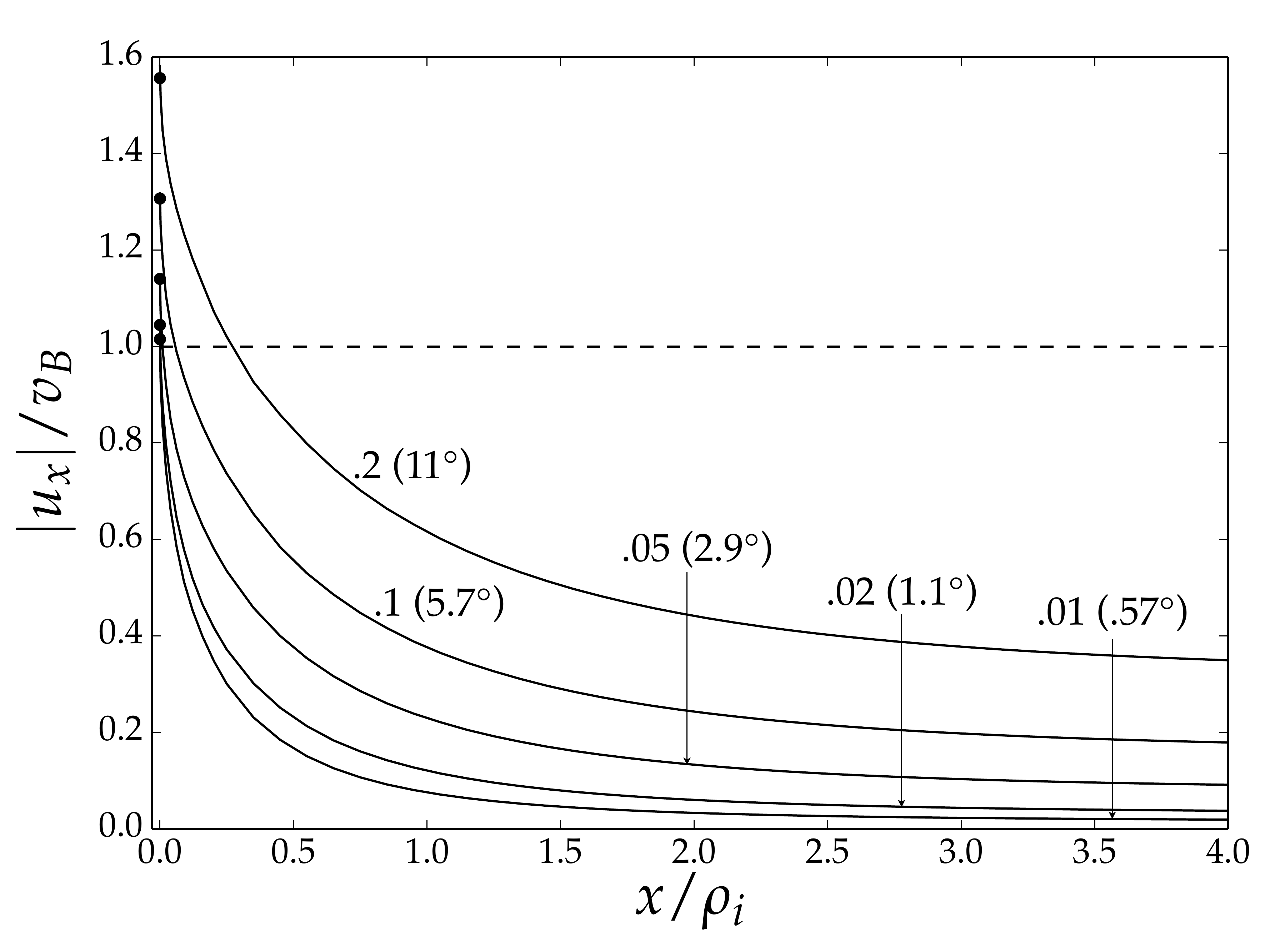}
\caption{The average ion velocity in the direction normal to the wall is shown at various angles $\alpha$, labelled in radians (degrees) above the corresponding curve or with an arrow. The flow velocity obtained via the integral (\ref{ux0-integral}) is shown with a black circle at $x=0$, which to lowest order coincides with the value we calculate from continuity. 
The usual cold ion Bohm limit is indicated by the dashed line $|u_x|/v_{\text{B}} = 1$. The ion flow lies above the cold ion Bohm limit at $x=0$ because the ions are ``warm'' ($T_{\text{i}} \neq 0$). However, at small angles $\alpha \lesssim 0.05$, the ion flow at $x=0$ approaches the cold ion Bohm limit. }
\label{fig-flux}
\end{figure}

The normalized electrostatic potential $e \phi ( x ) / T_{\text{e}}$ is shown in Figure \ref{fig-phiprofile} for a range of angles $\alpha$. 
A general property of the potential curves is that they rise very steeply near $x=0$, with the scaling $\phi(x) - \phi (0) \propto \sqrt{x}$ in that region (as can be seen explicitly in Figure \ref{fig-phires}). 
We have shown that this behaviour of $\phi(x)$ is expected, and it is connected with the marginal kinetic Bohm condition (\ref{Bohm-kinetic-marginal}) being satisfied. 
The value of $q_2$ that we calculate numerically from the distribution function at $x=0$, using equation (\ref{q2-def}), is consistent with the behaviour of the electrostatic potential near $x=0$.% given in equation (\ref{phi-near0-2}).
%Our results for the distribution function at $x=0$ lend further support to this connection.  %The approach to $x\rightarrow \infty$ obeys $\phi \propto 1/\left( x + C_{3/2} \right)^{4}$ because this is what we impose using the analytically derived boundary condition (\ref{phisol1}).

The ion density profiles for $\alpha=0.02$ and $\alpha = 0.1$ are shown in Figure \ref{fig-ni}. 
The open orbit density can be seen to initially increase and then quickly decrease with distance from the wall. 
This behaviour is consistent with the behaviour of $\Delta v_x$ for type II orbits (see Figure~\ref{fig-Deltavx} and the discussion following equation (\ref{ni-open-order})). 
The open orbit density is clearly the dominant contribution to the density in the neighbourhood of $x=0$, while for large $x$ approximately closed orbits give the largest contribution.
 
The flow velocity of ions across the magnetic presheath is commonly calculated in fluid models. 
Therefore, it is useful to calculate it to compare with previous results.
Here we calculate the flow by using the ion continuity equation.
The ion flux towards the wall across the magnetic presheath (which has no ion sources in our model) must be constant for steady state particle conservation,
\begin{align}
\frac{\partial }{\partial x} \left( n_{\text{i}} \left( x \right) u_x \left( x\right) \right) = 0 \text{,}
\end{align} 
where $ u_x \left( x\right) $ is the average velocity of ions in the $x$ direction.
  %Our formalism allows us to compute the lowest order ion density across the magnetic prsheath, but not the lowest order ion flow towrds the wall at every position $x$. 
 At the magnetic presheath entrance $x \rightarrow \infty$, the flow towards the wall is obtained from the average over the distribution function of the gyroaveraged motion of ions towards the wall, given by $\dot{\bar{x}}$ (note that, due to distortion of the orbits, this does not remain true across the magnetic presheath). 
 Using equations (\ref{vz-U-Uperp}) and (\ref{xbardot}), the flow in the $z$ direction, $u_{z\infty}$, is related to the flow in the $x$ direction, $u_{x\infty} $, via $u_{x\infty} =   - \alpha u_{z\infty}$. 
 This is equivalent to the boundary condition of flow being parallel to the magnetic field at $x \rightarrow \infty$ \cite{Riemann-1994}.
 The flow $u_{z\infty}$ is obtained as a moment of the incoming distribution function (see equation (\ref{finfty-flow}))
  \begin{align}
 u_{z\infty} = \frac{1}{n_{\infty}} \int f_{\infty} ( \vec{v} ) v_{z} d^3v \text{.}
 \end{align}
The flux of ions towards the wall is conserved and therefore given by $ n_{\text{i}} \left( x \right) u_x \left( x\right) = n_{\infty} u_{x,\infty} =  - \alpha n_{\infty} u_{z\infty} $. The average lowest order ion flow velocity towards the wall at a general position $x$ is therefore
 \begin{align} \label{ionflow}
u_x \left( x\right) \simeq - \frac{\alpha n_{\infty} u_{z \infty} }{n_{\text{i}} \left( x \right)} \text{.}
 \end{align}
 The function (\ref{ionflow}) evaluated at $x=0$ can be checked, for consistency, against the appropriate integral of the distribution function (\ref{f0-def}),
\begin{align} \label{ux0-integral}
u_{x0} = \frac{1}{n_{\text{i}} \left( 0 \right) } \int f_{0} \left( \vec{v} \right) v_x d^3v \text{.}
\end{align}
In Figure \ref{fig-flux}, we plot the average ion velocity profile $u_x\left( x\right)$, obtained using equation (\ref{ionflow}), for a range of angles $\alpha$. The magnetic presheath acceleration turns the ion flow from being (super)sonic in the direction parallel to the magnetic field to being (super)sonic in the $x$ direction normal to the wall. 
%At the smallest angles, acceleration of the flow in the $x$ direction is more noticeable because a smaller component of the parallel flow is directed normal to the wall, so the turning effect is larger. 
At $x=0$, the flow velocity is calculated in an alternative way, by taking the integral of the distribution function as in equation (\ref{ux0-integral}). The value thus obtained is marked on the curves for each value of $\alpha$, and it is consistent with the value obtained by using equation (\ref{ionflow}).

\begin{figure}[h]
\centering
\includegraphics[width=0.6\textwidth]{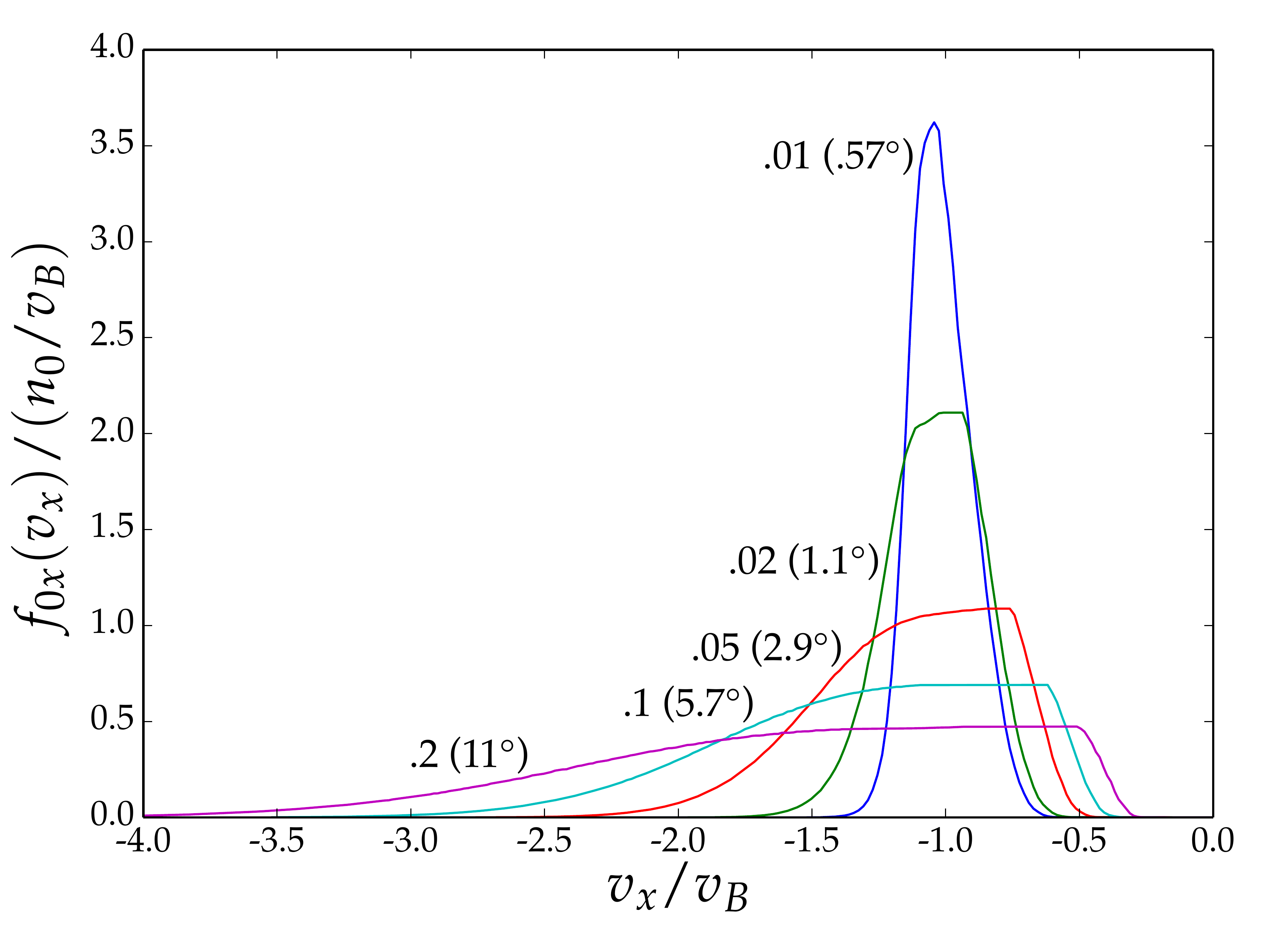}
\caption{The distribution function $f_{0x} \left(v_x\right) = \int_{-\infty}^{\infty}   dv_y \int_{-\infty}^{\infty}  f_{0}\left( v_x, v_y, v_z \right) dv_z $ for a range of angles $\alpha$, marked to the left of the corresponding curve in radians (degrees). }
\label{fig-f0}
\end{figure}
\begin{figure}[h]
\centering
\includegraphics[width=0.7\textwidth]{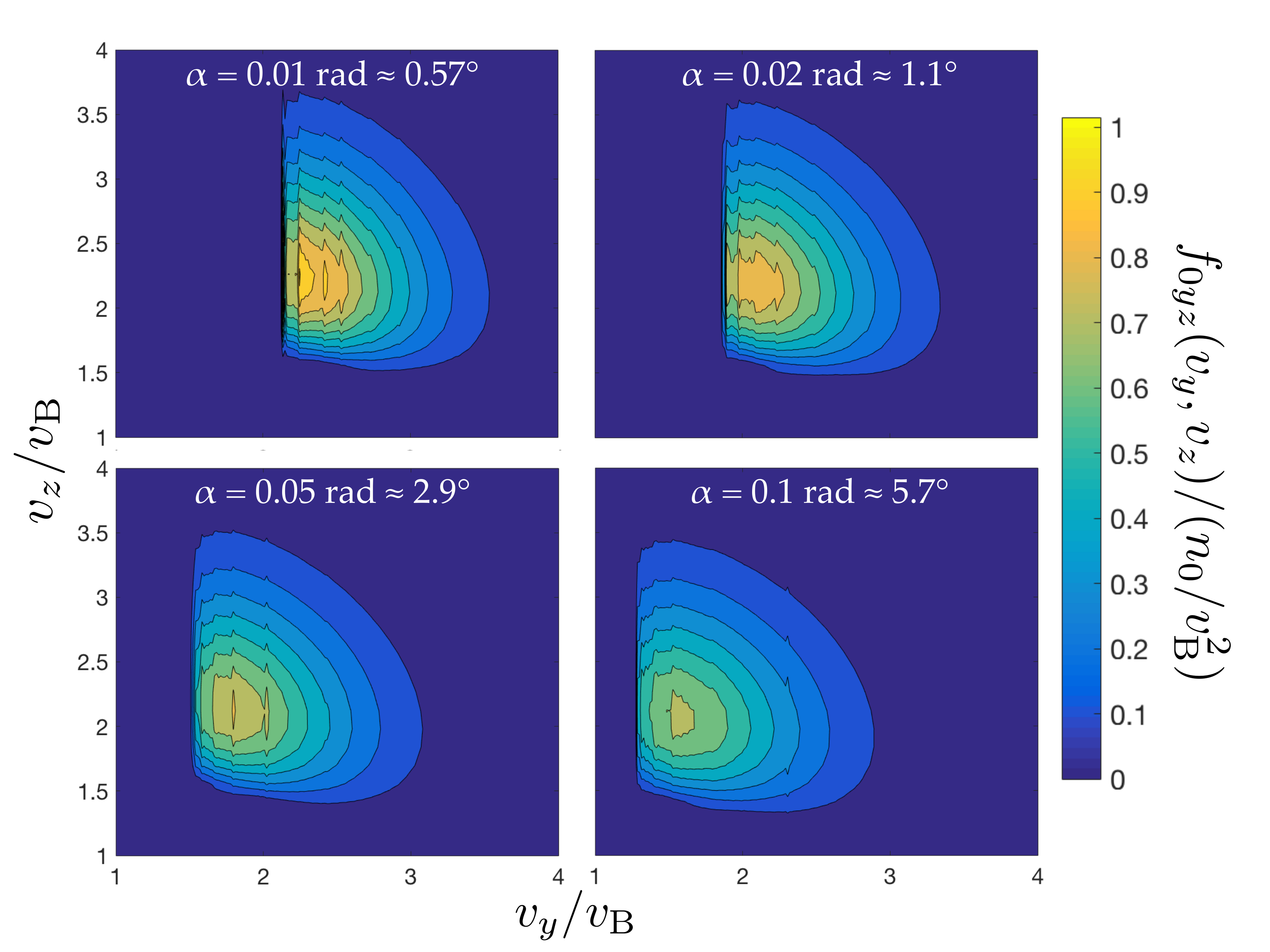}
\caption{The distribution function $f_{0yz} \left(v_y, v_z\right)  =  \int_{-\infty}^0   f_{0}\left( v_x, v_y, v_z \right) dv_x $ for a range of angles $\alpha$, marked on each panel in radians (rad) and degrees ($^{\circ}$). }
\label{fig-f0yz}
\end{figure}

By asymptotic matching, the distribution function in (\ref{f0-def}) is the distribution function entering the Debye sheath.
In the Debye sheath, electrostatic forces normal to the wall dominate over magnetic forces, hence $v_x$ is the only velocity component that changes significantly \cite{Riemann-review}. 
Therefore, only knowledge of the function %$f_{0x} \left( v_x \right) = \int \int f_{0} \left( \vec{v} \right) dv_y dv_z$, given by
%We obtain the function $f_{0x} ( v_x)$ by integrating $f_0 \left( v_x, v_y, v_z \right)$ in $v_y$ and $v_z$,
\begin{align} \label{f0x-def}
 f_{0x}\left( v_x \right)  = &  \int_{-\infty}^{\infty} dv_y \int_{-\infty}^{\infty}  f_{0}\left( v_x, v_y, v_z \right) dv_z  \nonumber \\
\simeq &  \int_{\bar{x}_{\text{m,o}}}^{\infty} \Omega d\bar{x} \int_{\chi_{\text{M}} \left( \bar{x} \right) }^{\infty} \frac{F_{\text{cl}}\left( \mu_{\text{gk}} \left( \bar{x},  \chi_M \left( \bar{x} \right) \right), U \right) }{V_{\parallel} \left( \chi_{\text{M}} \left( \bar{x} \right) , U \right) }  \nonumber \\
& \times \hat{\Pi} \left(  v_x ,  - V_x \left( 0, \bar{x}, \chi_{\text{M}} \right) - \Delta v_x  , - V_x \left( 0, \bar{x}, \chi_{\text{M}} \right)  \right) dU   \text{}
\end{align}
 is needed to solve for the electrostatic potential in the Debye sheath. 
The distribution $f_{0x} \left( v_x \right) $ is shown in Figure \ref{fig-f0} for a range of angles $\alpha$.  
A general feature of this function is that it is very close to zero near $v_x =0$. 
This is expected from the discussion in Section \ref{subsec-expansion-near0}, where we concluded that there is an exponentially small number of ions with small values of $v_x$ if the distribution function $F_{\text{cl}}$ exponentially decays at large energy $U$.
%We discussed in Section 5 how the flatness of the distribution function near $v_z =0$ influences the form of the electrostatic potential at the magnetic presheath entrance if the solvability condition is marginally satisfied. A distribution function which is flat near $v_x = 0$ similarly influences the electrostatic potential at the Debye sheath entrance \cite{Riemann-review}.  
Another pronounced feature of Figure \ref{fig-f0} is that the distribution function becomes narrower with decreasing $\alpha$. %, %which implies that a magnetic presheath with a very small angle $\alpha$ tends to make the ions entering the Debye sheath ``colder''. 
For the cases $\alpha = 0.01$ and $\alpha = 0.02$, the distribution function is thin, approximately symmetric and centred at the sonic speed $v_{\text{B}}$. %, consistent with the cold ion version of the Bohm condition. 
For all angles $\alpha$, the marginal form of the kinetic Bohm condition (\ref{Bohm-kinetic-marginal}) is found to be satisfied, as we predicted in Section \ref{subsec-expansion-near0}, with an error of $\lesssim 2\%$.
A thin distribution function implies that the distribution function must be centred at the sonic speed. 
If the ions entering the Debye sheath have a narrow velocity distribution, this can be approximated by a Dirac delta function, $f_{0x} \left( v_x \right) \simeq \delta_{\text{Dirac}} \left( v_x - u_{x0} \right) $. 
Substituting this approximation into (\ref{Bohm-kinetic-marginal}), we obtain the ``fluid'' marginal Bohm condition $u_{x0} = v_{\text{B}} $.

The broadening of the distribution function $f_{0x} \left( v_x \right) $ at larger values of $\alpha$ is due to typical values of $\Delta v_{x}$, given in equation (\ref{Deltavx-simpler}), becoming larger. The scaling $\Delta v_x \sim \sqrt{2\pi \alpha} v_{\text{t,i}}$ gives $\Delta v_x \sim v_{\text{t,i}} $ for $\alpha \sim 0.1$. Our expansion relies on $\Delta v_x$ being small, so one might question the validity of our results when $\Delta v_x \sim v_{\text{t,i}}$. 
While it is true that the accuracy of our expansion may to some extent be compromised at such large values of $\Delta v_x$, the broadening of the distribution function is expected to be physical. 
We expect our expansion to be accurate up to $\alpha \approx 0.1-0.12 \text{ rad} \approx 6-7^{\circ}$.
%Such broadening is due to the change of $D = U_{\perp} - \chi_{\text{M}} (\bar{x})$ after the last bounce of the ion in its orbit. 

In Figure \ref{fig-f0yz} we show a contour plot of $f_{0yz} \left( v_y, v_z \right) $, which is given by %The distribution function (\ref{f0-def}) integrated in $v_x$
\begin{align} \label{f0yz-def}
f_{0yz}\left( v_y, v_z \right)  =  \int   f_{0}\left( v_x, v_y, v_z \right) dv_x 
 \simeq  F_{\text{cl}} \left( \mu_{\text{gk}} \left( \bar{x},  \chi_M \left( \bar{x} \right) \right), U \right)  \Delta v_x \text{,}
\end{align} 
where (\ref{xbar-def}) and (\ref{U-open}) can be used to re-express $\bar{x}$ and $U$ in terms of $v_y$ and $v_z$ in equation (\ref{f0yz-def}). Comparing with the distribution function at the magnetic presheath entrance (shown in Figure \ref{fig-finfinitycomparison}), we see that the distribution function at $x=0$ is narrower (it occupies a smaller area in the $v_y-v_z$ plane of phase space) and that it has shifted to larger $v_z$ and to very large and positive $v_y$. The net motion of the ions in the $y$ direction can be explained by the fact that they acquire very large $\vec{E} \times \vec{B}$ velocities in the magnetic presheath (see Figure \ref{fig-iontraj}).  
From Figures \ref{fig-f0} and \ref{fig-f0yz}, we infer that ions entering the Debye sheath travel with a typical speed of $\sim 3v_{\text{B}}$, making an angle of $15-30^{\circ}$ with the plane parallel to the wall.
%There are many more ions that reach the Debye sheath travelling at angles above $30^{\circ}$ when $\alpha \gtrsim 0.05$. 
The ion speed and the angle that the ion trajectory makes to the wall are expected to increase in the Debye sheath as the electric field accelerates ions in the $x$-direction.
%In the context of fusion, these ions reach the divertor target at a less shallow angle, hence have a larger sputtering yield and cause more damage \cite{Eckstein-1993}.

 \begin{figure}[h!]
\centering
\includegraphics[width=0.6\textwidth]{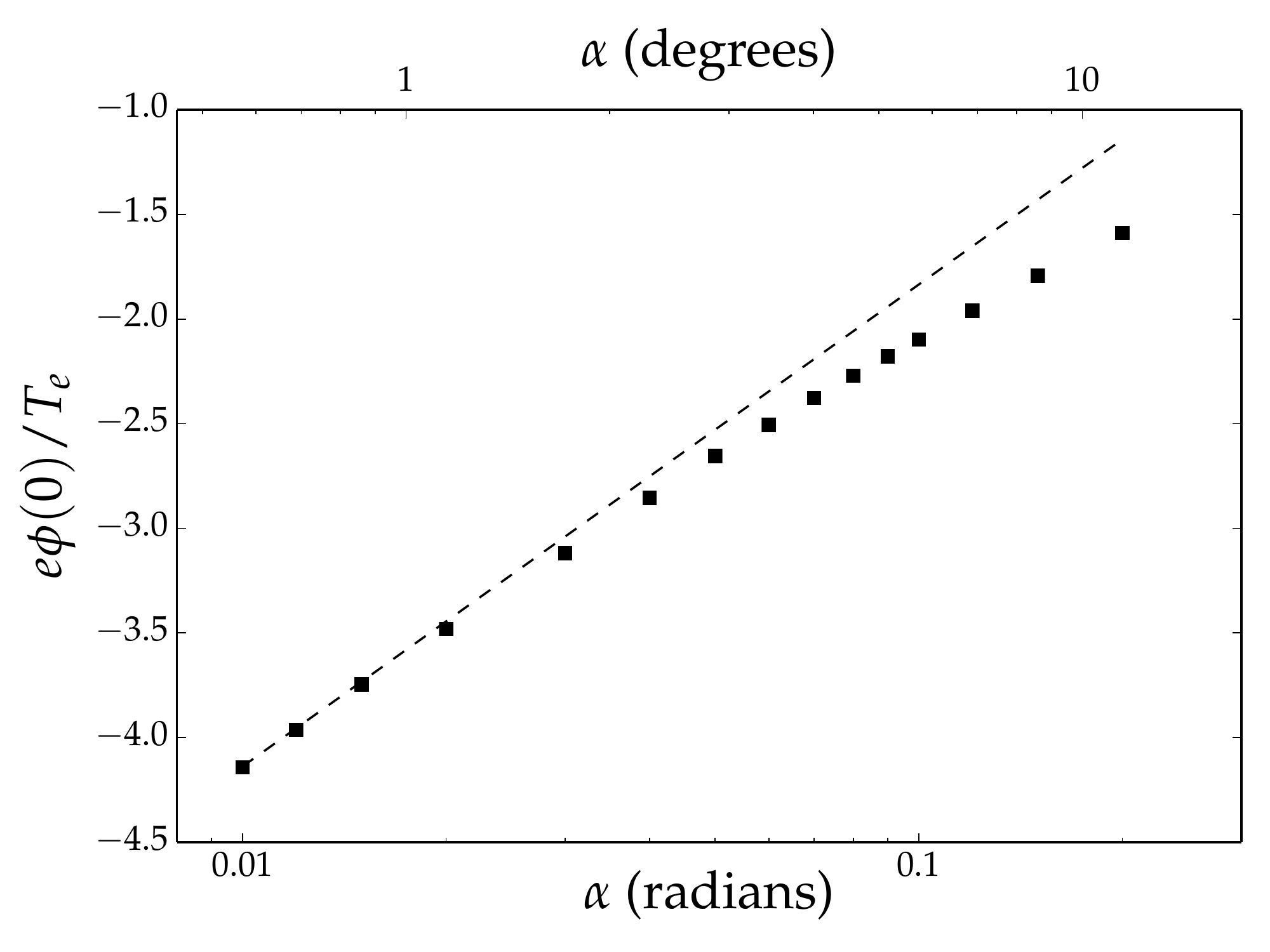}
\caption{The total electrostatic potential drop across the magnetic presheath for a range of angles $\alpha$ is shown with squares.
The dashed line represents the potential drop expected if the ions entering the Debye sheath are cold and the Bohm condition is marginally satisfied, $h(\alpha) = \ln \left(  \alpha u_{z \infty}   / v_{\text{B}} \right) $. For $\alpha \lesssim 0.05$, our results converge to the dashed line.
}
\label{fig-phidrop}
\end{figure}

The electrostatic potential drop across the magnetic presheath is shown in Figure \ref{fig-phidrop}. At small angles, $e\phi ( 0 )/T $ converges to the function
\begin{align} \label{phidrop-Bohm}
h \left( \alpha \right)   = \ln \left(  \frac{  \alpha u_{z \infty}  }{ v_{\text{B}} } \right) \text{,}
\end{align}
which is depicted using a dashed line in Figure \ref{fig-phidrop}. The reason is the following. % This convergence is connected with the property of a thin distribution function at smaller angles. We established that the kinetic version of Bohm's condition is marginally satisfied \cite{Riemann-review}. Our results for the potential drop approach the dashed line when $\alpha \lesssim 0.04$, which indicates that the ion flows into $x=0$ at almost exacly the sonic speed $v_{\text{B}}$ at such angles.
At $x=0$, the flow into the wall % is dominated by open orbits. The ion density is given, via quasineutrality, by the electron density, therefore the flux through $x=0$
is  $n_{\infty} \exp \left( e\phi\left(0\right)/T \right) | u_{x0} | $. Equating this to the flux through $x \rightarrow \infty$, equal to $\alpha n_{\infty} u_{z\infty} $, and rearranging, we obtain an expression for the potential drop in terms of the ion flow into and out of the magnetic presheath, 
\begin{align} \label{phidrop}
\frac{e\phi\left(0\right)}{T}   = \ln \left(  \frac{  \alpha u_{z \infty}  }{ | u_{x0} |}   \right) \text{.}
\end{align}
%Equation (\ref{phidrop}) with the calculated values of $u_{x,0}$ is consistent with the values of the electrostatic potential drop shown in Figure \ref{fig-phidrop}, as expected. 
Moreover, we previously found % by looking at the distribution function in Figure \ref{fig-f0} 
that for $\alpha \lesssim 0.05$ the cold ion Bohm condition is almost marginally satisfied, $|u_{x0}| \simeq v_{\text{B}}$, due to the thinness of the distribution function (see Figure \ref{fig-f0}). 
Then, the potential drop across the magnetic presheath can be predicted using equation (\ref{phidrop}) with $u_{x0} = v_{\text{B}}$, which is equation (\ref{phidrop-Bohm}), and therefore the potential drop converges to the dashed line in Figure \ref{fig-phidrop}.

\section{Conclusion}

We solved a collisionless and quasineutral magnetic presheath of characteristic thickness $\rho_{\text{i}}$ by expanding the ion trajectories for small $\alpha$. 
The contribution to the ion density due to ions in \emph{open} orbits was shown to be crucial and calculated. 
The quasineutrality equation (\ref{quasineutrality-compact}), with the closed and open orbit pieces of the ion density given by equations (\ref{ni-closed}) and (\ref{ni-open-Deltavx}), was solved numerically for the boundary condition (\ref{f-infty}) for a number of angles $\alpha$.
The method of solution is valid for any distribution function at the magnetic presheath entrance. 
We have also derived the solvability condition (\ref{solvability-vz}) by expanding the quasineutrality equation at the magnetic presheath entrance. 
This condition is the generalization of  Chodura's condition, first derived in reference \cite{Chodura-1982}, to include the effect of kinetic ions at small $\alpha$. 

Our numerical results for electrostatic potential, ion density and flow are qualitatively consistent with the picture of the magnetic presheath that emerges using fluid equations \cite{Chodura-1982}. We find a decrease in density as the ions approach the wall (Figure \ref{fig-ni}), and a corresponding increase in the ion fluid velocity towards the wall (Figure \ref{fig-flux}). %This is consistent with the usual picture of a magnetic presheath whose function is to ``turn'' the ion velocities from parallel to the magnetic field to normal to the wall  
The fluid velocity $u_x$ is equal to or exceeds the Bohm limit $ v_{\text{B}}$ at the entrance of the Debye sheath ($x=0$), as expected. 
In addition, our kinetic treatment explains several features of the potential and flow profiles.
For example, we numerically observe a scaling $\phi \left( x \right) - \phi \left( 0 \right) \propto \sqrt{x}$ near $x=0$ (see Figures \ref{fig-phires} and \ref{fig-phiprofile}) and find that the distribution of ion velocities at $x=0$ marginally satisfies the kinetic Bohm condition.
We demonstrate that these two features of the numerical results are necessary for a self-consistent solution of the system (Section \ref{subsec-expansion-near0}).
Moreover, we observe the distribution $f_{0x} (v_x)$ of the component of the velocity normal to the wall (Figure \ref{fig-f0}) to be substantially narrower at smaller values of $\alpha$. As a consequence, for small $\alpha$ the ``fluid'' velocity tends to the Bohm limit at the Debye sheath entrance (as observed in Figure \ref{fig-flux}), which can be used to predict the potential drop across the magnetic presheath using equation (\ref{phidrop-Bohm}). This is confirmed by the potential drop converging to the dashed line, given by equation (\ref{phidrop-Bohm}), for $\alpha \lesssim 0.05 \simeq 3^{\circ}$ in Figure~\ref{fig-phidrop}. 
 
By providing the equations and a numerical procedure to obtain the velocity distribution of ions entering the Debye sheath after travelling through the magnetic presheath, this work is a step towards advancing our knowledge of how energy is deposited by ions onto divertor targets in the fusion-relevant regime $\alpha \ll 1$.
Moreover, the numerical scheme provided here is computationally cheap: using a laptop, it takes less than one minute for the iteration procedure to converge.
The Debye sheath equations \cite{Riemann-review} can be solved using our magnetic presheath results to obtain the velocity distribution of ions reaching the target.  
Knowledge of how damage to the target material depends on the projectile velocity and angle of incidence \cite{Eckstein-1993} could, together with the tools provided here, help to quantitatively predict the damage made by ions to divertor targets of a fusion device. 
An important general conclusion that we can make is that there are substantially fewer ions reaching the Debye sheath with a large component of the velocity normal to the wall when $\alpha$ is small. 
Our work can also be used to predict the sputtering caused by impurities, which typically has a lower kinetic energy threshold \cite{Mellet-2016}. 
In the limit of small (trace) impurity density, the electrostatic potential obtained in this work can be used to obtain the impurity distribution function at the Debye sheath entrance from the impurity distribution function at the magnetic presheath entrance.
%If the impurity cannot be treated as a trace in the plasma, the impurity density can be included on the right hand side of the quasineutrality equation (\ref{quasineutrality-compact}).
%Hence, decreasing $\alpha$ is expected to reduce the average sputtering yield of ions reaching the divertor target, therefore reducing damage made by ions to the divertor material. 

\ack{
This work has benefited from the input of several people. In particular, the authors would like to thank Dmitri Ryutov, Greg Hammett, John Omotani and Ian Abel for some very helpful discussions and useful suggestions. 
AG would also like to thank Adwiteey Mauriya, Nicolas Christen and Michael Hardman for help in finding mistakes in the numerical work.
This work has received funding from the RCUK Energy Programme [grant number EP/P012450/1].
%This work has been carried out within the framework of the EUROfusion Consortium and has received funding from the Euratom research and training programme 2014-2018 under grant agreement No 633053 and from the RCUK Energy Programme [grant number EP/I501045]. 
%The views and opinions expressed herein do not necessarily reflect those of the European Commission.
}

\appendix

\section{ Drift-kinetic expansion of the ion density near $x\rightarrow \infty$ }  \label{app-quasi-expansion}

Here we derive equation (\ref{niclosedinfty}) in the following steps. 
First, in \ref{subapp-mu-expansion} we expand the adiabatic invariant (\ref{mu-Uperp-xbar}) as a function of $\bar{x}$ and $U_{\perp}$ for small electrostatic potential, $e\phi (x) / T_{\text{e}} \ll 1$, and small gradients of the electrostatic potential, $\epsilon = \rho_{\text{i}} \phi'(x) / \phi(x) \ll 1 $.
Then, in \ref{subapp-varphi-expansion} we expand equation (\ref{varphi-def}) to obtain an expression for $\bar{x}$ as a function of $\varphi$, $x$ and $\mu$. 
We also obtain an expression for $U_{\perp}$ as a function of $\varphi$, $x$ and $\mu$. 
Then, by making the change of variables $(x, \bar{x}, U_{\perp}, U)  \rightarrow (x, \varphi, \mu, U) $, we obtain an expression for the ion density in \ref{subapp-ni-expansion}.
Finally, this is carefully expanded in \ref{subapp-ni-finalexpansion}. 
The results of this appendix are valid to lowest order in $\alpha$. % because the adiabatic invariant, as defined in equation (\ref{mu-Uperp-xbar}), is used. In practice, the expansions near $x \rightarrow \infty$ can be obtained for exact $\alpha$ if a co-oordinate system in which $z$ is aligned with the magnetic field is used (but this is not a convenient co-ordinate system in the bulk of the magnetic presheath).

\subsection{Adiabatic invariant expansion} \label{subapp-mu-expansion}

We proceed to derive an expression for $\mu$ as a function of $\bar{x}$ and $U_{\perp}$ by expanding equation (\ref{mu-Uperp-xbar}) near $x\rightarrow \infty$, where $e\phi (x) / T_{\text{e}} \ll 1$. In addition, we assume that the length scale of changes in the electrostatic potential is much larger than the ion gyroradius $\rho_{\text{i}}$, defining the small parameter $\epsilon$ of equation (\ref{epsilonsmall}). We first expand the expression inside the square root of equation (\ref{vx-Uperp-xbar-x}) around $x=\bar{x}$ to second order in $\epsilon$, obtaining
\begin{align}
v_x = \sigma_x V_x\left(x, \bar{x}, U_{\perp} \right) =  \sigma_x \sqrt{2} \left[  U_{\perp} - \frac{1}{2}\Omega^2 \left(x-\bar{x} \right)^2 - \frac{\Omega  \phi\left(\bar{x}\right)}{B} \right.  \nonumber \\  \left. -  \frac{\Omega  \phi'\left(\bar{x}\right)}{B} \left( x - \bar{x} \right) -   \frac{\Omega  \phi''\left(\bar{x}\right)}{2B} \left( x - \bar{x} \right)^2  + O \left( \epsilon^3 \hat{\phi} v_{\text{t,i}}^2  \right) \right]^{1/2} \text{.}
\end{align}
Note that the electric field is locally approximated as linearly varying \cite{Geraldini-2017}.
Completing the square in the square root and dropping small terms gives
\begin{align} \label{Vx-infty}
V_x\left(x, \bar{x}, U_{\perp} \right) = & A \Omega \sqrt{1 + \frac{\phi''\left( \bar{x}\right)}{\Omega B}} \nonumber
\\ & \times \sqrt{ 1 - \frac{1}{A^2} \left[ x -\bar{x} + \frac{ \phi'\left( \bar{x} \right)}{ \Omega B}  \right]^2 + O \left(  \hat{\phi} \epsilon^3, \hat{\phi}^2 \epsilon^2 \right)  }  \text{,}
\end{align}
where we have defined the orbit amplitude
\begin{align} \label{A-def}
A = \frac{1}{\Omega}  \left( 1 + \frac{\phi''(\bar{x})}{\Omega B} \right)^{-1/2} \sqrt{ 2U_{\perp} - \frac{ 2\Omega \phi\left( \bar{x} \right) }{B}    }  \text{.}
\end{align}
The bounce points of the closed orbit are obtained by solving $V_x \left( x, \bar{x}, U_{\perp}  \right) = 0$, leading to
\begin{align} \label{xb-infty}
x_{\text{b}} = \bar{x} - \frac{\phi'\left( \bar{x} \right)}{ \Omega B}  - A \text{,}
\end{align}
\begin{align} \label{xt-infty}
x_{\text{t}} = \bar{x} - \frac{\phi'\left( \bar{x} \right)}{ \Omega B}  + A \text{.}
\end{align}
By substituting (\ref{Vx-infty}) into equation (\ref{Omegabar-def}) and using (\ref{xb-infty}) and (\ref{xt-infty}) for the integration limits, we have
\begin{align} \label{varphi-explicit}
\frac{\pi}{\overline{\Omega}} = & \int_{x_{\text{b}}}^{x_{\text{t}}} \left( A \Omega  \sqrt{1 + \frac{\phi''\left( \bar{x}\right)}{\Omega B}}   \sqrt{ 1 - \frac{1}{A^2} \left[ x -\bar{x} + \frac{\phi'\left( \bar{x} \right)}{ \Omega B}  \right]^2  } \right)^{-1} dx  \nonumber \\ & + O \left( \frac{\hat{\phi}  \epsilon^3}{\Omega} , \frac{\hat{\phi}^2  \epsilon^2}{\Omega} \right)  \text{,}
\end{align}
which leads to the modified gyrofrequency
\begin{align} \label{Omegabar-infty}
\overline{\Omega} =  \Omega \sqrt{1 + \frac{\phi''\left( \bar{x}\right)}{\Omega B}} + O \left( \hat{\phi}  \epsilon^3 \Omega ,  \hat{\phi}^2  \epsilon^2 \Omega  \right)   =  \Omega \left( 1 + \frac{\phi''\left( \bar{x}\right)}{2\Omega B} +  O \left( \hat{\phi}  \epsilon^3 ,  \hat{\phi}^2  \epsilon^2  \right) \right) \text{.}
\end{align}
We exploit (\ref{Omegabar-infty}) to simplify equation (\ref{Vx-infty}),
\begin{align} \label{Vx-infty-simpler}
V_x\left(x, \bar{x}, U_{\perp} \right) =   \overline{\Omega} A \sqrt{ 1 - \frac{1}{A^2} \left[ x -\bar{x} + \frac{\phi'\left( \bar{x} \right)}{\Omega B}\right]^2 + O \left( \hat{\phi} \epsilon^3 ,  \hat{\phi}^2 \epsilon^2   \right)  }  \text{.}
\end{align}
By inserting (\ref{Vx-infty-simpler}) into expression (\ref{mu-Uperp-xbar}) for the adiabatic invariant we have
\begin{align}
\mu = \frac{1}{\pi} \int_{x_{\text{b}}}^{x_{\text{t}}} \overline{\Omega} A \sqrt{ 1 - \frac{1}{A^2} \left[ x -\bar{x} + \frac{ \phi'\left( \bar{x} \right)}{ \Omega B}  \right]^2  } dx + O \left( \hat{\phi }  \epsilon^3 \frac{v_{\text{t,i}}^2}{\Omega} , \hat{\phi }^2  \epsilon^2 \frac{v_{\text{t,i}}^2}{\Omega}  \right) \text{,}
\end{align}
which evaluates to
\begin{align} \label{mu-Omega}
\mu & = \frac{1}{2}\overline{\Omega} A^2 +  O \left(  \hat{ \phi } \epsilon^3 \frac{v_{\text{t,i}}^2}{\Omega},  \hat{\phi}^2 \epsilon^2  \frac{v_{\text{t,i}}^2}{\Omega} \right)  \nonumber \\
& =  \frac{1}{\overline{\Omega}} \left( U_{\perp} - \frac{ \Omega \phi\left( \bar{x} \right) }{B}     \right)  +  O \left(  \hat{ \phi } \epsilon^3 \frac{v_{\text{t,i}}^2}{\Omega},  \hat{\phi}^2 \epsilon^2  \frac{v_{\text{t,i}}^2}{\Omega} \right) \text{.}
\end{align}
Rearranging equation (\ref{mu-Omega}) and using (\ref{Omegabar-infty}), we obtain
\begin{align} \label{mu-expanded1}
U_{\perp}  & = \overline{ \Omega } \mu  +  \frac{\Omega \phi\left( \bar{x} \right) }{ B }  + O \left(  \hat{ \phi } \epsilon^3 v_{\text{t,i}}^2,  \hat{\phi}^2 \epsilon^2  v_{\text{t,i}}^2 \right) \nonumber \\ 
& = \Omega \mu  +   \frac{\Omega \phi\left( \bar{x} \right) }{B } +  \frac{\mu \phi'' \left( \bar{x} \right) }{  2B } + O \left(  \hat{ \phi } \epsilon^3 v_{\text{t,i}}^2,  \hat{\phi}^2 \epsilon^2  v_{\text{t,i}}^2 \right)  \text{.}
\end{align}
At $x\rightarrow \infty$, the zeroth order in $\hat{\phi}$ of all the equations in this Appendix is valid exactly. 
Then, we have $\overline{\Omega} = \Omega$ from equation (\ref{Omegabar-infty}) and $\mu = U_{\perp}/\Omega$ from equation (\ref{mu-Omega}). Hence, the equations $2\Omega \mu = v_x^2 + v_y^2$ and $2\left( U - \Omega \mu \right) = v_z^2$ are valid at $x \rightarrow \infty$.
These equations are used to obtain equation (\ref{changevar-infty}) and to obtain $F_{\text{cl}}(\mu, U)$ from $f_{\infty}(\vec{v})$ via equation (\ref{F-closed}).

%From equation (\ref{mu-expanded1}) we obtain the important result
%\begin{align} \label{dUperpdmu}
%\left( \frac{ \partial U_{\perp} }{\partial \mu } \right)_{\bar{x}} = \overline{\Omega} \text{,}
%\end{align}
%which is valid for arbitrary $\hat{\phi}$ and $\epsilon$.

\subsection{Gyrophase expansion} \label{subapp-varphi-expansion}

We require an expression for $U_{\perp}$ as a function of $\mu$, $\varphi $ and $x$. 
To obtain it from equation (\ref{mu-expanded1}), we need an equation for $\bar{x}$ as a function of $\mu$, $\varphi$ and $x$, which we proceed to derive.
First, we insert equation (\ref{Vx-infty-simpler}) into the definition of the gyrophase $\varphi$, equation (\ref{varphi-def}), and use the top bounce point in (\ref{xt-infty}) as the lower integration limit to obtain
\begin{align} \label{varphi-app}
\varphi = \sigma_x \int^x_{x_{\text{t}}} \left( A  \sqrt{ 1 - \frac{1}{A^2} \left[ s -\bar{x} + \frac{ \phi'\left( \bar{x} \right)}{ \Omega  B}  \right]^2  } \right)^{-1} ds + O \left(  \hat{\phi} \epsilon^3, \hat{\phi}^2 \epsilon^2  \right)   \text{.}
\end{align}
Note that $\varphi >0$ when $\sigma_x = -1$. 
Using equation (\ref{varphi-app}) and $A=\sqrt{2\mu/\overline{\Omega}}$ (from equation (\ref{mu-Omega})) we obtain the relation
\begin{align} \label{x-xbar-varphi}
x -\bar{x} + \frac{\phi'\left( \bar{x} \right)}{ \Omega B} = \sqrt{\frac{2\mu}{\overline{\Omega}}} \cos \varphi + O \left(  \hat{\phi}  \epsilon^3 \rho_{\text{i}},  \hat{\phi}^2 \epsilon^2 \rho_{\text{i}}  \right) \text{.}
\end{align}
%Now we obtain $\bar{x}$ as a function of gyrophase by expanding 
%\begin{align}
%\bar{x} & = x - \left( 1 +  \frac{\Omega \phi'' (x)}{B}  \right) A \cos \varphi + \frac{\Omega \phi' (x)}{B} + O \left(  \hat{\phi} \epsilon^3,  \hat{\phi}^2 \epsilon^2 \rho_{\text{i}} \right) \nonumber \\
%& = x - \frac{\overline{\Omega}^2}{\Omega^2} A \cos \varphi + \frac{\Omega \phi' (x)}{B} + O \left(  \hat{\phi} \epsilon^3,  \hat{\phi}^2 \epsilon^2 \rho_{\text{i}} \right)
% \text{.}
%\end{align}
%By substituting (\ref{x-xbar-varphi}) into (\ref{Vx-infty-simpler}) and using (\ref{vx-Uperp-xbar-x}), $v_x$ is given by
%\begin{align} \label{vx-varphi}
%v_x = - \overline{\Omega} A \sin \varphi +  O \left(  \hat{\phi} \epsilon^3 v_{\text{t,i}},  \hat{\phi}^2 \epsilon^2 v_{\text{t,i}}  \right)  \text{.}
%\end{align}
Then, we expand equation (\ref{x-xbar-varphi}) around the lowest order $\bar{x} = x - \sqrt{2\mu/\overline{\Omega}} \cos \varphi $ to obtain
\begin{align} \label{xbar-expanded}
\bar{x} & = x - \left( 1 +  \frac{3 \Omega \phi'' (x)}{4B}  \right) \sqrt{ \frac{2\mu}{\Omega} } \cos \varphi + \frac{ \phi' (x)}{\Omega B} + O \left(  \hat{\phi} \epsilon^3 \rho_{\text{i}},  \hat{\phi}^2 \epsilon^2 \rho_{\text{i}} \right)  \text{.}
%& = x - \frac{\overline{\Omega}^2}{\Omega^2} \sqrt{\frac{2\mu}{\overline{ \Omega}} } \cos \varphi + \frac{ \phi' (x)}{\Omega B} + O \left(  \hat{\phi} \epsilon^3 \rho_{\text{i}},  \hat{\phi}^2 \epsilon^2 \rho_{\text{i}} \right)
% \text{.}
\end{align}
Similarly, we expand equation (\ref{mu-expanded1}) around $\bar{x} = x - \sqrt{2\mu/\Omega} \cos \varphi $ to obtain
\begin{align} \label{Uperp-expanded}
U_{\perp} = \Omega \mu +\frac{\Omega \phi\left( x \right) }{B} - \frac{\Omega \phi'\left( x \right) }{B} \sqrt{\frac{2\mu}{\Omega} } \cos \varphi + \frac{\mu \phi'' \left( x \right) }{2 B} \left( 1 + 2 \cos^2 \varphi \right) \nonumber  \\
+ O \left( \hat{\phi} \epsilon^3 v_{\text{t,i}}^2, \hat{\phi}^2 \epsilon^2 v_{\text{t,i}}^2 \right) \text{.}
\end{align}
Defining $ \delta U_{\perp} = \Omega \mu - U_{\perp} $, equation (\ref{Uperp-expanded}) leads to equation (\ref{deltaUperp}).
%Differentiating (\ref{xbar-expanded}) with respect to $\varphi$ and substituting equation (\ref{vx-varphi}) we get
%\begin{align} \label{dxbardvarphi}
%\left( \frac{\partial \bar{x}}{\partial \varphi} \right)_{x, \mu, U }  =  - \frac{\overline{\Omega}}{\Omega^2} v_x  + O \left(  \hat{\phi} \epsilon^3 \rho_{\text{i}},  \hat{\phi}^2 \epsilon^2 \rho_{\text{i}} \right) \text{.}
%\end{align}

\subsection{Change of variables in the ion density integral}  \label{subapp-ni-expansion}

\begin{figure} 
\includegraphics[width=0.9\textwidth]{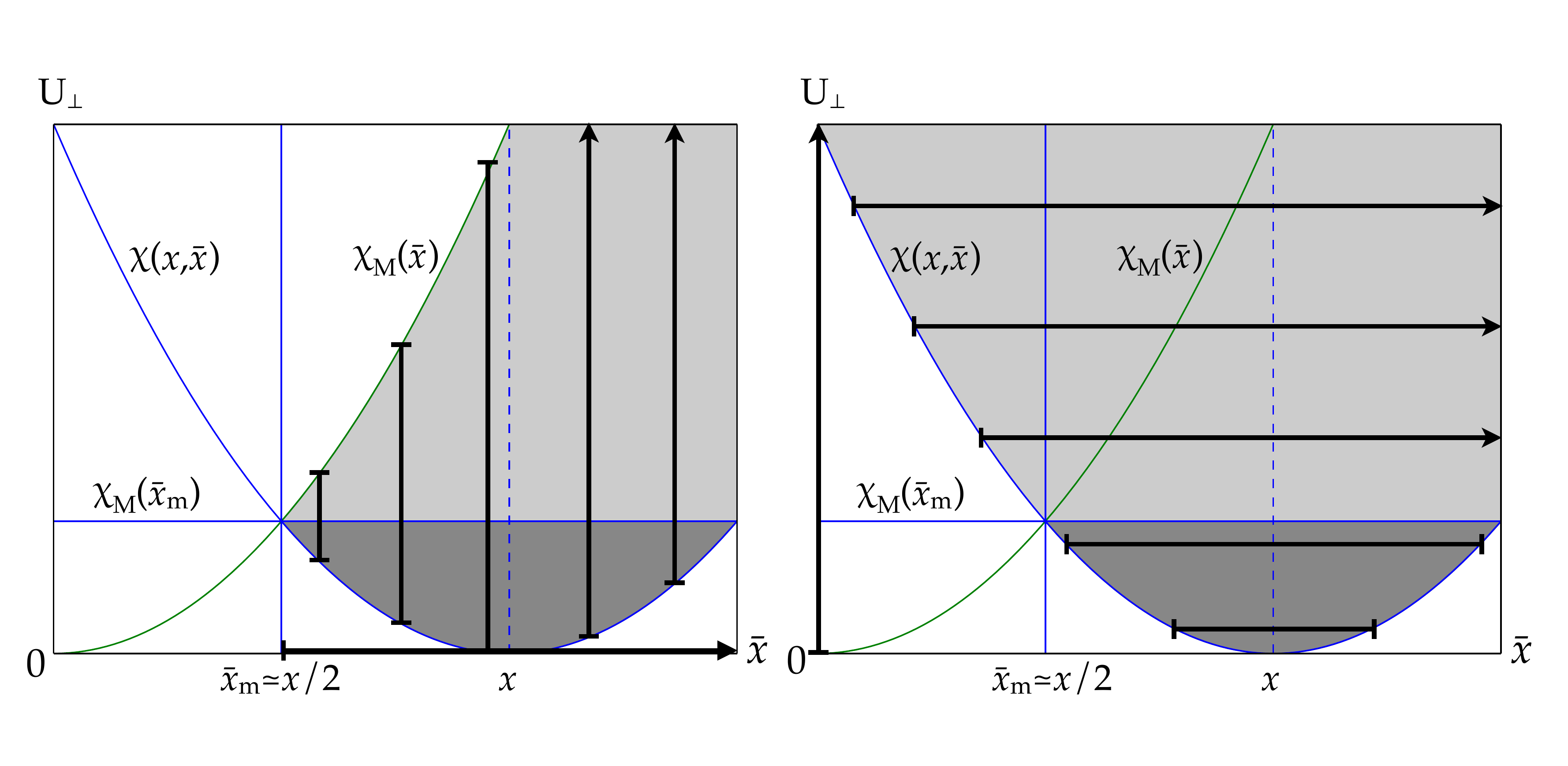}
\caption{The integration domain in $(\bar{x}, U_{\perp})$ of equation (\ref{ni-closed-change}) consists of both shaded regions on the left hand side drawing. When we exchange the integration order, the integration limits (bold lines) are picked such that the integration domain coincides in the dark grey region but not in the light grey one. The light grey region satisfies $U_{\perp} > \chi_{\text{M}} (x/2) = \Omega^2 x^2 / 8 \gg v_{\text{t,i}}^2 $ near $x\rightarrow\infty$, and at such large energies we expect the distribution function to be exponentially small. Thus, the contribution to the integral from this region of phase space is negligible and the limits of integration of equation (\ref{ni-closed-orderexchange}) are appropriate. }
\label{fig-integrationchange}
\end{figure}

For sufficiently large $x$, the open orbit density is zero and the closed orbit density is given by equation (\ref{ni-closed}).
The value of $\bar{x}_{\text{m}}(x)$ is given by equation (\ref{xbarm-general}) evaluated near $x \rightarrow \infty$,
\begin{align}
\bar{x}_{\text{m}} (x) \simeq \frac{1}{2} x \text{.}
\end{align}
The effective potential maximum at large $\bar{x}$ is, from equation (\ref{xbar-infty-limit}), $\chi_{\text{M}} (\bar{x}) \simeq  \Omega^2 \bar{x}^2/2$.
Hence, equation (\ref{ni-closed}) becomes
\begin{align} \label{ni-closed-change}
n_{\text{i,cl}} (x) \simeq \int_{x/2}^{\infty} \Omega d\bar{x} \int_{\chi\left( x, \bar{x} \right)}^{ \Omega^2 \bar{x}^2/2 } \frac{2dU_{\perp}}{\sqrt{2\left(U_{\perp} - \chi\left(x, \bar{x} \right) \right)}} \int_{U_{\perp}}^{\infty} \frac{F_{\text{cl}} \left( \mu_{\text{gk}} \left( \bar{x}, U_{\perp} \right), U \right) }{\sqrt{2\left( U - U_{\perp} \right)}} dU  \text{.}
\end{align}
For large values of $x$, we can exchange the integrals over $\bar{x}$ and $U_{\perp}$ in (\ref{ni-closed-change}) to get %and use $|v_x| = \sqrt{ 2\left(U_{\perp} - \chi (x, \bar{x}) \right) }$ 
\begin{align} \label{ni-closed-orderexchange}
n_{\text{i,cl}} (x) \simeq \int_0^{\infty} dU_{\perp} \int_{\bar{x}_{\text{b}}}^{\bar{x}_{\text{t}}} \frac{2\Omega d\bar{x}}{\sqrt{2\left( U_{\perp} - \chi (x, \bar{x}) \right) } } \int_{U_{\perp}}^{\infty} \frac{F_{\text{cl}} \left( \mu_{\text{gk}} \left( \bar{x}, U_{\perp} \right), U \right) }{\sqrt{2\left( U - U_{\perp} \right)}} dU  \text{,}
\end{align}
where 
\begin{align}
\bar{x}_{\text{b}} = x - \frac{1}{\Omega} \sqrt{ 2\left( U_{\perp} - \frac{\Omega \phi(x)}{B} \right) } \text{,}
\end{align}
\begin{align}
\bar{x}_{\text{t}} = x + \frac{1}{\Omega} \sqrt{ 2\left( U_{\perp} - \frac{\Omega \phi(x)}{B} \right) } \text{.}
\end{align}
The change in the integration limits is explained in Figure \ref{fig-integrationchange}. 
%Then, from equations (\ref{vx-Uperp-xbar-x}), (\ref{dUperpdmu}) and (\ref{dxbardvarphi}) we obtain
%\begin{align} \label{jacobian-infty}
%\frac{1}{\sqrt{2\left( U_{\perp} - \chi (x, \bar{x}) \right) } } \left|  \left( \frac{\partial U_{\perp}}{\partial \mu} \right)_{\bar{x}} \left( \frac{\partial \bar{x}}{\partial \varphi} \right)_{x, \mu }  \right|  & = \frac{  \overline{\Omega}^2  }{ \Omega^2  } + O\left( \hat{\phi} \epsilon^3 n_{\infty},  \hat{\phi}^2 \epsilon^2 n_{\infty} \right) \nonumber \\
%& = 1 + \frac{\phi''(x)}{\Omega B} + O\left( \hat{\phi} \epsilon^3 ,  \hat{\phi}^2 \epsilon^2  \right)
% \text{,}
%\end{align}
%Equation (\ref{jacobian-infty}) can be used to make the change of variables $(\bar{x}, U_{\perp} ) \rightarrow (\varphi, \mu ) $ in equation (\ref{ni-closed-orderexchange}),
Equations (\ref{xbar-expanded}) and (\ref{Uperp-expanded}) can be used to make the change of variables $(x, \bar{x}, U_{\perp}, U ) \rightarrow (x, \varphi, \mu, U ) $ in equation (\ref{ni-closed-orderexchange}).
Using equations (\ref{xbar-expanded}) and (\ref{Uperp-expanded}), the Jacobian
\begin{align}
\left| \frac{ \partial \left( \bar{x}, U_{\perp} \right) }{\partial \left( \varphi, \mu \right) } \right| = \left( 1 + \frac{5\phi''(x)}{4B\Omega} \right) \sqrt{2\Omega \mu} \left| \sin \varphi \right| + O \left(  \hat{\phi} \epsilon^3 v_{\text{t,i}}, \hat{\phi}^2 \epsilon^2 v_{\text{t,i}}   \right) \text{}
\end{align}
and the result
\begin{align}
\frac{1}{\sqrt{2\left( U_{\perp} - \chi (x, \bar{x}) \right)}} = \left( 1 - \frac{\phi''(x)}{4B\Omega } \right) \frac{1}{\sqrt{2\Omega\mu} \left| \sin \varphi \right|} + O\left( \frac{\hat{\phi}\epsilon^3}{v_{\text{t,i}}}, \frac{\hat{\phi}^2\epsilon^2}{v_{\text{t,i}}} \right) \text{,}
\end{align}
we obtain
\begin{align} \label{niclosedinftynotexp}
n_{\text{i,cl}} \left(x \right) = & \left( 1 +  \frac{\phi''(x)}{\Omega B}  \right) \int_{-\pi}^{\pi} d\varphi \int_0^{\infty} \Omega d\mu \int_{\Omega \mu}^{\infty} \frac{ F_{\text{cl}}(\mu, U)  }{\sqrt{2\left( U - \Omega  \mu + \delta U_{\perp} \right)}}   dU \nonumber \\
& + O\left( \hat{\phi} \epsilon^3 n_{\infty},  \hat{\phi}^2 \epsilon^2 n_{\infty} \right) \text{,}
\end{align}
where $\delta U_{\perp}$ is defined in equation (\ref{deltaUperp}). 
 Note that we changed the lower limit of the integral over $U$ from $U_{\perp}$ to $\Omega \mu$ in going from equation (\ref{ni-closed-orderexchange}) to (\ref{niclosedinftynotexp}). The distribution function is zero for $U < \Omega \mu$. Therefore, the integrand is zero in the region $U_{\perp} < U < \Omega \mu $ and $U_{\perp}$ can be replaced by $\Omega \mu$ in the integration limit of the integral in $U$. 

\subsection{Expansion of the integral over $U$ in equation (\ref{niclosedinftynotexp})} \label{subapp-ni-finalexpansion}

We begin by changing variables from $U$ to $U_{\star} = U - \Omega \mu  + \delta U_{\perp} $,
\begin{align}
\int_{\Omega \mu}^{\infty} \frac{  F_{\text{cl}}(\mu, U) dU}{\sqrt{2\left(U - \Omega\mu + \delta U_{\perp} \right) }}   =  \int_{\delta U_{\perp}}^{\infty} \frac{ F_{\text{cl}}(\mu, U_{\star} +  \Omega \mu - \delta U_{\perp} ) }{\sqrt{2U_{\star}  }}  dU_{\star} \text{.}
\end{align}
Note that $\delta U_{\perp} >0$ for typical values of $\mu$. We Taylor expand the distribution function
\begin{align} \label{Taylor-dist}
& \int_{\delta U_{\perp}}^{\infty} \frac{ F_{\text{cl}}(\mu, U_{\star} + \Omega \mu - \delta U_{\perp} ) }{\sqrt{2U_{\star}  }}  dU_{\star}   =  \int_{\delta U_{\perp}}^{\infty}  \frac{F_{\text{cl}}(\mu, U_{\star} + \Omega \mu )}{\sqrt{2U_{\star}}} dU^{\star} \nonumber \\  
& -  \int_{\delta U_{\perp}}^{\infty} \frac{\delta U_{\perp}}{\sqrt{2U_{\star}}}  \frac{\partial  F_{\text{cl}} }{\partial U} \left( \mu, U_{\star} + \Omega \mu \right)  dU_{\star}  + \frac{1}{2}  \int_{\delta U_{\perp}}^{\infty}  \frac{\delta U_{\perp}^2}{\sqrt{2U_{\star}}}   \frac{ \partial^2 F_{\text{cl}}}{\partial U^2}(\mu, U_{\star} + \Omega \mu )  dU_{\star} + \ldots \text{.}
\end{align}
Each of the terms of equation (\ref{Taylor-dist}) can then be split into two separate integrals over $U_{\star}$
\begin{align}
n_{\text{i}} (x) = & \int_{0}^{\infty}  \frac{dU_{\star}}{\sqrt{2U_{\star}}}   \left( F_{\text{cl}}(\mu, U_{\star} + \Omega \mu ) - \delta U_{\perp} \frac{\partial F_{\text{cl}}}{\partial U} (\mu, U_{\star} + \Omega \mu ) \right.  \nonumber \\
& \left.  + \frac{1}{2} \delta U_{\perp}^2 \frac{ \partial^2 F_{\text{cl}}}{\partial U^2} (\mu, U_{\star} + \Omega \mu )  \right) 
 -   \int_{0}^{\delta U_{\perp}}  \frac{dU_{\star}}{\sqrt{2U_{\star}}}   \left(  F_{\text{cl}}(\mu, U_{\star} + \Omega \mu )  \phantom{ \frac{\partial F}{\partial \mu} } \right.  \nonumber \\
& \left.  - \delta U_{\perp} \frac{\partial F_{\text{cl}}}{\partial U} (\mu, U_{\star} + \Omega \mu )    \right) + \ldots \text{.}
\end{align}
Then, for small $\delta U_{\perp}$, we Taylor expand the distribution function near $U_{\star}=0$ in the integrals between $0$ and $\delta U_{\perp}$ (and we neglect terms of order $\delta U_{\perp}^{5/2}$)
\begin{align}
n_{\text{i}} (x) = & \int_{0}^{\infty}  \frac{dU_{\star}}{\sqrt{2U_{\star}}}   \left( F_{\text{cl}}(\mu, U_{\star} + \Omega \mu ) - \frac{\partial F_{\text{cl}}}{\partial U} (\mu, U_{\star} + \Omega \mu )  \delta U_{\perp}  \right. \nonumber \\
& \left. + \frac{1}{2} \frac{\partial^2 F_{\text{cl}}}{\partial U^2} (\mu, U_{\star} + \Omega \mu ) \delta U_{\perp}^2 \right)   -   \int_{0}^{\delta U_{\perp}}  \frac{dU_{\star}}{\sqrt{2U_{\star}}}   \left( F_{\text{cl}}(\mu, \Omega \mu ) \phantom{ \frac{\partial F}{\partial \mu} }  \right. \nonumber \\
& \left. + \left( U_{\star} - \delta U_{\perp}  \right) \frac{\partial F_{\text{cl}}}{\partial U} (\mu, \Omega \mu ) \right)  + \ldots
\end{align}
Carrying out the integrals between $0$ and $\delta U_{\perp}$, we obtain
\begin{align} \label{int-expanded}
n_{\text{i}} (x) = & \int_{0}^{\infty}  \frac{dU_{\star}}{\sqrt{2U_{\star}}}   \left( F_{\text{cl}}(\mu, U_{\star} + \Omega \mu ) - \frac{\partial F_{\text{cl}}}{\partial U} (\mu, U_{\star} + \Omega \mu )  \delta U_{\perp}  \right. \nonumber  \\
& \left. + \frac{1}{2} \frac{\partial^2 F_{\text{cl}}}{\partial U^2} (\mu, U_{\star} + \Omega \mu ) \delta U_{\perp}^2 \right)   -   \sqrt{2\delta U_{\perp}}   F_{\text{cl}}(\mu, \Omega \mu ) \nonumber \\
&  +\frac{1}{3} \left( 2\delta U_{\perp}  \right)^{3/2} \frac{\partial F_{\text{cl}}}{\partial U} (\mu, \Omega \mu ) + \ldots \text{.}
\end{align}
Then, inserting (\ref{int-expanded}) into equation (\ref{niclosedinftynotexp}) and changing the dummy integration variable to $U = U_{\star} + \Omega \mu $, we are left with the result of equation (\ref{niclosedinfty}).

\section{Cold ion limit and fluid Chodura condition} \label{app-coldion}

In the cold ion limit, $T_{\text{i}} = 0$, the ions in the magnetic presheath are mono-energetic and should thus be well-described by the fluid equations used by Chodura in reference \cite{Chodura-1982}. 
%our quasineutrality equation (\ref{quasineutrality-compact}) must have the same solution as the set of fluid equations used by Chodura in reference \cite{Chodura-1982}. 
We show here that equations (\ref{solvability-vz}) and (\ref{phisol2}) are consistent with two of the main results found in Chodura's paper \cite{Chodura-1982} when we take $T_{\text{i}} = 0$.

Setting $T_{\text{i}} = 0$, we expect the ion distribution function at the magnetic presheath entrance to be
\begin{align} \label{fcold-def}
f_{\infty\text{,cold}} \left( \vec{v} \right) = \frac{n_{\text{e}\infty}}{Z} \delta_{\text{Dirac}} \left( v_x \right) \delta_{\text{Dirac}} \left( v_y \right) \delta_{\text{Dirac}} \left( v_z - u_z \right) \text{,}
\end{align}
where $\delta_{\text{Dirac}}$ is the Dirac delta function. With the distribution function (\ref{fcold-def}), the solvability condition (\ref{solvability-vz}) is
\begin{align}
\frac{n_{\text{e}\infty}}{Zv_{\text{B}}^2} \geqslant  \int \frac{f_{\infty\text{,cold}}\left( \vec{v} \right)}{v_z^2} d^3 v = \frac{n_{\text{e}\infty}}{Z u_z^2} \text{,}
\end{align}
which leads to 
\begin{align}
u_{z} \geqslant v_{\text{B}} \text{.}
\end{align}
Therefore, the incoming ion flow must be at least sonic in the $+z$ direction, which to lowest order is the direction parallel to the magnetic field towards the wall. 
This condition is the small-$\alpha$ limit of the condition derived by Chodura in reference \cite{Chodura-1982}.

When the incoming distribution is given by equation (\ref{fcold-def}) and the Chodura condition is marginally satisfied, $u_z = v_{\text{B}}$, the term $\partial F_{\text{cold}} \left( \mu, \Omega \mu \right) / \partial U $ that appears in the numerator of $k_{3/2}$ is equal to zero, which means that $k_{3/2} =0$ and the correct form of the potential at $x\rightarrow \infty$ is given by equation (\ref{phisol2}). The value of $k_2$ in the cold ion limit, $k_{2\text{,cold}}$, is obtained from (\ref{k2}) re-expressed using the set of variables $(v_x, v_y, v_z)$. 
From (\ref{changevar-infty}), equation (\ref{k2}) becomes
\begin{align} \label{k2-cold}
k_{2,\text{cold}} = \frac{\Omega^2 e}{2v_{\text{B}}^2T_{\text{e}}}   \frac{ 3v_{\text{B}}^4   \int \frac{ f_{\infty\text{,cold}} \left( \vec{v} \right) }{v_z^4}  d^3 v  - \frac{n_{e\infty}}{Z} }{  \frac{n_{e\infty}}{Z} + \int f_{\infty\text{,cold}} \left( \vec{v} \right) \frac{v_x^2 + v_y^2 }{2v_z^2} d^3v  } \text{.}
\end{align}
Using (\ref{fcold-def}) and $u_z = v_{\text{B}}$, the second term in the denominator evaluates to zero and the first term in the numerator is evaluated using the result
\begin{align} \label{k2-cold-integralnum}
v_{\text{B}}^4 \int \frac{ f_{\infty\text{,cold}} \left( \vec{v} \right) }{v_z^4}  d^3 v = \frac{n_{e\infty}}{Z}  \text{.}
\end{align}
Inserting equation (\ref{k2-cold-integralnum}) into (\ref{k2-cold}), we obtain
\begin{align} \label{k2-cold-final}
k_{2,\text{cold}} = \frac{\Omega^2  e}{v_{\text{B}}^2 T_{\text{e}}} \text{.}
\end{align}
Inserting (\ref{k2-cold-final}) into equation (\ref{phisol2}), the electrostatic potential near $x\rightarrow \infty$ in the cold ion limit is
\begin{align}
\frac{e\phi \left( x \right) }{T_{\text{e}}} = - \frac{6 v_{\text{B}}^2/\Omega^2}{ \left( x+C_{2,\text{cold}} \right)^2} \text{.}
\end{align}
At sufficiently large $x$, $C_{2,\text{cold}} $ can be neglected and Chodura's result for the scaling of the potential at the magnetic presheath entrance $x\rightarrow \infty$ is recovered (this scaling is obtained from Chodura's paper \cite{Chodura-1982} by combining equations (22), (23), and the equation immediately after (24), and noting that Chodura's notation is $\psi = \pi/2 - \alpha$ and his derivation is valid for general $\psi$).

\section{Proof that $k_2 > 0$ and $q_2>0$} \label{app-k2>0}

In order to show that $k_2 > 0$, it is sufficient to show that
\begin{align} \label{k2>0equivalent}
 6\pi  \int_0^{\infty} \Omega d\mu  \int_{\Omega \mu}^{\infty}  \frac{F_{\text{cold}}(\mu, U) v_{\text{B}}^4 }{\left( 2\left( U - \Omega \mu \right) \right)^{5/2} } dU - \frac{ n_{e\infty} }{Z}  > 0 \text{.}
\end{align}
Remembering $v_z = \sqrt{2\left( U - \Omega \mu \right)}$ and equation (\ref{changevar-infty}), the integral in the first term can be recast as
\begin{align} \label{Reexpressed}
2\pi \int_0^{\infty} \Omega d\mu \int_0^{\infty}  \frac{F_{\text{cold}}(\mu, U) dU}{\left( 2\left( U - \Omega \mu \right) \right)^{5/2} } =  \int_0^{\infty}  \frac{ f_{\infty z} \left( v_z \right) }{v_z^{4} } dv_{z} \text{,}
\end{align}
where 
\begin{align} \label{Reexpressed-vz}
f_{\infty z} \left( v_z \right) = \int_{-\infty}^{\infty} dv_x  \int_{-\infty}^{\infty}  f_{\infty} \left( \vec{v} \right) dv_y \text{.}
%\int_{-\infty}^{\infty} dv_x \int_{-\infty}^{\infty}   f_{\infty} \left( \vec{v} \right) dv_y \text{.} %commented out
\end{align}

The marginal form of Chodura's condition (\ref{solvability}) can be expressed as
\begin{align} \label{Schwarz-Chodura}
 n_{e\infty} =  Z v_{\text{B}}^2 \int_0^{\infty}  \frac{ f_{\infty z} \left( v_z \right) }{v_z^2}   dv_z \text{.}
\end{align}
Then, by application of Schwarz's inequality we have the relation
\begin{align} \label{Schwarz-application}
\int_0^{\infty}  \frac{ f_{\infty z} \left( v_z \right) }{v_z^4} dv_z  \int_0^{\infty}  f_{\infty z} \left( v_z \right)  dv_z   \geqslant  \left( \int_0^{\infty}  \frac{ f_{\infty z} \left( v_z \right) }{v_z^{2} }  dv_z \right)^2  \text{,}
\end{align}
and from quasineutrality we have
\begin{align} \label{Schwarz-quasineutrality}
Z \int_0^{\infty}  f_{\infty z} \left( v_z \right)  dv_z   = n_{e\infty} \text{.}
\end{align}
Substituting (\ref{Schwarz-Chodura}) and (\ref{Schwarz-quasineutrality}) in (\ref{Schwarz-application}), we obtain 
\begin{align}
Z v_{\text{B}}^4 \int_0^{\infty}  \frac{ f_{\infty z} \left( v_z \right)  }{v_z^4} dv_z   \geqslant  n_{e \infty}  \text{.} 
\end{align}
Re-expressing the left hand side of the inequality in terms of $F(\mu, U)$ and $U$ by using (\ref{Reexpressed}), we obtain  
\begin{align} \label{k2>0-almost}
2\pi \int_0^{\infty} \Omega d\mu \int_{\Omega \mu }^{\infty}  \frac{ F_{\text{cl}}(\mu, U) v_{\text{B}}^4 }{\left( 2\left( U - \Omega \mu \right) \right)^{5/2} } dU \geqslant \frac{n_{e\infty}}{Z}    \text{.}
\end{align}
From (\ref{k2>0-almost}) we see that
\begin{align}
6\pi \int_0^{\infty} \Omega d\mu \int_{\Omega \mu }^{\infty}  \frac{ F_{\text{cl}}(\mu, U) v_{\text{B}}^4  }{\left( 2\left( U - \Omega \mu \right) \right)^{5/2} } dU - \frac{n_{e\infty}}{Z}    \geqslant    \frac{2n_{e\infty}}{Z}       >   0 \text{,}
\end{align}
from which (\ref{k2>0equivalent}) immediately follows. 

This proof can be straightforwardly adapted to show that $q_2 >0$, where $q_2$ is defined in equation (\ref{q2-def}). Again, it suffices to show that the numerator of equation (\ref{q2-def}) is positive,
\begin{align} \label{q2-equivalent}
v_{\text{B}}^4 \int_{\bar{x}_{\text{c}}}^{\infty}   \Omega d\bar{x}  \int_{\chi_{\text{M}} (\bar{x} )}^{\infty}  \frac{  F_{\text{cl}} \left( \mu_{\text{gk}} \left( \bar{x}, \chi_{\text{M}} \left( \bar{x} \right) \right) , U \right)   }{\sqrt{2\left( U - \chi_{\text{M}} (\bar{x}) \right)}} \Delta \left[ \frac{1}{v_{x0}^3} \right]  dU - \frac{n_{\text{e}0}}{Z} > 0 \text{.}
\end{align}
The integral can be re-expressed as
\begin{align} 
& \int_{\bar{x}_{\text{c}}}^{\infty}   \Omega d\bar{x}  \int_{\chi_{\text{M}} (\bar{x} )}^{\infty}  \frac{  F_{\text{cl}} \left( \mu_{\text{gk}} \left( \bar{x}, \chi_{\text{M}} \left( \bar{x} \right) \right) , U \right)   }{\sqrt{2\left( U - \chi_{\text{M}} (\bar{x}) \right)}} \Delta \left[ \frac{1}{v_{x0}^3} \right]  dU \nonumber \\
& = \int_{\bar{x}_{\text{c}}}^{\infty}   \Omega d\bar{x}  \int_{\chi_{\text{M}} (\bar{x} )}^{\infty}  \frac{ F_{\text{cl}} \left( \mu_{\text{gk}} \left( \bar{x}, \chi_{\text{M}} \left( \bar{x} \right) \right) , U \right) }{\sqrt{2\left( U - \chi_{\text{M}} (\bar{x}) \right)}} dU \nonumber \\
& ~ \times \int_{-\infty}^{0}  \frac{3}{v_x^4} \hat{\Pi} \left(  v_x ,  - V_x \left( 0, \bar{x}, \chi_{\text{M}} \right) - \Delta v_{x0}  , - V_x \left( 0, \bar{x}, \chi_{\text{M}} \right)  \right) dv_x \nonumber \\
 & = 3 \int  \frac{f_0 ( \vec{v} ) }{v_{x}^4}  d^3v 
 \nonumber \\
 & = 3 \int_{-\infty}^0 \frac{f_{0x} ( v_x ) }{v_{x}^4}  dv_x  \text{,}
\end{align}
where $f_{0x}(v_x)$ is defined in equation (\ref{f0x-def}).
The marginal form of Bohm's condition is
\begin{align}
Z v_{\text{B}}^2 \int_{-\infty}^0 \frac{f_{0x} ( v_x ) }{v_{x}^2}  dv_x = n_{\text{e}0}  \text{}
\end{align}
and quasineutrality is
\begin{align}
Z  \int_{-\infty}^0 f_{0x} ( v_x )  dv_x = n_{\text{e}0}  \text{.}
\end{align}
Proceeding in an analogous way to the previous derivation, we conclude that
\begin{align}
v_{\text{B}}^4 \int_{\bar{x}_{\text{c}}}^{\infty}  \Omega d\bar{x} \int_{\chi_{\text{M}}(\bar{x})}^{\infty}  \frac{F_{\text{cl}} \left( \mu_{\text{gk}} (\bar{x}, \chi_{\text{M}}(\bar{x}) , U \right)}{\sqrt{ 2\left( U - \chi_{\text{M}} ( \bar{x} ) \right)  } } \Delta \left[ \frac{1}{v_{x0}^{3}}\right] dU - \frac{n_{\text{e}0}}{Z}   \geqslant    \frac{2n_{\text{e}0}}{Z}       >   0 \text{,}
\end{align}
from which (\ref{q2-equivalent}) immediately follows.

\section{Neglecting the contribution of type II closed orbits near $x=0$} \label{app-notypeIIclosed}

The expansion of the closed orbit density near $x=0$ relies on distinguishing type I and type II effective potential curves. In Section \ref{subsec-expansion-near0} we omitted the contribution of closed orbits associated with type II curves, denoted $n_{\text{i,cl,II}} (x) $. 
We proceed to show that this contribution is negligible.

From equation (\ref{ni-closed-expanded}), and using the expansion (\ref{VxII}) of $V_x$ near the stationary maximum $x_{\text{M}}$, we obtain an expression for the contribution to the density near $x=0$ due to ions in approximately closed type II orbits,
\begin{align} \label{ni-closed-near0}
n_{\text{i,cl,II}} (x) \simeq   2\int_{\bar{x}_{\text{m}}(x)}^{\bar{x}_{\text{m,I}}} \Omega \sqrt{| \chi_{\text{M}}'' |}  \left| x - x_{\text{M}} \right|  d\bar{x} 
 \int_{\chi_{\text{M}} (\bar{x})}^{\infty} \frac{ F\left(\mu_{\text{gk}} \left(\bar{x}, \chi_{\text{M} } \right), U \right) }{\sqrt{2\left( U - \chi_{\text{M}}(\bar{x}) \right) }} dU  \text{.}
\end{align}
The upper limit of integration in $\bar{x}$ is $\bar{x}_{\text{m,I}}$, which is the value of $\bar{x}$ above which the effective potential is a type I curve.
%In equation (\ref{ni-closed-near0}) we integrate over all possible values of $\bar{x}$ for which $\chi$ is a type II curve, and explicitly include the Heaviside function that is present in the definition of $f_{\text{cl}}$ in equation (\ref{fclosed}). 
%From Section \ref{subsec-effpot-types} we expect $x_{\text{c}} \neq 0$ and, for sufficiently small $x$, $x < x_{\text{c}} < x_{\text{t,M}} $, therefore $\Theta \left( (x-x_{\text{M}} )(x_{\text{t,M}} - x) \right) = \Theta ( x-x_{\text{M}} )$.
It is easier to express the integral in (\ref{ni-closed-near0}) by changing variables from $\bar{x}$ to $x_{\text{M}}$ (for type II curves, $x_{\text{M}}$ depends on the value of $\bar{x}$). The Jacobian of this change of variables can be obtained using the equation for a stationary maximum, which is $\chi'(x_{\text{M}}, \bar{x} ) = 0$. Rearranging equation (\ref{stationary-points}) evaluated at the stationary point $x_{\text{M}}$, we have
\begin{align} \label{xbar-xM}
\bar{x} = x_{\text{M}} + \frac{\phi' ( x_{\text{M}} )}{\Omega B}  \text{.}
\end{align}
Differentiating this equation with respect to $x_{\text{M}}$, we obtain $\left| \partial \bar{x} / \partial x_{\text{M}} \right| =  |\chi_{\text{M}}''| / \Omega^2$. Then, the integral (\ref{ni-closed-near0}) can be written in terms of $x_{\text{M}}$. The integration limit $ \bar{x} = \bar{x}_{\text{m,I}}$ corresponds to $x_{\text{M}} = 0$, while the integration limit $\bar{x} = \bar{x}_{\text{m}}(x)$ corresponds to $x_{\text{M}} = x$.

For small $x$, we can Taylor expand the integrand near $\bar{x} = \bar{x}_{\text{m,I}}$ (which corresponds to $x_{\text{M}} = 0$) and retain only the leading order,
\begin{align} \label{ni-closed-near0-expanded}
n_{\text{i,cl,II}} (x)  \simeq &  2\int_0^x   ( x - x_{\text{M}} )  \frac{|\chi'' (0, \bar{x}_{\text{m,I}} ) |^{3/2}}{\Omega} d x_{\text{M}} 
  \int_{\chi_{\text{M}} (\bar{x})}^{\infty} \frac{ F_{\text{cl}} \left(\mu_{\text{gk}} \left(\bar{x}_{\text{m,I}}, \chi_{\text{M}} \left(\bar{x}_{\text{m,I}} \right) \right), U \right) }{\sqrt{2\left( U - \chi_{\text{M}} (\bar{x}_{\text{m,I}}) \right) }} dU \nonumber \\
 \simeq & x^2  \frac{|\chi'' (0, \bar{x}_{\text{m,I}} ) |^{3/2}}{\Omega}     
 \int_{\chi_{\text{M}} \left(\bar{x}_{\text{m,I}} \right) }^{\infty} \frac{ F_{\text{cl}} \left(\mu_{\text{gk}} \left(\bar{x}_{\text{m,I}}, \chi_{\text{M}} \left(\bar{x}_{\text{m,I}} \right) \right), U \right) }{\sqrt{2\left( U - \chi_{\text{M}} (\bar{x}_{\text{m,I}}) \right) }} dU 
   \text{.}
\end{align}
Hence, the contribution from type II closed orbits near $x=0$ is proportional to $ x^2 $ and therefore subdominant compared to $x$, making it negligible.
In fact, when $\bar{x}_{\text{m,I}} \rightarrow \infty$, we expect the contribution to be even smaller than (\ref{ni-closed-near0-expanded}) because the distribution function is exponentially small for $U \rightarrow \infty$.

\section*{References}
\bibliography{gyrokineticsbibliography}{}
\bibliographystyle{unsrt}

\end{document}